\documentclass[11pt,a4paper]{article}
\pdfoutput=1
\usepackage{jheppub}
\usepackage{cite}
\usepackage{physics}
\usepackage[normalem]{ulem}
\usepackage{braket}
\usepackage{caption}
\usepackage{subcaption}

\usepackage{multirow, graphicx,amssymb,url,mathrsfs,amsmath}
\usepackage{wrapfig,boxedminipage,epsfig}
\usepackage{amsxtra,amstext,latexsym,dsfont,amsfonts}
\usepackage{color}
\usepackage{float}
\usepackage{slashed}
\usepackage{calligra}
\usepackage{enumerate}
\DeclareFontShape{T1}{calligra}{m}{n}{<->s*[2.2]callig15}{}
\DeclareMathAlphabet{\mathcalligra}{T1}{calligra}{m}{n}

\graphicspath{{./figs/}}

\newcommand{\be}{\begin{equation}}
\newcommand{\ee}{\end{equation}}
\newcommand{\bea}{\begin{eqnarray}}
\newcommand{\eea}{\end{eqnarray}}

\title{Black Hole Multi-Entropy Curves\\
{\it \Large - Secret entanglement between Hawking particles -}}

 \author[a, b]{Norihiro Iizuka}
 
 \author[c]{Simon Lin} 
 
 \author[d]{and Mitsuhiro Nishida}
 
\affiliation[a]{\it Department of Physics, National Tsing Hua University, Hsinchu 300044, Taiwan}

\affiliation[b]{\it Yukawa Institute for Theoretical Physics, Kyoto University, Kyoto 606-8502, Japan}

\affiliation[c]{\it New York University Abu Dhabi, P.O. Box 129188, Abu Dhabi, United Arab Emirates}

\affiliation[d]{\it National Institute of Technology, Yuge College, Ehime 794-2593, Japan}
 
\emailAdd{iizuka@phys.nthu.edu.tw}
\emailAdd{simonlin@nyu.edu}
\emailAdd{mnishida124@gmail.com}

\abstract{
We investigate the multi-partite entanglement structure of an evaporating black hole and its Hawking radiation by dividing the radiation into finer subsystems. We approximate an evaporating black hole and its radiation with a Haar-random state for this purpose. Using the multi-entropy of these configurations, we define a {\it black hole multi-entropy curve}, which describes how the multi-entropy changes during the black hole evaporation. This black hole multi-entropy curve is a natural generalization of the Page curve since the multi-entropy reduces to the entanglement entropy for the bi-partite case. The multi-entropy curve keeps increasing in the early time. It reaches the maximum value at the {\it multi-entropy time}, which is later than the Page time, and starts to decrease. However, it does not decrease to zero at the end of the black hole evaporation. This non-zero value of the multi-entropy represents the secret entanglement between Hawking particles. }

\begin{document}
\maketitle

\section{Introduction}

Completing a quantum theory of gravity, consisting of the consistency of quantum mechanics and gravity is a major unsolved problem. In particular, the black hole information paradox \cite{Hawking:1976ra} is a touchstone for our understanding of quantum gravity theory.
In the black hole information paradox, the Page curve \cite{Page:1993df, Page:1993wv}, which represents how the entanglement between Hawking radiation and an evaporating black hole changes, plays a key role in our understanding of physics since it deviates significantly from Hawking's prediction \cite{Hawking:1975vcx} even when an evaporating black hole is large enough such that semi-classical approximation is expected to hold. 
Note that the Page curve is a measure of {\it bi-partite} entanglement, {\it i.e.}, an entanglement between an evaporating black hole and total Hawking radiation. 

\begin{figure}[t]
    \centering
    \includegraphics[width=9cm]{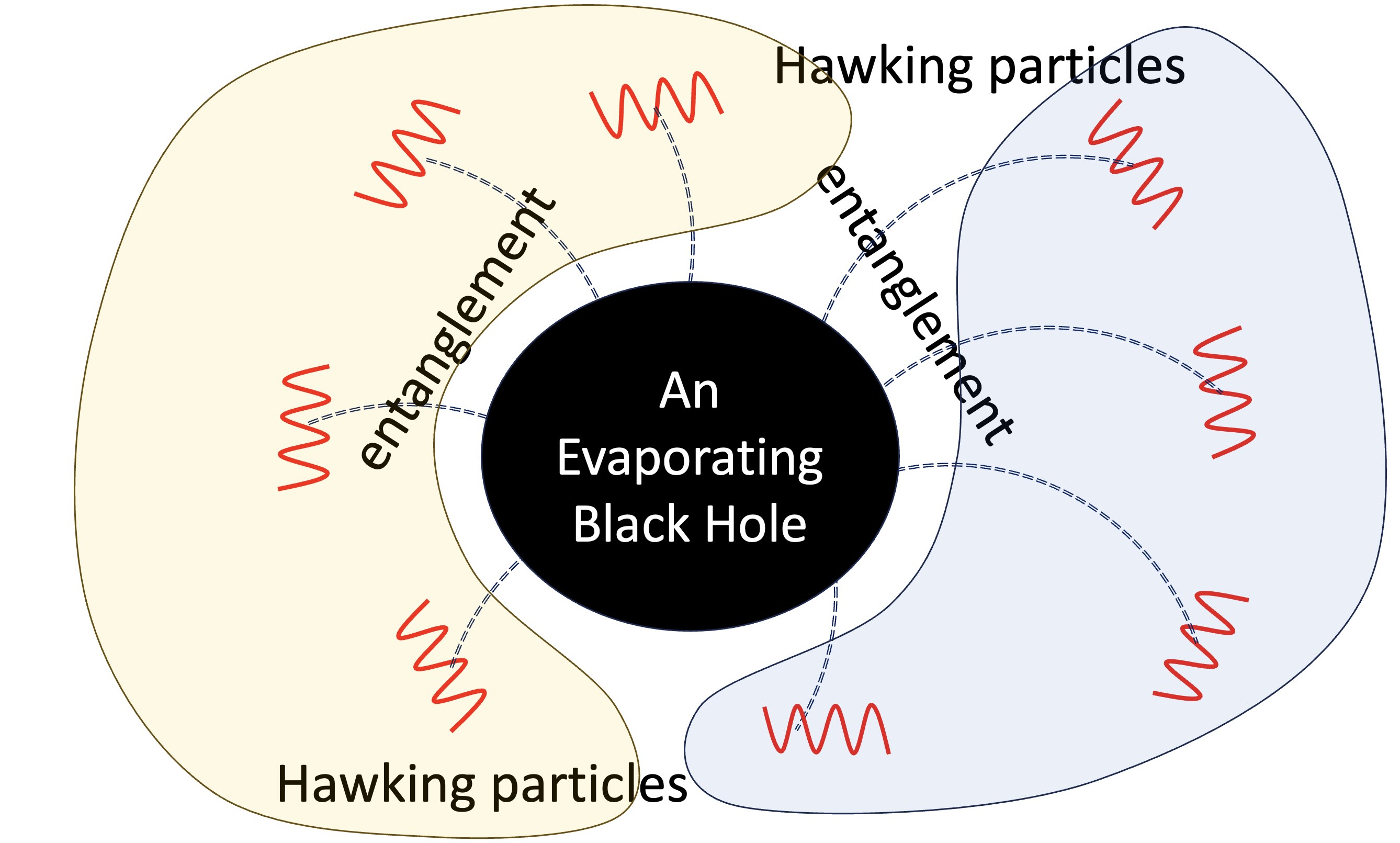}
    \caption{In the Page curve \cite{Page:1993df, Page:1993wv}, all the Hawking radiation is considered as one system, but it is possible to consider muti-partite quantum entanglement by dividing Hawking radiation into multiple subsystems according to the direction of the flying Hawking particles.}
    \label{fig:multi-rad}
\end{figure}

On the other hand, a system with many degrees of freedom can be divided into multiple subsystems, in which they exhibit richer entanglement properties that are not captured by simple bi-partite measures\footnote{See \cite{Walter:2016lgl} for a review on multi-partite entanglement.}. For example, Hawking radiation can be divided into different subsystems based on the direction of the emitted particles, rather than as a whole. 
This division of the radiation degrees of freedom by the direction where they flew is quite natural in terms of locality in the theory of gravity. See Fig.~\ref{fig:multi-rad}.  By dividing Hawking radiation in this way, 
and by carefully studying the multi-partite entanglement structure of them, 
it is expected that one can obtain more detailed information of the total quantum state, a feature not available with a bi-partite measure such as von Neumann entropy considered in the Page curve\footnote{From a bi-partite viewpoint, using mutual information, this question is studied in \cite{Iizuka:2013ria}. See also \cite{Hollowood:2021nlo, Hollowood:2021lsw}.}. 
Especially the time evolution of such multi-partite measures should tell how the entanglement between Hawking radiation changes by time.

Hence, in this paper, we investigate the multi-partite quantum entanglement of Hawking radiation and an evaporating black hole by dividing the radiation into finer subsystems. The goal of this paper is to reveal a more detailed structure of quantum entanglement among different subsystems of the radiation, which are considered as a single system in the Page curve. 

Motivated by holographic settings, several tri-partite entanglement measures have been studied, which include entanglement negativity \cite{Vidal:2002zz}, reflected entropy \cite{Dutta:2019gen}, entanglement of purification \cite{Takayanagi:2017knl} and odd entanglement entropy \cite{Tamaoka:2018ned}. They are tri-partite measures in that they diagnose entanglement between three parties \cite{Zou:2020bly}. 
In addition, recently progress has been made toward the systematic classification of entanglement multi-partite measures in \cite{Gadde:2022cqi, Gadde:2023zzj}\footnote{See also \cite{Balasubramanian:2014hda, Cui:2018dyq, Bao:2018gck, Bao:2019zqc, Balasubramanian:2024ysu} for other progresses on multi-partite entanglement measures in holography.}. It is now clear that many multi-partite measures can be obtained by taking the density matrix contraction in some generic ways. 
Thus, above mentioned measures, such as entanglement negativity, are only part of them. Among these many new muti-partite entanglement measures, a special new measure called multi-entropy \cite{Gadde:2022cqi} is of special interest since it symmetrically treats all its subsystems, and it reduces to entanglement entropy in the bi-partite case.

\begin{figure}[t]
    \centering
    \includegraphics[width=9cm]{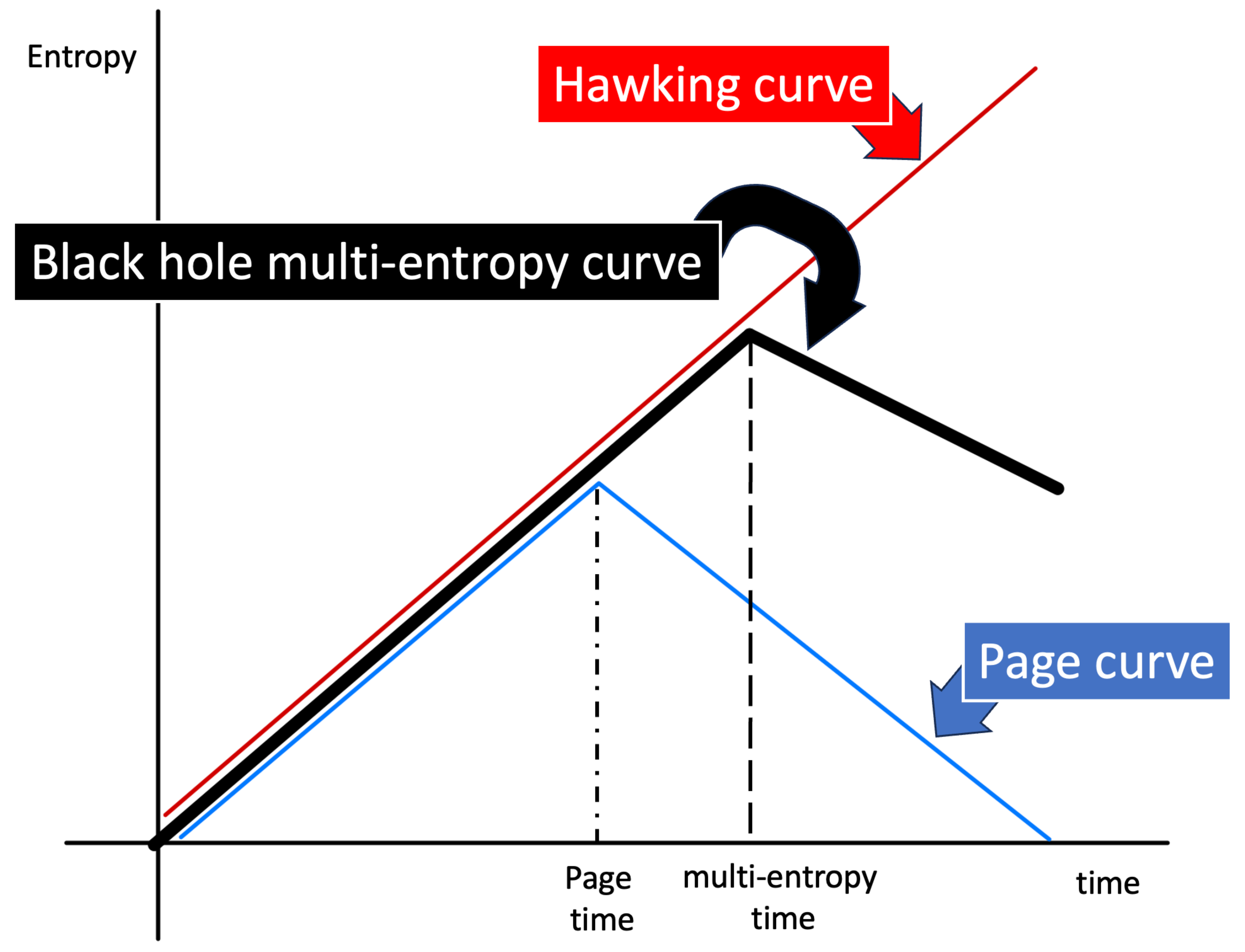}
    \caption{Typical behavior of a black hole multi-entropy curve for $\mathtt{q} \ge 3$ is shown in the black curve. The red curve represents the Hawking curve and the blue one represents the Page curve. }
    \label{fig:BHmultiScurve}
\end{figure}

In this paper, we focus on the multi-entropy by dividing Hawking radiation into $\mathtt{q}-1$ subsystems. An evaporating black hole itself forms one subsystem, thus we have $\mathtt{q}$ subsystems in total. We study the time-evolution of $\mathtt{q}$-partite multi-entropy during Hawking radiation as a generalization of the Page curve. Especially we will focus on the following two points:
\begin{enumerate}
\item {\it Black hole multi-entropy curves}, which describe how the above-mentioned multi-entropy changes as a function of time during the entire time evolution of an evaporating black hole, from early radiation to the complete evaporation of a black hole. This multi-entropy curve reduces to the Page curve if $\mathtt{q}=2$. Typical behavior of a black hole multi-entropy curve for $\mathtt{q} \ge 3$ is shown Fig.~\ref{fig:BHmultiScurve}.   Generic properties of them are described in detail in section \ref{q=3n=2multicurve}. Fig.~\ref{fig:multiq=3n=2}, \ref{fig:MultiandPage}, \ref{fig:multiq=3n=3}, \ref{fig:multiq=4n=2} are other examples of black hole multi-entropy curves. More details, including the precise definition of the ``time'' and ``entropy'' in Fig.~\ref{fig:BHmultiScurve}, are given in section \ref{sec:subleadinggeneral}.

\item In the above multi-entropy curves, especially at the very early stage limit of the evaporation, we expect Hawking's prediction \cite{Hawking:1975vcx} to be correct. This means that in the limit where the black hole dimension is large, $d_{\rm BH} \to \infty$, there is no correlation between Hawking radiation particles.
 However, we also expect if we look at subleading corrections in $1/d_{\rm BH} \sim e^{-S_{\rm BH}}$ expansion,  we will start to see a deviation from Hawking's prediction and the correlation between radiation particles.
  These correlations are the secret relationship between Hawking particles which is not observed at $d_{\text{BH}}\to\infty$ \cite{Hawking:1975vcx}. 
\end{enumerate}

For the actual computations, we will model the evaporating black hole using a single multi-partite random tensor. By using a random tensor model, we will work out the multi-entropy in detail to see the difference between the generic random state and Hawking's semiclassical predictions. Let us add a few comments on why we approximate an evaporating black hole and radiation with a generic random state. We do not know how quantum gravity theory works in general. However, we do know through the gauge/gravity correspondence \cite{Maldacena:1997re} that quantum gravity in asymptotic AdS is equivalent to ordinary quantum mechanical systems living on the boundary, which do not contain gravity. Furthermore, the boundary theories, which are, in general, a large $N$ Yang-Mills theories, are expected to be very chaotic \cite{Sekino:2008he, Shenker:2013pqa, Maldacena:2015waa} at the deconfinement phase which is dual to a black hole dominant phase \cite{Witten:1998qj, Witten:1998zw}. Thus, such chaotic boundary states can be very well approximated by random states \cite{Srednicki:1994mfb, Cotler:2016fpe} and thus, the degrees of freedom corresponding to an evaporating black hole and Hawking radiation must be very well approximated by a random state. This is the main reason why we approximate an evaporating black hole and radiation by a random state. 

In terms of random tensors, the work of Page \cite{Page:1993df, Page:1993wv} can be interpreted as the entanglement entropy of a single random tensor with two subscripts, one subscript for a black hole and the other one for Hawking radiation. 
A single random tensor with three or more subscripts can have a multi-partite entanglement structure between subsystems, and one can compute entanglement negativity \cite{Shapourian:2020mkc, Kudler-Flam:2021efr} and reflected entropy \cite{Akers:2021pvd, Akers:2022max, Akers:2022zxr, Akers:2023obn, Akers:2024pgq} in random tensor models.

In this paper, we calculate multi-entropies in a single random tensor model for black hole multi-entropy curves. For that purpose, let us consider the following setting: we interpret a single random tensor with $\mathtt{q}$ subscripts as a system consisting of $\mathtt{q}-1$ Hawking radiation and a black hole. We fix the bond dimension of total Hilbert space and vary the bond dimensions for Hawking radiation to draw the black hole multi-entropy curves. Two important properties of the black hole multi-entropy curves, as seen in Fig.~\ref{fig:BHmultiScurve}, are as follows. 1) Multi-entropy time, which is a point when the multi-entropy becomes maximal, is later than Page time for entanglement entropy. 2) The multi-entropy is not only nonzero but large even if a black hole evaporates completely. This is due to the entanglement between Hawking radiation. Note that the entanglement between Hawking radiation is not visible in the semi-classical approximation \cite{Hawking:1975vcx}.

The structure of this paper is as follows. After reviewing random tensor techniques and multi-entropy in Section \ref{sec2}, in Section \ref{secRenyimultientropyn2q3} we compute $n=2$ R\'enyi multi-entropy for a single tri-partite random tensor model with $\mathtt{q}=3$ and draw a black hole multi-entropy curve. This is the simplest example of the R\'enyi multi-entropy. Two other examples of black hole multi-entropy curves are also shown in Section \ref{sec:org1e539bc}. In Section \ref{sec:analytic}, we give analytic expressions for the mutli-entropy with arbitrary $n$ and $\mathtt{q}$, focusing on the limit where either the dimension of the black hole or the radiation is large, which corresponds to the very early stage and the very late stage of black hole evaporation, respectively. We end in Section \ref{sec:Conclusion} with conclusion and discussion.  In Appendix~\ref{negativityappendix} we compute averaged replica partition functions for R\'enyi entanglement negativity and R\'{e}nyi reflected entropy in a single tri-partite random state and compare them with the R\'{e}nyi multi-entropy curve. In Appendix~\ref{app:conjecturecoefs} we provide our reasoning for the conjectured formulae and partial results regarding two coefficients that capture the subleading behavior of the late-time multi-entropy. In Appendix~\ref{app:proof} we present proof about the replica partition function we use in the main text.

\section{Prerequisites}\label{sec2}

In this section, we review the necessary background materials that are relevant for our calculation, setting up some of our notations at the same time.
Our main results utilizes techniques from random tensor networks \cite{Hayden:2016cfa} in the replica trick of (R\'enyi) multi-entropy \cite{Gadde:2022cqi}.

\subsection{Random tensor and the permutation group method}
\label{sec:RT-group}
Here we give a quick review of random tensor techniques and notations used throughout this paper. For more detailed review please refer to \cite{Hayden:2016cfa,Akers:2021pvd}.
Consider a rank-\(k\) tensor \(T_{\mu_1,\mu_2,\cdots,\mu_k}\). This tensor defines a vector in the product Hilbert space \(\mathcal{H} = \mathcal{H}_1 \otimes \mathcal{H}_2 \otimes \cdots \otimes \mathcal{H}_k\):
\begin{equation}
\ket{\psi} = \sum_{\{\mu_k\}} T_{\mu_1,\mu_2,\cdots,\mu_k} \ket{\mu_1} \ket{\mu_2} \cdots \ket{\mu_k}.
\end{equation}
The dimensions of the individual Hilbert spaces, denoted 
\begin{align}
\label{dfordimension}
d_i = \dim(\mathcal{H}_i)
\end{align} 
are called the {\it bond dimensions} of \(T\).
By a {\it random tensor} we mean that \(T_{\mu_1,\mu_2,\cdots,\mu_k}\) is sampled from a uniform random ensemble on the product Hilbert space.
This can be achieved by acting with a unitary \(U\) drawn from a Haar-random measure on some fixed anchor state \(\ket{\phi}\).

In the following we will introduce a standard diagrammatic notation from random tensor networks that will aid our computation of entropies.
For example, suppose that we start with a bi-partite state \(\ket{\psi}_{AB}= \sum_{i}^{d_A}\sum_{j}^{d_B}c_{ij}\ket{A_i}\ket{B_j}\).
The density matrix \(\rho_{AB}=\ket{\psi}_{AB}\bra{\psi}_{A^*B^*}\) can be represented as a four-legged tensor
\begin{equation}
  \begin{split}
     \includegraphics[scale=0.3]{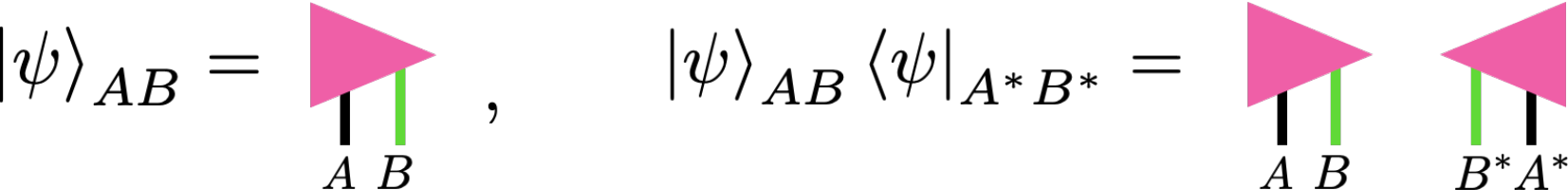} 
  \end{split}\quad ,
\end{equation}
where we use purple triangles to represent the random state \(\ket{\psi}_{AB}\) and we use black (green) lines to represent the subspaces associated to subsystem \(A \, (B)\).
Contraction over subspaces is indicated by connecting corresponding legs.
For example, the reduced density matrix \(\rho_A = \Tr_B \ket{\psi}_{AB}\bra{\psi}_{A^*B^*}\) can be expressed as
\begin{equation}
  \begin{split}
     \includegraphics[scale=0.3]{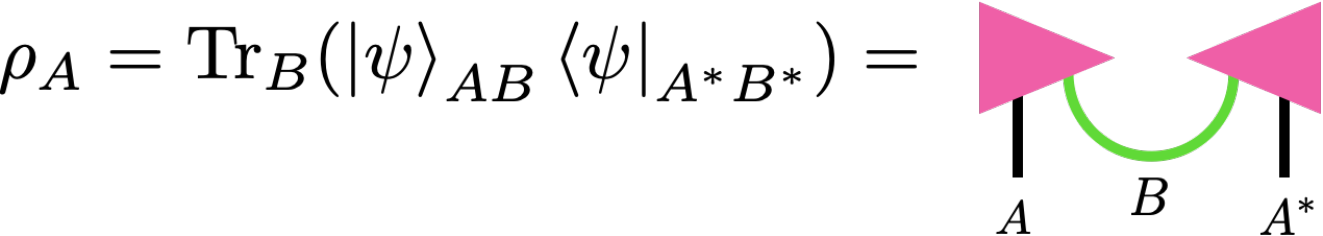} 
  \end{split} \quad.
\end{equation}

In the setting of quantum information, one is often interested in entanglement measures that can be expressed as contractions between multiple copies of density matrices \cite{Gadde:2023zzj}.
Let's say we have \(n\) copies of the density matrix \(\rho=\ket{\psi}\bra{\psi}\). The ensemble average over the Haar measure can be readily done using Schur's lemma \cite{Harrow:2013nib}:
\begin{equation}
\overline{\left(\ket{\psi}\bra{\psi}\right)^{\otimes n}} = \int [DU] \left(U  \ket{\phi} \bra{\phi}   U\right)^{\otimes n} \propto \sum_{g\in S_n} \Sigma(g) ,
\end{equation}
where  \(g\) is an element of the symmetry group \(S_n\) and \(\Sigma(g)\) is the ``twist operator" whose action permutes the replica Hilbert space according to \(g\).
In the diagrammatic notation, we write this as
\begin{equation}
\label{bluenorm}
  \begin{split}
   \includegraphics[scale=0.3]{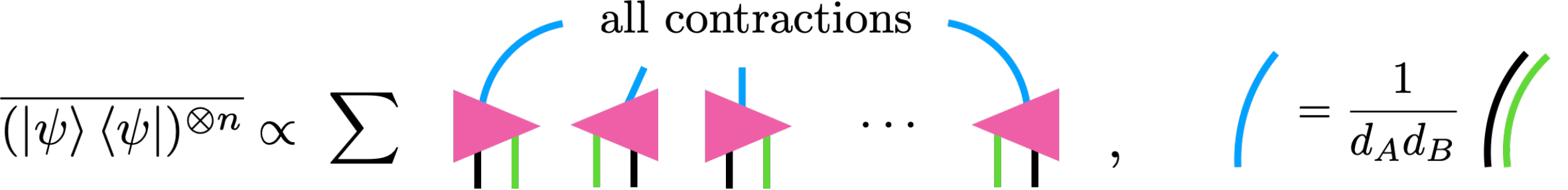} 
  \end{split}\quad,
\end{equation}
where we represent the summation over the group elements using blue lines that goes on top of the tensors. Note that we have to sum over all pairwise contractions between bras and kets.\footnote{Effectively this is the same as treating $\ket{\psi}$ as a uniform Gaussian random state. The pairwise contractions between bras and kets corresponds to the Wick contractions of the Gaussian random variable. The Haar random state is then obtained by normalizing the Gaussian random state $\ket{\psi}$. See also Footnote \ref{ft:Gaussian}.}
Note also that the state \(\ket{\psi}\) is not normalized. Normalization can be achieved ``on average" by dividing out the average of the norm \(\ket{\psi}\).\footnote{For large bond dimensions this is sufficient for our purposes. See the analysis at the end of this subsection for the finite \(d\) corrections.}
This is achieved for by including a factor of \(1/d_A d_B\) to each blue contraction line when summing over \(S_n\), as shown above.

The lower end of the diagram is determined by the contraction pattern of the entanglement measure we wish to compute.
For example, to compute the purity of the reduced density matrix \(\Tr \rho_A^2\), we use
\begin{equation}
\label{puritydiagram}
  \begin{split}
   \includegraphics[scale=0.3]{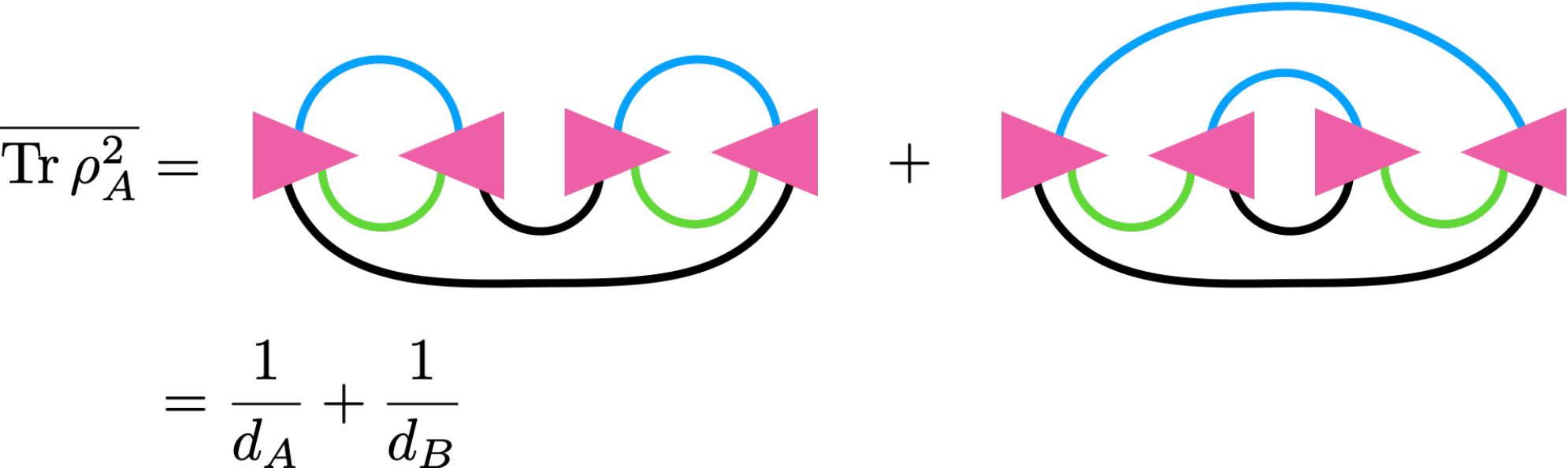} 
  \end{split}\quad , 
\end{equation}
where one obtains a factor of \(d_A \, (d_B)\) from tracing out \(\mathcal{H}_A \, (\mathcal{H}_B)\) for each completed black (green) loop. The first diagram contains one black (two green) loops and thus yielding $d_A d_B^2$, and the second diagram contains two black (one green) loops, thus yielding $d_A^2 d_B$. Taking into account the normalization of \eqref{bluenorm}, we obtain the result of \eqref{puritydiagram}. 

Note that this matches with the exact result \cite{Lubkin:1978nch} 
\be
\overline{\Tr \rho_A^2} = \frac{d_A+d_B}{1+d_Ad_B}
\ee in the limit of large bond dimensions, $d_Ad_B \to \infty$. 

The above diagrammatic procedure can be repackaged into the following ``replica partition functions''
\begin{equation}
\overline{Z_n} \equiv \overline{\Bra{\psi}^{\otimes n}_{AB}\Sigma_A(g_A)\Sigma_{B}(g_B)\Ket{\psi}^{\otimes n}_{AB}} = \sum_{g\in S_n} d_A^{-d(g,g_A)} d_B^{-d(g,g_B)},
\end{equation}
where the \(\Sigma_{A,B}\) are the twist operators implementing the permutation action of \(g_{A/B}\) on the respective Hilbert spaces.
The elements \(g_{A/B}\) are determined by the specific entanglement measure we wish to compute and diagrammatically, the elements \(g_{A/B}\) specifies the lower end of the diagrams, {\it i.e.,} how black, green, etc. lines are contracted. 
For example, to compute \(\Tr \rho_A^n\) (which arises from R\'enyi entropies), we use \(g_A = \tau_n \equiv (12\cdots n)\), the full cyclic permutation in \(S_n\) and \(g_B=e\), the identity element.
The exponential weights of the summand is given by the {\it Cayley distance} of \(S_n\)
\begin{equation}
d(g,h) = n - \#(gh^{-1}),
\end{equation}
where \(\#(\cdot)\) is a function that counts the total number of cycles (including trivial cycles) in a group element\footnote{Note that the Cayley distance $d(g,h)$ is different from dimensions of Hilbert spaces (\ref{dfordimension}).}. 
\(d(g,h)\) can be thought of as a measure of distance on the symmetric group $S_n$, in the sense that it is equal the minimal number of swaps (2-element transpositions) it takes to go from \(g\) to \(h\).\footnote{For more detailed review of the properties of permutation group and Cayley distance we refer the reader to Appendix A of \cite{Akers:2021pvd}.}

The averaged R\'enyi entropy of \(\rho_A\) can be written as
\begin{align}
\overline{S_n(\rho_A)} &\equiv \frac{1}{1-n}\overline{\log\frac{\Tr \rho^n_A}{(\Tr \rho_A)^n}} = \frac{1}{1-n}\overline{\log \frac{Z_n}{(Z_1)^n}} \label{eq:Sn_avg1} \\
&= \frac{1}{1-n}\left[\log \frac{\overline{Z_n}}{\overline{Z^n_1}}+\sum_{m=1}^\infty\frac{(-1)^m}{m}\left(\frac{\overline{(\delta Z^n_1)^m}}{(\overline{Z^n_1})^m}-\frac{\overline{(\delta Z_n)^m}}{(\overline{Z_n})^m}\right)\right] \label{eq:Sn_avg2},
\end{align}
where \(\delta Z \equiv Z-\overline{Z}\) is the deviation.
What we have done passing from \eqref{eq:Sn_avg1} to \eqref{eq:Sn_avg2} is replacing the averaged entropy as the ratio of the averaged partition functions \(\overline{Z_n}\).
There will be corrections resulting from this replacement, as indicated as the series expansion in \eqref{eq:Sn_avg2}.
These corrections are controlled by the variance (or its higher order equivalent), which are suppressed in large bond dimensions for random tensors \cite{Hayden:2016cfa,10.1093/acprof:oso/9780199535255.001.0001}.
Hence, in the limit of large bond dimensions \(d_\mathcal{H}\gg 1\), it suffices to approximate the R\'enyi entropy by
\begin{equation}
\overline{S_n(\rho_A)} \underset{d_\mathcal{H} \gg 1}{\approx} \frac{1}{1-n}\log \frac{\overline{Z_n}}{\overline{(Z_1)^n}}.
\end{equation}

\subsection{Multi-entropy}
\label{sec:multi-entropy}
\begin{figure}
  \centering
  \includegraphics[width=.9\textwidth]{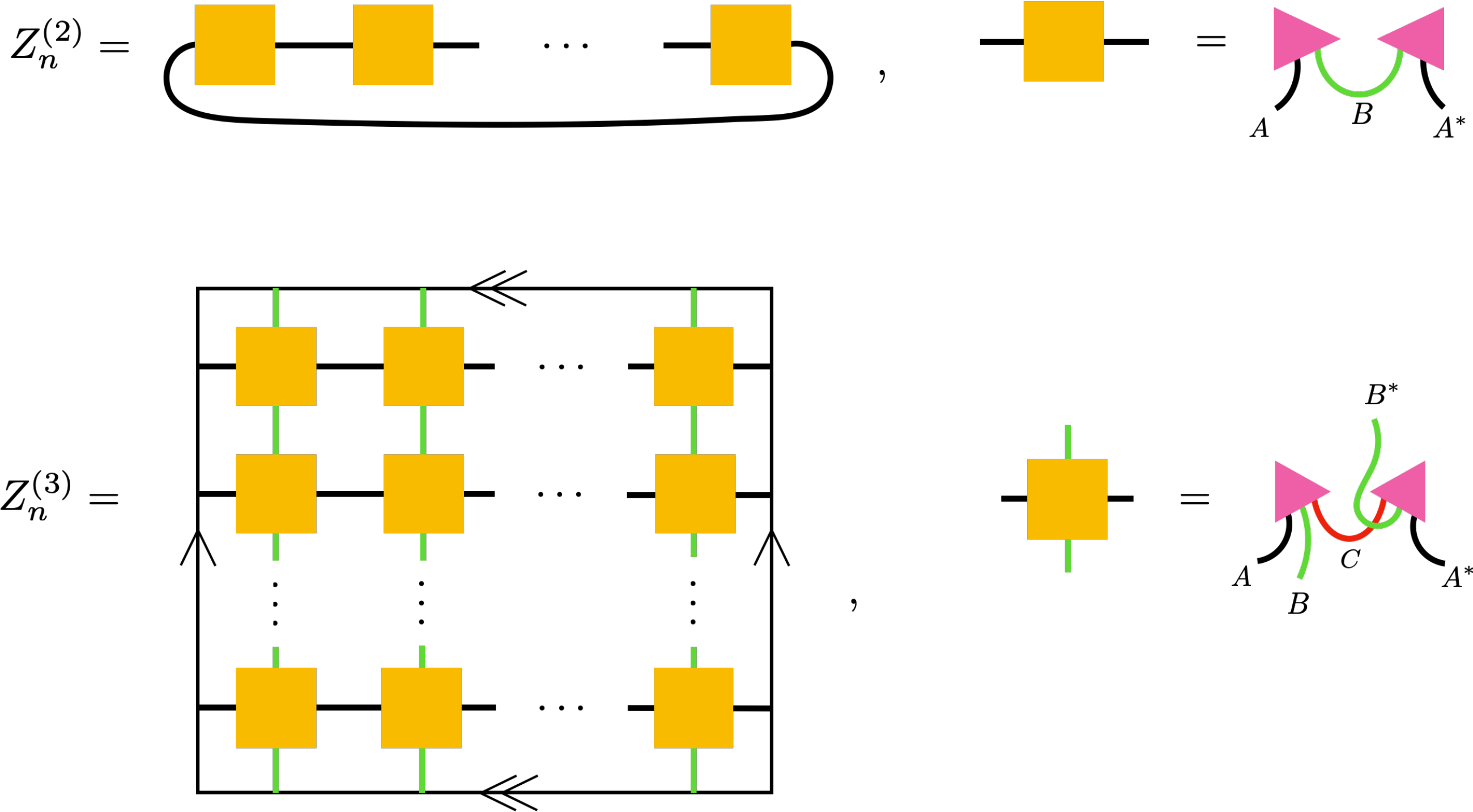}
  \caption{The contraction pattern that leads to \(\mathtt{q}=2\) (top) and \(\mathtt{q}=3\) (bottom) R\'enyi multi-entropies. We use the orange squares to represent the reduced density matrices \(\rho\). The setup can be described as follows: We lay out a \((\mathtt{q}-1)\)-dim hypercube lattice and we identify the opposing ends of each spatial direction (so it is really a \((\mathtt{q}-1)\)-torus). We place on each lattice point  reduced density matrix \(\rho\). Since \(\rho\) is obtained from tracing out one party from a \(\mathtt{q}\)-partite state, it has \((\mathtt{q}-1)\) pairs of free legs. We orient each pair of legs along a specific axis of the hypercube and we contract the density matrices along the corresponding direction (recall that the ends are identified so it is a cyclic contraction). This process is repeated for each axis and we obtain the figure at hand.}
  \label{fig:contraction-multi}
\end{figure}

The mutli-entropy \cite{Gadde:2022cqi,Gadde:2023zzj} is a symmetric multi-partite entanglement measure that can be thought of as a generalization of the entanglement entropy.
Consider a \(\mathtt{q}\)-partite state \(\ket{\psi_{i_1 i_2\cdots i_\mathtt{q}}}\) with bond dimensions \(d_1,d_2,\cdots,d_\mathtt{q}\).
The multi-entropy is characterized by a specific total contraction over \(n^{\mathtt{q}-1}\) density matrices defined in the following way:
We start with a $(\mathtt{q}-1)$-dimensional hypercube of length \(n\) and we assign each integer lattice point \(\vec{x}=(x_1,x_2,\cdots,x_{\mathtt{q}-1})\) a density matrix.
The action of \(g_k\), the twist operator on the \(k\)-th party, is defined as the cyclic permutation of the elements along the \(k\)-th coordinate axis, {\it i.e.}, 
\begin{align}
\label{gkdefinition}
g_{k} & \cdot (x_1,\cdots,x_k,\cdots,x_{\mathtt{q}-1}) = (x_1,\cdots,x_{k}+1,\cdots,x_{\mathtt{q}-1}), \quad 1\le k \le \mathtt{q}-1, \\
g_\mathtt{q} &= e .
\end{align}
We identify \(x_i= n + 1 \) with \(x_i=1\) for all $ 1 \le i \le \mathtt{q}-1$. See Fig.~\ref{fig:contraction-multi} for examples of the contraction pattern for \(\mathtt{q}=2,3\).

The contraction pattern is symmetric across different subsystems in the sense that the expression is invariant under any reorientation of indices \(g_i\to g_{i'}\). %\footnote{\color{red}More specifically, given any \(\sigma\in S_{\mathtt{q}}\), there exists an element \(\lambda\in S_{n^{\mathtt{q}-1}}\) such that \(g_i = \lambda\cdot g_{\sigma(i)} \cdot\lambda^{-1}\) for all \(i\in\{1,\cdots,\mathtt{q}\}\).}. 
As a result, the Cayley distance\footnote{In this case, the symmetric group is $S_{n^{ \mathtt{q}-1 }}$ because there are \(n^{\mathtt{q}-1}\) density matrices.} between any pair \((g_i,g_j)\) is identical: 
\begin{equation}
d(g_i,g_j) = n^{\mathtt{q}-1} - \#(g_i g_j^{-1}) =  n^{\mathtt{q}-2}(n-1),\quad { 1 \le i \ne j \le \mathtt{q}} \,.
\end{equation}

We now define the \((\mathtt{q},n)\)-R\'enyi multi-entropy \(S^{(\mathtt{q})}_n\) in an analogous fashion as the bi-partite R\'enyi entropy:
\begin{align}
S^{(\mathtt{q})}_n &= \frac{1}{1-n}\log \frac{Z^{(\mathtt{q})}_n}{(Z^{(\mathtt{q})}_1)^{n^{\mathtt{q}-1}}},\\
Z^{(\mathtt{q})}_n &= \bra{\psi}^{\otimes n^{\mathtt{q}-1}} \Sigma_1(g_1)\Sigma_2(g_2)\cdots\Sigma_\mathtt{q}(g_\mathtt{q})\ket{\psi}^{\otimes n^{\mathtt{q}-1}}.
\end{align}
One can check that this expression reduces to the usual R\'enyi entropy in the bi-partite case, \(\mathtt{q}=2\).

The von Neumann \(\mathtt{q}\)-multi-entropy, or simply the \(\mathtt{q}\)-multi-entropy, is defined as the analytic continuation of the R\'enyi multi-entropy at \(n\to 1\):
\begin{equation}
S^{(\mathtt{q})} \equiv \lim_{n\to 1}  S^{(\mathtt{q})}_n.
\end{equation}
The act of actually performing the analytic continuation is quite subtle.
For example one might expect an application of Carlson's theorem given conditions on the growth rate of the function at complex infinity.
However, one does not know whether the partition functions of the system in the limit of large bond dimensions \(d_{\rm Total} \to\infty\) remains an analytic function --- Indeed it is well known that such statment is false in the thermal dynamical limit at phase transitions due to condensation of zeros\footnote{This issue was pointed out in the context of random tensor networks and AdS/CFT in \cite{Akers:2024pgq}, for example.}.
We will not tackle the difficult problem of analytic continuing the multi-entropy rigorously in this paper.
Instead, we will provide exact results for tractable integer $(\mathtt{q},n)$-R\'enyi multi-entropies via direct calculations. Base on these results, we observe qualitative features of the multi-entropy curves that persist independent of $n$ and we conjecture that these feature survive as one take $n\to1$.
In parallel, we obtain analytical expressions for arbitrary-$(\mathtt{q},n)$ in the limit $d_{\rm BH} \to\infty$, where the system is sufficiently far away from the phase transition, with the expectation that the entropy function is analytic. These results serve as additional corroboration for our claims.

A very illuminating example of how multi-entropy can be useful is as follows \cite{Gadde:2022cqi,PhysRevLett.86.5184}. Let us consider the following two states, 
\begin{align}
\ket{\rm W} &= \frac{1}{\sqrt{3}} \Bigl( \ket{001} + \ket{010} + \ket{100} \Bigr) \,, \\ 
\ket{\rm generalized\, GHZ} &=   \frac{1}{\sqrt{3}} \Bigl( \ket{000} + \sqrt{2} \ket{111}  \Bigr) \,,
\end{align}
These states have the property that the corresponding reduced density matrices (for any subsystems) of either state have the same eigenvalues.
As a result, no bi-partite entanglement measure can differentiate between $\ket{\rm W}$ and $\ket{\rm generalized\, GHZ}$.
Nevertheless, it is straightforward to show that the R\'enyi multi-entropy of these two states are different \cite{Gadde:2022cqi}.
In this way, multi-entropy contains broader information than simple bi-partite entanglement measures.

In the rest of this paper, we will take \(\ket{\psi}\) to be a \(\mathtt{q}\)-partite Haar-random state.
Using techniques from Sec.~\ref{sec:RT-group}, the averaged partition function \(Z^{(\mathtt{q})}_n\) can be expressed as a sum over the replica group \(S_{n^{\mathtt{q}-1}}\). The  \((\mathtt{q},n)\)-R\'enyi multi-entropy for the \(\mathtt{q}\)-partite Haar-random state \(\ket{\psi}\) reads
\begin{align}
  \label{eq:mutliSn-sum}
\overline{S^{(\mathtt{q})}_n} &\equiv \frac{1}{1-n}\overline{\log \frac{Z^{(\mathtt{q})}_n}{(Z^{(\mathtt{q})}_1)^{n^{\mathtt{q}-1}}}} \approx \frac{1}{1-n}\log \frac{\overline{Z^{(\mathtt{q})}_n}}{\overline{(Z^{(\mathtt{q})}_1)^{n^{\mathtt{q}-1}}}}, \\
\overline{Z^{(\mathtt{q})}_n} &= \sum_{g\in S_{n^{\mathtt{q}-1}}} d_1^{-d(g,g_1)} d_2^{-d(g,g_2)} \cdots d_\mathtt{q}^{-d(g,g_\mathtt{q})}.
\end{align}
Note that similar to \eqref{eq:Sn_avg1}, we have replaced the ensemble average of the entire entropy by the quantity constructed by averaged partition functions.
This approximation is valid in the limit of large bond dimensions.
We will return to this problem when we investigate $1/d$ corrections in in Sec.~\ref{sec:orge89da24} and \ref{sec:org704fc05}.

In the rest of this paper, we only focus on $\overline{S^{(\mathtt{q})}_n}$ for \(\mathtt{q}\)-partite Haar-random state. Thus, for notation simplicity, we omit $\overline{S^{(\mathtt{q})}_n}$ and write simply it as ${S^{(\mathtt{q})}_n}$. 

\section{Black hole multi-entropy curves and the simplest example, $\mathtt{q}=3$ and $n=2$}\label{secRenyimultientropyn2q3}

We start this section with a new definition of the {\it black hole multi-entropy curve}, a natural generalization of the Page curve that captures entanglement between multiple Hawking radiation subsystems.

\subsection{Definition of a black hole multi-entropy curve}
\label{defmulticurve}
Here we define the black hole multi-entropy curves for generic $\mathtt{q}$. 
Let us divide the whole Hawking radiation into $\mathtt{q}-1$ subsystems, where we take the dimensions of all radiation subsystems the same. Furthermore, let us call that dimension $d_{\rm R}$, {\it i.e.,} 
\begin{align}
\label{alldrthesame}
d_{\rm R1} = \cdots = d_{\rm R \mathtt{q}-1} \equiv d_{\rm R} \,.
\end{align}
Here $d_{\rm R1}$, $\cdots$ $d_{\rm R \mathtt{q}-1}$ represents the dimension of radiation 1,$\cdots$,  radiation $\mathtt{q} -1$ subsystems.

This setting is equivalent to the setting where we define a subsystem of each radiation by dividing all angles equally in the $q-1$ direction, considering, for example, the case of isotropic Hawking radiation such as the Schwarzschild black hole. In a 4-dim Schwarzschild black hole, each subsystem would occupy a surface area of $4 \pi/(\mathtt{q}-1)$ of the unit sphere. 

We also set the dimensions of whole $\mathtt{q}$ subsystems, $d_{\rm Total}$,  fixed, 
\begin{align} 
\label{Totalconstraint}
d_{\rm Total} = d_{\rm R}^{\mathtt{q}-1} d_{\rm BH} \mbox{ = fixed} .
\end{align} 
where $d_{\rm BH}$ is the dimension of a black hole. 
Then all dimensions $d_{\rm R1}$, $\cdots$, $d_{\rm R\mathtt{q}-1}$, $d_{\rm BH}$ are determined by $d_{\rm R}$ and the $n$-th R\'enyi multi-entropy become unique functions of $d_{\rm R}$. 

Black hole multi-entropy curves are, for any $n$, the $n$-th R\'enyi multi-entropy drawn by varying $d_{\rm R}$,  from $d_{\rm R} =1$, the very starting point of the Hawking radiation, till $d_{\rm BH} = 1$, the complete evaporation point of a black hole, where 
\begin{align}
\label{BHevapotime}
\qquad \qquad d_{\rm R} = \left( d_{\rm Total} \right)^{\frac{1}{\mathtt{q}-1}} \quad  \mbox{(black hole evaporation time)}
\end{align}
according to the constraint \eqref{alldrthesame} and \eqref{Totalconstraint}. 

This black hole multi-entropy curve is a natural generalization of the Page curve and it reduces to the Page curve for $\mathtt{q} = 2$. This black hole multi-entropy curve describes how the multi-entropy changes during the whole time evolution of a black hole evaporation.

\subsection{Explicit example: $\mathtt{q} =3$ tri-partite subsystems}
\begin{figure}[t]
    \centering
    \includegraphics[width=2cm]{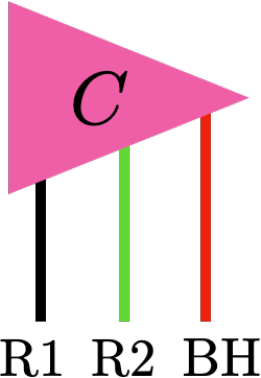}
    \caption{Cartoon of the random tensor representing the state \eqref{StatewithCorrelation}.}
    \label{fig:random-tensor}
\end{figure}

As a first example, we consider the following pure state on a 3-party ($\mathtt{q} = 3$) system $\mathcal{H}_{\rm R1}\otimes\mathcal{H}_{\rm R2}\otimes\mathcal{H_{\text{BH}}}$
\begin{align}
|\psi\rangle=\sum_{i=1}^{d_{\rm R1}}\sum_{j=1}^{d_{\rm R2}}\sum_{k=1}^{d_{\text{BH}}}c_{ijk}| {\rm R1}_i\rangle\otimes | {\rm R2}_j\rangle\otimes |\text{BH}_k\rangle,\label{StatewithCorrelation}
\end{align}
where $c_{ijk}$ defines a uniform Haar-random state, which behaves as complex Gaussian random variables\footnote{\label{ft:Gaussian} To see this, recall a Haar-random matrix over $U(N)$ can be generated by first consider a $N\times N$ matrix with entries as i.i.d. complex Gaussian variables, then apply Gram-Schmidt algorithm to its columns \cite{2006math.ph...9050M}. We then use the ancilla state $\ket{\phi}=(1,0,0,\cdots,0)$, which simply picks out the first column of the matrix.},
and $|{\rm R1}_i\rangle, |{\rm R2}_j\rangle, |\text{BH}_k\rangle$ are orthonormal bases in the Hilbert space $\mathcal{H}_{\rm R1}\otimes\mathcal{H}_{\rm R2}\otimes\mathcal{H_{\text{BH}}}$. 
Symbolically this is represented as Fig.~\ref{fig:random-tensor}. 

The dimensions of each Hilbert space are
\begin{align}\label{dimHilbert}
    \dim\mathcal{H}_{\text{R1}}=d_{\rm R1},\;\;\;\dim\mathcal{H}_{\text{R2}}=d_{\rm R2},\;\;\;\dim\mathcal{H}_{\text{BH}}=d_{\text{BH}}.
\end{align}
We interpret $\mathcal{H}_{\rm R1}$ and $\mathcal{H}_{\rm R2}$ as two subsystems of Hawking radiation and $\mathcal{H_{\text{BH}}}$ as a system of black hole microstates. These three subsystems are correlated via the Gaussian random variables $c_{ijk}$.
Suppose we lay out \(n^2\) reduced density matrices in a \(n\times n\) square lattice, then \(g_1\) and \(g_2\) can be thought of as the cyclic permutations for the rows and columns.

To make our calculation concrete, we will now specialize to $n=2$ case in the rest of this section. We have $2^2=4$ replicas and \eqref{gkdefinition} reduces to
\begin{align}
g_1 &= (12)(34), \\
g_2 &= (13) (24).
%g_1 &= (12\cdots n)((n+1)(n+2)\cdots2n)\cdots((n^2-n+1)\cdots n^2), \\
%g_2 &= (1 (n+1) (2n+1) \cdots (n^2-n+1)) (2 (n+2) (2n+2)\cdots (n^2-n+2)) \cdots (n (2n) (3n) \cdots n^2).
\end{align}
These are permutation group elements of \(S_{n^{2}}\), and the Cayley distance on \(S_{2^{2}}\) is 
\begin{align}
d(g,h) = 2^2 - \#(gh^{-1}) \label{Cayleyd} \,.
\end{align}

\begin{figure}[t]
    \centering
    \includegraphics[width=.9\textwidth]{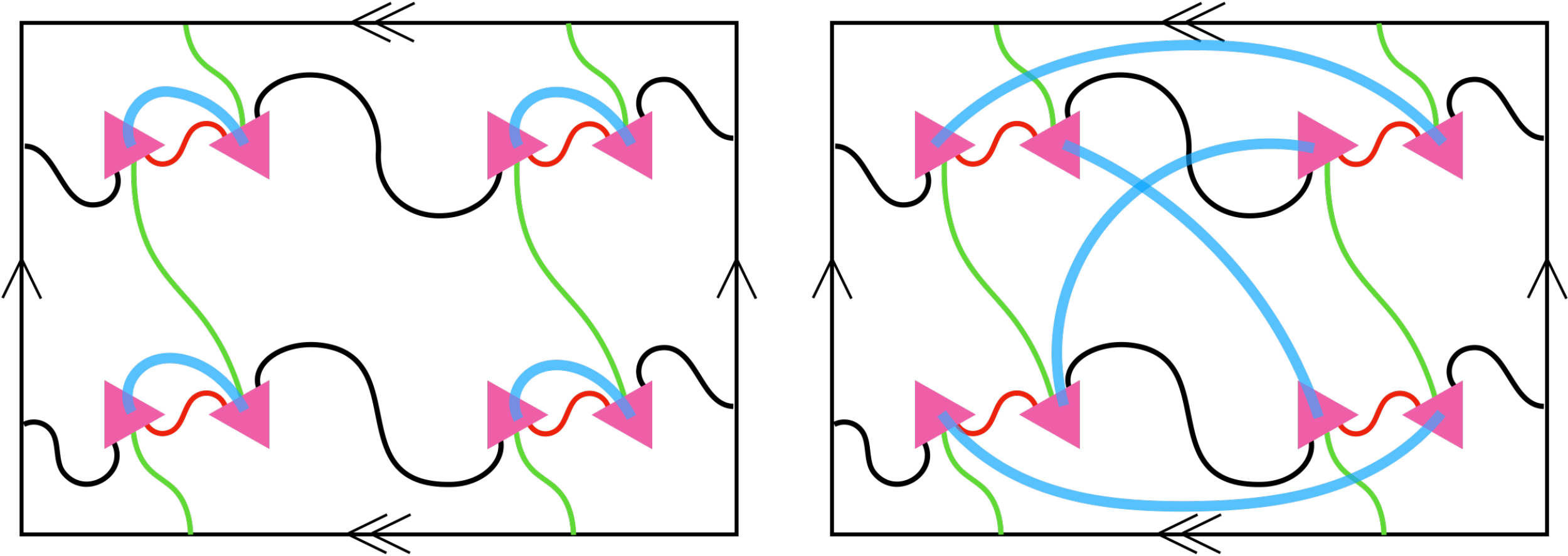}
    \caption{Two examples of Wick contraction (bold blue) out of a total of 24 Wick contractions. The black horizontal line is for index R1 and the green vertical line is for index R2. The red one is for index BH. We have flipped the tensors corresponding to the bra $\bra{\psi}$ for a cleaner  diagrammatic presentation. left figure, there are (2, 2, 4) loops for (R1, R2, BH) indices, yielding $d_{\rm R1}^2 d_{\rm R2}^2 d_{\rm BH}^4$. On the other hand, in the right figure,  there are (3, 1, 1) loops for (R1, R2, BH) indices, yielding $d_{\rm R1}^3 d_{\rm R2}^1 d_{\rm BH}^1$.}
    \label{fig:twoexamples}
\end{figure}

The \((\mathtt{q}=3,n=2)\)-R\'enyi multi-entropy reads
\begin{align}
    S_2^{(3)}\approx-\log \frac{\overline{{Z}_2^{(3)}}}{\overline{({Z}_1^{(3)})^{4}}}=-\log\left(\frac{\sum_{g\in S_{4}}d^{-d(g,g_1)}_{\text{R1}}d^{-d(g,g_2)}_{\text{R2}}d^{-d(g,e)}_{\text{BH}}}{\sum_{g\in S_{4}}(d_{\text{R1}}d_{\text{R2}}d_{\text{BH}})^{-d(g,e)}}\right)\label{men2q3}
\end{align}
of the pure state (\ref{StatewithCorrelation}). Here $\overline{\cdots}$ represents averaging over the Haar measure\footnote{Equivalently $\overline{\cdots}$ represents the average of Gaussian random variables $c_{ijk}$.}. 
The replica partition function ${Z}_2^{(3)}$ (\ref{men2q3}) consists of $n^{\mathtt{q}-1}=2^2$ reduced density matrices, and there are $n^{\mathtt{q}-1}!=(2^2)!=24$ ways to take Wick contraction of $c_{ijk}$, where each contraction can be represented by the group elements of \(S_{4}\). By using the diagrammatic techniques we reviewed in Sec. \ref{sec:RT-group}, we can compute the 24 contributions to $\overline{{Z}_2^{(3)}}$ and $\overline{({Z}_1^{(3)})^{4}}$ explicitly. 
See Fig.~\ref{fig:twoexamples}, which shows two examples of Wick contraction out of 24.   

Note that in writing \eqref{men2q3} down we have approximated \(\overline{\log \left(Z_2^{(3)}/(Z_1^{(3)})^4\right)}\) by \(\log \left(\overline{Z_2^{(3)}}/\overline{(Z_1^{(3)})^4}\right)\), c.f. \eqref{eq:mutliSn-sum}, which is valid in the limit of large total dimensions $d_{\rm Total} \to \infty$ \cite{Hayden:2016cfa,10.1093/acprof:oso/9780199535255.001.0001}. As a more direct evidence for this claim, later in Sec. \ref{sec:orge89da24}, we show that to the first order in \(1/d_{\text{BH}}\) where  $d_{\rm BH} \to \infty$, these kinds of corrections do not contribute.

By counting the number of each loop, 24 Wick contractions give
\begin{align}
\label{24Wick}
 \overline{{Z}_2^{(3)}} &= d_{\rm R1}d_{\rm R2}d_{\text{BH}} \Bigl( d_{\rm R1}d_{\rm R2}d_{\text{BH}}(9+d_{\rm R1}^2+d_{\rm R2}^2+d_{\text{BH}}^2)  \nonumber \\
    & \quad  +2(d_{\rm R1}^2+d_{\rm R2}^2+d_{\text{BH}}^2+d_{\rm R1}^2d_{\rm R2}^2+d_{\rm R2}^2d_{\text{BH}}^2+d_{\text{BH}}^2d_{\rm R1}^2) \Bigr) /(d_{\rm R1}d_{\rm R2}d_{\text{BH}})^4,
\end{align}
where the normalization factor $1/(d_{\rm R1}d_{\rm R2}d_{\text{BH}})^4$ comes from $n^2$ in $d(g,h)$ in (\ref{Cayleyd}). 
The replica partition function in the denominator is 
\begin{align}
\overline{ \left( {Z}_1^{(3)} \right)^4 } = \frac{d_{\rm R1}d_{\rm R2}d_{\text{BH}}(d_{\rm R1}d_{\rm R2}d_{\text{BH}}+1)(d_{\rm R1}d_{\rm R2}d_{\text{BH}}+2)(d_{\rm R1}d_{\rm R2}d_{\text{BH}}+3)}{(d_{\rm R1}d_{\rm R2}d_{\text{BH}})^4}.
\end{align}
This result can also be obtained easily by setting $d_{\rm R1} = d_{\rm R2}=1$, and $d_{\rm BH} \to d_{\rm R1} d_{\rm R2} d_{\rm BH}$ in \eqref{24Wick}. 

Thus, the final result becomes 
\begin{align}
  \hspace{-10mm}  S_2^{(3)}=-\log \left[\frac{d_{\rm R1}d_{\rm R2}d_{\text{BH}}(9+d_{\rm R1}^2+d_{\rm R2}^2+d_{\text{BH}}^2)+2(d_{\rm R1}^2+d_{\rm R2}^2+d_{\text{BH}}^2+d_{\rm R1}^2d_{\rm R2}^2+d_{\rm R2}^2d_{\text{BH}}^2+d_{\text{BH}}^2d_{\rm R1}^2)}{(d_{\rm R1}d_{\rm R2}d_{\text{BH}}+1)(d_{\rm R1}d_{\rm R2}d_{\text{BH}}+2)(d_{\rm R1}d_{\rm R2}d_{\text{BH}}+3)}\right].\label{Formulame}
\end{align}
This expression is totally symmetric in $d_{\rm R1}$, $d_{\rm R2}$, and $d_{\text{BH}}$. 

Let us consider the limit of $d_{\text{BH}}\to\infty$. In that limit, \eqref{Formulame} becomes 
\begin{align}
  \hspace{-10mm}  S_2^{(3)} \approx -\log \left[ \frac{1}{d^2_{\rm R1} d^2_{\rm R2}}  \left(1 + \frac{2 \left(d^2 _{\rm R1} + d^2_{\rm R2} -2 \right) }{ d_{\rm R1} d_{\rm R2} d_{\rm BH} } \right)  + \order{\frac{1}{d^2_{\rm BH}}}\right].
\end{align}
Thus 
\begin{align}
\lim_{d_{\text{BH}}\to\infty}S_2^{(3)}=2\log d_{\rm R1}+2\log d_{\rm R2} -  \frac{2 \left(d^2 _{\rm R1} + d^2_{\rm R2} -2 \right) }{ d_{\rm R1} d_{\rm R2}  } \frac{1}{d_{\rm BH}}  + \order{\frac{1}{d^2_{\rm BH}}},
\label{MElargeBHlimit}
\end{align}
which decouples into the contributions from $\mathcal{H}_{\rm R1}\otimes\mathcal{H_{\text{BH}}}$ and $\mathcal{H}_{\rm R2}\otimes\mathcal{H_{\text{BH}}}$ only in the leading order.
However, the subleading order gives the correlations between Hawking radiation. In Hawking's calculation \cite{Hawking:1975vcx}, the entropy of Hawking radiation keeps increasing, which would be true if $d_{\text{BH}}$ is large enough. Thus, we define $S_{2 \; \text{Hawking}}^{(3)}$ as the leading term in eq.~(\ref{MElargeBHlimit})
\begin{align}
    S_{2 \; \text{Hawking}}^{(3)}:= \lim_{d_{\rm BH} \to \infty}S_{2}^{(3)} =2 \log d_{\rm R1} d_{\rm R2}.\label{HawkingCalc}
\end{align}

Just for comparison in Appendix \ref{negativityappendix}, we also evaluate the averaged $n$ replica partition functions for the {\it R\'enyi entanglement negativity} in $n=3$ and $n=4$ cases. Note that in $n=4$ case, there are four replicas, and thus, the number of replicas for $n=4$ entanglement negativity coincides with one for the $n=2$, $\mathtt{q} =3$ multi-entropy. However, it is clear that even though the number of replicas is the same, the $(d_{\rm R1}$,  $d_{\rm R2}$,  $d_{\rm BH})$ dependence are completely different. In particular, for the entanglement negativity, the results in \eqref{negativityfor4replica} are not symmetric between $d_{\rm R1}$,  $d_{\rm R2}$ and  $d_{\rm BH}$ but for the multi-entropy, the result \eqref{24Wick} (and thus, the entropy \eqref{Formulame}) is symmetric. 

Another interesting tri-partite entanglement measure is the $(m,n)$-R\'enyi reflected entropy \cite{Dutta:2019gen}, which is defined as a total contraction over $m\times n$ copies of reduced density matrices ($m$ is an even integer). It turns out the contraction pattern for $(m=2,n=2)$ reflected entropy coincides with the $(\mathtt{q}=3,n=2)$ R\'enyi multi-entropy considered here (Fig.~\ref{negativitymulti4replica} (b)). However, the difference of normalization between the multi-entropy and the reflected entropy ends up with different behaviors. For more details, see Appendix~\ref{negativityappendix}.

\subsection{Black hole multi-entropy curves}
\label{q=3n=2multicurve}

\begin{figure}[t]
    \centering
    \includegraphics[width=9cm]{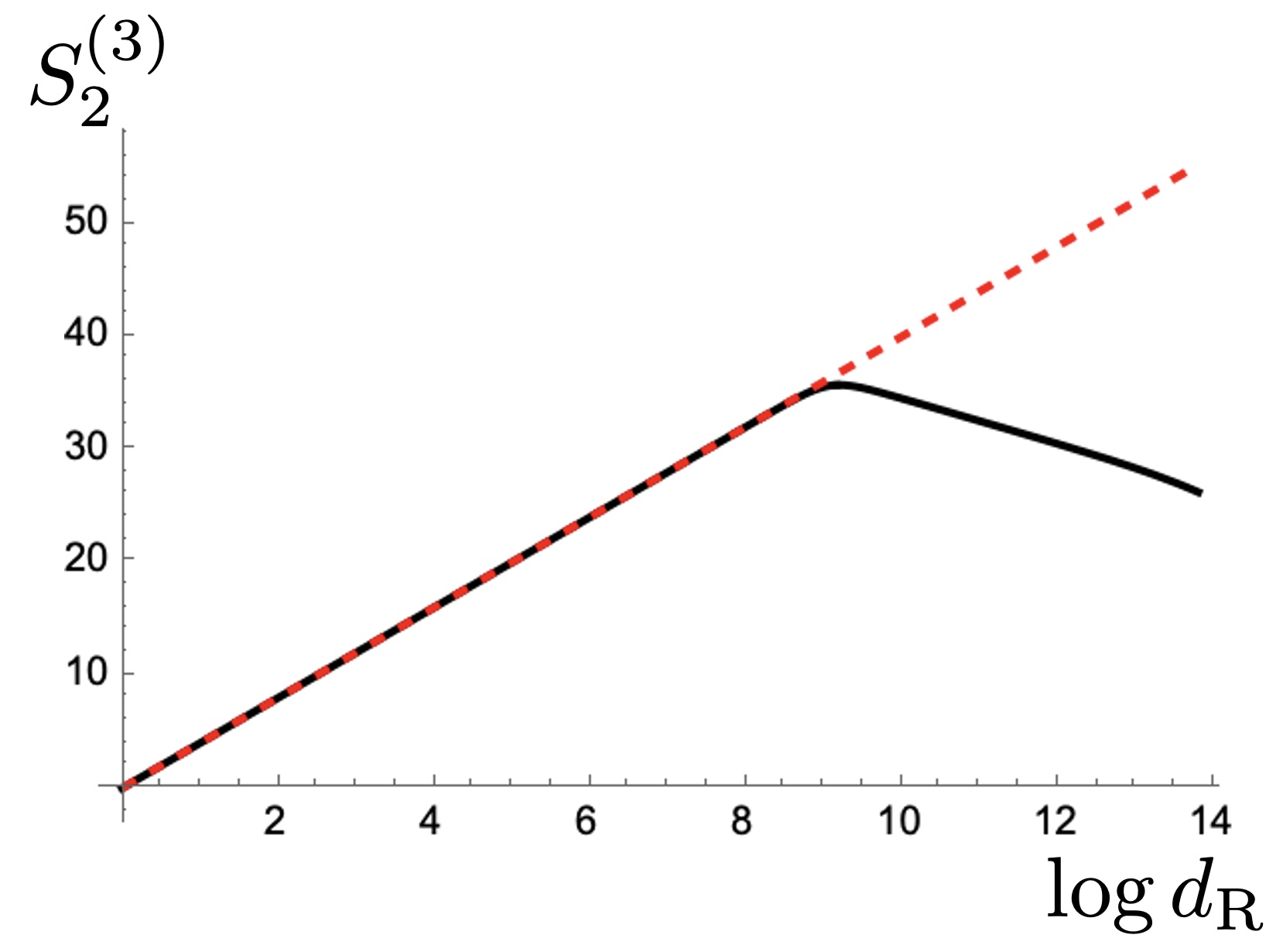}
    \caption{A black solid curve is a black hole multi-entropy curve for the $n=2$ R\'enyi multi-entropy with $\mathtt{q}=3$ case, $S^{(3)}_{2}$, as a function of $\log d_{\rm R}$ for the parameter choice $d_{\rm Total} =d_{\rm R1} d_{\rm R2} d_{\rm BH} =  10^{12}$. This curve is obtained from \eqref{BHmultiScurveq=3n=2}. Origin is $d_{\rm R}=1$ and it ends at the black hole evaporation time \eqref{BHevapotime}, which is $\log d_{\rm R} = \log \left( d_{\rm Total}\right)^{1/2} \approx 13.8$. On the other hand, a red dashed curve is Hawking's prediction given by \eqref{HawkingCalc}. They begin to deviate at multi-entropy time \eqref{multiStime}, $\log d_{\rm R} = \log \left( d_{\rm Total}\right)^{1/3} = \log 10^4 \approx 9.21$. On the other hand, the Page time \eqref{Pagetime} is $\log d_{\rm R} =\log \left( d_{\rm Total}\right)^{1/4} \approx 6.91$.}
    \label{fig:multiq=3n=2}
\end{figure}

We defined a black hole multi-entropy curve in Sec. \ref{defmulticurve}. For $\mathtt{q} =3$ case, we can obtain $S^{(3)}_2$ as a function of $d_{\rm R}$ and it becomes 
\begin{align}
\label{BHmultiScurveq=3n=2}
  \begin{split}
S^{(3)}_2 &= -\log \left[ \frac{ d_{\rm R}^2 d_{\rm BH}  (9 + 2 d_{\rm R}^2 + d_{\rm BH}^2 ) +  2 ( 2 d_{\rm R}^2 + d_{\rm BH}^2 + d_{\rm R}^4 + 2 d_{\rm R}^2 d_{\rm BH}^2 )}{(1 + d_{\rm BH} d_{\rm R}^2) (2 + d_{\rm BH} d_{\rm R}^2) (3 +  d_{\rm BH} d_{\rm R}^2)} \right] \,, \\
&   \qquad \mbox{where} \quad d_{\rm BH} = \frac{d_{\rm Total}}{d_{R}^2} \,,
  \end{split}
\end{align}
it is straightforward to plot the curve and the result is shown in Fig~\ref{fig:multiq=3n=2}. 

Recall that the black hole multi-entropy curve reduces to the Page curve \cite{Page:1993df, Page:1993wv} for $\mathtt{q}=2$ case. What we present here (Fig.~\ref{fig:multiq=3n=2}) is for $\mathtt{q}=3$, and they exhibit interesting different behaviors. We comment on the differences below. 

Before we make any comments, we would like to point out that even though so far we have shown a black hole multi-entropy curve only for $n=2$ and $\mathtt{q} = 3$ case, the comments we will make govern common properties of black hole multi-entropy curves and are valid for {\it generic values of $n$ and $\mathtt{q}$}.
In later sections, we will see other examples of black hole multi-entropy curves and that everything stated here is consistent with the numerical results. 

\begin{enumerate}
  
\item At early time, Hawking's prediction and generic multi-entropy behave almost the same way. In other words, Hawking's prediction gives a very good approximation. This is very similar to the Page curve.
  
\item However, the point when they begin to deviate significantly is different. For the Page curve ($\mathtt{q}=2$ case), this point is $d_{\rm R} = d_{\rm BH}$, which is in the middle of the horizontal axis (so-called the Page time), where half of the degrees of freedom are emitted by radiation. Rather, for generic black hole multi-entropy curves with $\mathtt{q}\ge 3$ case, they begin to deviate at 
\begin{align}
d_{\rm R} = d_{\rm BH} = \frac{d_{\rm Total}}{\left( d_{\rm R} \right)^{\mathtt{q}-1}} \quad \Leftrightarrow \quad d_{\rm R} = \left( d_{\rm Total} \right)^{1/\mathtt{q}} \quad \mbox{(Multi-entropy time)}
\label{multiStime}
\end{align} 
We call this time point {\it multi-entropy time}. It is a point when the multi-entropy becomes maximal and the dimensions of all of the degrees of freedom are the same. Thus, the multi-entropy time is determined solely by $\mathtt{q}$ and is independent of the R\'enyi index $n$.
After the multi-entropy time, multi-entropy decreases as we continue to increase $d_{\rm R}$. On the other hand, according to Hawking's prediction \cite{Hawking:1975vcx}, multi-entropy simply keeps increasing.

\item The Page time is where the total radiation dimension is equal to the black hole dimension and where entanglement entropy between total radiation and a black hole becomes maximal. Note that since the dimension of each radiation subsystem is taken to be the same, {\it i.e.,} $d_{\rm R1} = \cdots = d_{\rm R\mathtt{q}-1} = d_{\rm R}$, the dimension of all radiation is $\left( d_{\rm R} \right)^{\mathtt{q}-1}$. 
Thus, the Page time is defined as 
\begin{align}
\label{Pagetime}
\left( d_{\rm R} \right)^{\mathtt{q}-1} = d_{\rm BH} = \frac{d_{\rm Total}}{\left( d_{\rm R} \right)^{\mathtt{q}-1}} \quad \Leftrightarrow  \quad d_{\rm R} = \left( d_{\rm Total} \right)^{\frac{1}{2(\mathtt{q}-1)}} \quad \mbox{(Page time)}
\end{align}
Since 
\begin{align}
\frac{1}{\mathtt{q}} - \frac{1}{2 (\mathtt{q}-1)} 
%= \frac{2 (\mathtt{q}-1) - \mathtt{q}}{2 \mathtt{q} (\mathtt{q}-1)} 
= \frac{\mathtt{q}-2}{2 \mathtt{q} (\mathtt{q}-1)}  > 0 \quad \mbox{for $\mathtt{q} \ge 3$} \,,
\end{align}
the multi-entropy time is always greater than the Page time for $\mathtt{q} \ge 3$. 
\begin{align}
\log \left( d_{\rm Total} \right)^{1/\mathtt{q}} > \log \left( d_{\rm Total} \right)^{1/2(\mathtt{q}-1)}  \,.
\end{align}
and they coincide at $\mathtt{q}=2$.
Thus, multi-entropy time reduces to the Page time for the case $\mathtt{q}=2$.

\item For the Page curve ($\mathtt{q}=2$ case), entanglement entropy reduces to zero at the end when black hole evaporates completely. On the other hand, for generic black hole multi-entropy curves ($\mathtt{q} \ge 3$ case) they decrease at later time but they do not reduce to zero when a black hole evaporates completely. Because $d_{\rm BH} =1$, there can be no entanglement between an evaporated black hole and radiation. The contribution to non-zero multi-entropy must come from (secret) entanglement between Hawking's particles. 
  Note that these entanglement between Hawking particles are not visible in semi-classical approximations \cite{Hawking:1975vcx}.

\item In the limit $\mathtt{q} \to \infty$, the black hole multi-entropy curve behaves the same as the Hawking curve during the whole period of black hole evaporation \cite{Hawking:1975vcx}. This can be seen since the multi-entropy time \eqref{multiStime} and the black hole evaporation time \eqref{BHevapotime} coincide in the $\mathtt{q} \to \infty$ limit.
  In this limit, the multi-entropy will not miss any entanglement of the random state due to the infinite division of the system\footnote{Since we currently do not have a complete understanding of the operational or quantum information theoretic meaning of the multi-entropy, this statement involves speculation. However, since the multi-entropy is: 1) a natural generalization of the von Neumann entropy, 2) treats all subsystems symmetrically, 3) our random state setup treats all radiation subsystems symmetric, we believe the multi-entropy plays a special role in capturing the final state entanglement of the Hawking particles.}.

\end{enumerate}

\subsection{Multi-entropy curve for decoupled radiation state}

To compare the above generic results with the case of uncorrelated Hawking radiation, let us consider the following pure state on a 3-party system $\mathcal{H}_{\text{R1}}\otimes\mathcal{H}_{\text{R2}}\otimes\mathcal{H_{\text{BH}}}$
\begin{align}
|\psi\rangle_{\rm \bf decoupled \, rad}=\left(\sum_{i_1=1}^{d_{\rm R1}}\sum_{j_1=1}^{d_{\text{BH}1}}c_{i_1j_1}|\text{R1}_{i_1}\rangle\otimes |\text{BH1}_{j_1}\rangle\right)\otimes\left(\sum_{i_2=1}^{d_{\rm R2}}\sum_{j_2=1}^{d_{\text{BH}2}}d_{i_2j_2}|\text{R2}_{i_2}\rangle\otimes |\text{BH2}_{j_2}\rangle\right),\label{StatewithoutCorrelation}
\end{align}
where $c_{i_1j_1}$ and $d_{i_2j_2}$ define two uniform random states. Note that  we divide $\mathcal{H_{\text{BH}}}$ into $\mathcal{H_{\text{BH}}}=\mathcal{H}_{\text{BH1}}\otimes\mathcal{H}_{\text{BH2}}$, and $|\text{BH1}_{j_1}\rangle$, $|\text{BH2}_{j_2}\rangle$ are orthonormal bases of $\mathcal{H}_{\text{BH1}}, \mathcal{H}_{\text{BH2}}$, respectively. 
The dimension of each Hilbert space is
\begin{align}
\dim\mathcal{H}_{\text{BH1}}=d_{\text{BH1}},\;\;\;\dim\mathcal{H}_{\text{BH2}}=d_{\text{BH2}},\;\;\;\dim\mathcal{H}_{\text{BH}}= d_{\rm BH}=d_{\text{BH}1}d_{\text{BH}2} 
\end{align}
We call $|\psi\rangle_{\rm \bf decoupled \, rad}$ as ``decoupled radiation'' state. 
Since $|\psi\rangle_{\rm \bf decoupled \, rad}$ is a tensor product of two pure states on $\mathcal{H}_{\text{R1}}\otimes\mathcal{H}_{\text{BH1}}$ and $\mathcal{H}_{\text{R2}}\otimes\mathcal{H}_{\text{BH2}}$, there is no correlation between two Hawking radiation in this setup. 

Since R\'enyi multi-entropy $S_2^{(3)}$ is additive under tensor products of pure states \cite{Penington:2022dhr}, 
$S_2^{(3)}$ of the decoupled radiation state, $S_{2 \, {\rm \bf decoupled \, rad}}^{(3)}$ is given by a sum of two contributions from $\sum_{i_1=1}^{d_{\rm R1}}\sum_{j_1=1}^{d_{\text{BH}1}}c_{i_1j_1}|\text{R1}_{i_1}\rangle\otimes |\text{BH1}_{j_1}\rangle$ and $\sum_{i_2=1}^{d_{\rm R2}}\sum_{j_2=1}^{d_{\text{BH}2}}d_{i_2j_2}|\text{R2}_{i_2}\rangle\otimes |\text{BH2}_{j_2}\rangle$. Each contribution can be obtained from the formula (\ref{Formulame}) by setting one of $d$ (either $d_{\rm R2}$ or $d_{\rm R1}$) to be 1. Thus, we obtain
\begin{align} 
S_{2 \, {\rm \bf decoupled \, rad}}^{(3)} =-&\log \left[\frac{d_{\rm R1}d_{\text{BH1}}(10+d_{\rm R1}^2+d_{\text{BH1}}^2)+2(1+2d_{\rm R1}^2+2d_{\text{BH1}}^2 +2d^2_{\rm R1}d^2_{\text{BH1}})}{(d_{\rm R1}d_{\text{BH1}}+1)(d_{\rm R1}d_{\text{BH1}}+2)(d_{\rm R1}d_{\text{BH1}}+3)}\right]\notag\\
-&\log \left[\frac{d_{\rm R2}d_{\text{BH2}}(10+d_{\rm R2}^2+d_{\text{BH2}}^2)+2(1+2d_{\rm R2}^2+2d_{\text{BH2}}^2 +2d^2_{\rm R2}d^2_{\text{BH2}})}{(d_{\rm R2}d_{\text{BH2}}+1)(d_{\rm R2}d_{\text{BH2}}+2)(d_{\rm R2}d_{\text{BH2}}+3)}\right].\label{Formulame2}
\end{align}
In the limit of $d_{\text{BH1}},d_{\text{BH2}}\to\infty$, $S_{2 \, {\rm \bf decoupled \, rad}}^{(3)}$ is simplified as
\begin{align}
\lim_{d_{\text{BH1}},d_{\text{BH2}}\to\infty}  S_{2 \, {\rm \bf decoupled \, rad}}^{(3)}
&= -  \log \left[ \frac{1}{d^2_{\rm R1}} \left(1 + \frac{2 (d^2_{\rm R1} - 1)}{d_{\rm R1}} \frac{1}{d_{\rm BH1}} +\order{ \frac{1}{d^2_{\rm BH1}} }\right) \right] \nonumber \\
&\quad -  \log \left[ \frac{1}{d^2_{\rm R2}} \left(1 + \frac{2 (d^2_{\rm R2} - 1)}{d_{\rm R2}} \frac{1}{d_{\rm BH2}} +\order{ \frac{1}{d^2_{\rm BH2}} }\right) \right] \nonumber \\
& =2 \log d_{\rm R1}+2\log d_{\rm R2} - \frac{2 (d^2_{\rm R1} - 1)}{d_{\rm R1}} \frac{1}{d_{\rm BH1}} - \frac{2 (d^2_{\rm R2} - 1)}{d_{\rm R2}} \frac{1}{d_{\rm BH2}}  \nonumber \\ &\qquad \qquad + \order{ \frac{1}{d^2_{\rm BH1}} }
+ \order{ \frac{1}{d^2_{\rm BH2}} }
 \label{MElargeBHlimit2}
\end{align}
which agrees with eq.~(\ref{MElargeBHlimit}) only in the leading order. 

When the dimension $d_{\text{BH}}$ of black hole microstates is much larger than the dimensions $d_{\rm R1}, d_{\rm R2}$ of Hawking radiation, R\'enyi multi-entropy $S_2^{(3)}$ does not depend much on whether there is a correlation between Hawking radiation or not, as shown in eqs.~(\ref{MElargeBHlimit}) and (\ref{MElargeBHlimit2}). However, as $d_{\rm R1}$ and $d_{\rm R2}$ increase, $S_2^{(3)}$ would depend on the presence of a correlation on $\mathcal{H}_{\text{R1}}\otimes\mathcal{H}_{\text{R2}}$ between Hawking radiation. 
This is the secret relationship between Hawking radiation which one misses in the leading order at $d_{\rm BH} \to \infty$.

\begin{figure}[t]
    \centering
    \includegraphics[width=9cm]{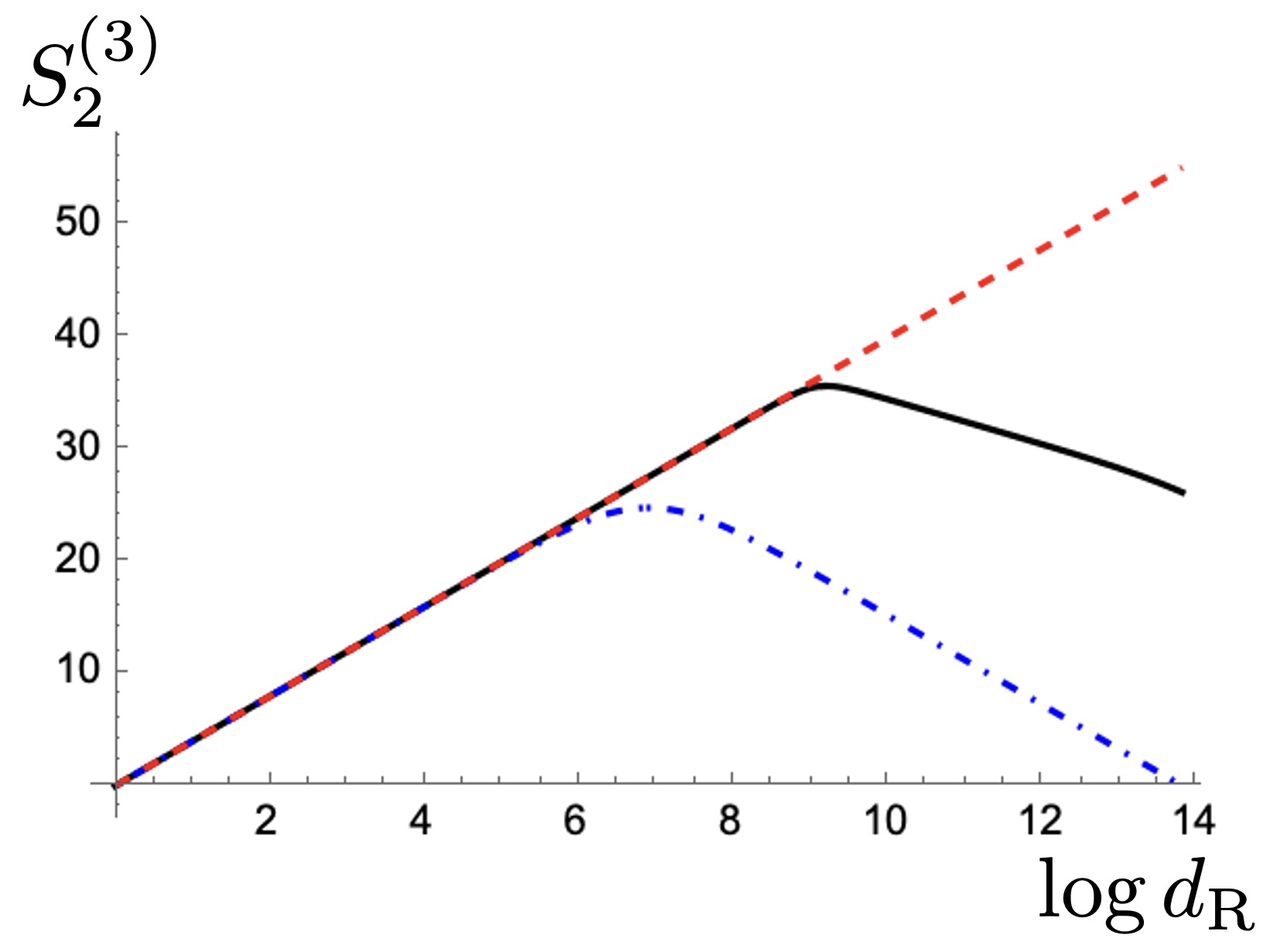}
%    \put(5,0){$d_{\text{R}}$}
%    \put(-200,135){$S_2^{(3)}$}
    \caption{Numerical plots of $S_2^{(3)}$ with $d_{\rm Total}=10^{12}$. The black curve is for generic state (\ref{BHmultiScurveq=3n=2}), the blue dashed curve is $S_2^{(3)}$ without the correlation (\ref{FormulaMEwithoutCorrelation}), and the red dashed curve is $S_{2 \; \text{Hawking}}^{(3)}$ (\ref{HawkingCalc2}). The blue dashed curve is essentially the same as the ordinarily Page curve.}
    \label{fig:MultiandPage}
\end{figure}

We numerically compare $S_{2 \; \text{Hawking}}^{(3)}$ from (\ref{HawkingCalc}), $S_2^{(3)}$
for Haar random states from (\ref{Formulame}), and $S_{2 \, {\rm \bf decoupled \, rad}}^{(3)}$ for the decoupled radiation state \eqref{StatewithoutCorrelation}
from (\ref{Formulame2}). To simplify the comparison, we set $d_{\rm R1}=d_{\rm R2}=d_{R}$ and $d_{\text{BH}1}=d_{\text{BH}2}=\sqrt{d_{\text{BH}}}$. Then, the formula (\ref{Formulame2}) becomes
\begin{align}
S_{2 \, {\rm \bf decoupled \, rad}}^{(3)} =-2\log \left[\frac{d_{R}\sqrt{d_{\text{BH}}}(10+2d_{R}\sqrt{d_{\text{BH}}}+d_{R}^2+d_{\text{BH}})+2(1+2d_{R}^2+2d_{\text{BH}})}{(d_{R}\sqrt{d_{\text{BH}}}+1)(d_{R}\sqrt{d_{\text{BH}}}+2)(d_{R}\sqrt{d_{\text{BH}}}+3)}\right].\label{FormulaMEwithoutCorrelation}
\end{align}
Taking the leading order at $d_{\text{BH}}\to\infty$, we obtain
\begin{align}
    S_{2 \; \text{Hawking}}^{(3)}=2 \log d_{\rm R}^2.\label{HawkingCalc2}
\end{align}

Fig.~\ref{fig:MultiandPage} shows numerical plots of the above three $S_2^{(3)}$ with $d_{\text{Total}}=10^{12}$. The black curve is the black hole multi-entropy curve of $S_2^{(3)}$ (\ref{BHmultiScurveq=3n=2}) for a random state \eqref{StatewithCorrelation}, and the blue curve is $S_{2 \, {\rm \bf decoupled \, rad}}^{(3)}$. 
Since decoupled radiation state (\ref{StatewithoutCorrelation}) is a direct product of the bi-partite system, 
there is no correlation between radiation 1 and 2. Thus, the multi-entropy curve for such a decoupled state reduces to the entanglement entropy curve between a) radiation 1 and 2 combined, and b) black hole 1 and 2 combined. The red curve is $S_{2 \; \text{Hawking}}^{(3)}$ (\ref{HawkingCalc2}).

One can see that the three curves have a similar behavior at small $d_{\text{R}}$. However, at large $d_{\text{R}}$, their behaviors are different. In particular, the blue curve (\ref{FormulaMEwithoutCorrelation}) for $|\psi\rangle_{\rm \bf decoupled \, rad}$ becomes zero when the black hole evaporates at $d_\text{BH}=1$. In contrast, the black hole multi-entropy curve (\ref{BHmultiScurveq=3n=2}) is nonzero when the black hole evaporates as
\begin{align}
\lim_{d_\text{BH}\to1}S_2^{(3)}=-\log \left[ \frac{ 2 \left(2 d_{\rm R}^4+9 d_{\rm R}^2+1\right)}{(1 + d_{\rm R}^2) (2 +  d_{\rm R}^2) (3 +   d_{\rm R}^2)} \right],  \quad \mbox{where} \quad d_{\rm Total} = d_{\rm R}^2 \label{divergent} 
\end{align}
due to the entanglement between radiation 1 and 2.

\subsection{Corrections from the variance}
\label{sec:orge89da24}

We now examine in more detail the approximation that leads to \eqref{men2q3}, i.e. by replacing \(\overline{\log \left(Z_2^{(3)}/(Z_1^{(3)})^4\right)}\) by \(\log \left(\overline{Z_2^{(3)}}/\overline{(Z_1^{(3)})^4}\right)\). Following \eqref{eq:Sn_avg2}, by Taylor expanding the logarithm we obtain
\begin{align} \label{avblog}
\overline{\log \frac{Z_2^{(3)}}{(Z_1^{(3)})^4}}=\log \frac{\overline{Z_2^{(3)}}}{\overline{(Z_1^{(3)})^4}}+\sum_{m=1}^\infty\frac{(-1)^{m-1}}{m}\left(\frac{\overline{(\delta Z_2^{(3)})^m}}{\overline{Z_2^{(3)}}^m}-\frac{\overline{(\delta (Z_1^{(3)})^4)^m}}{\overline{(Z_1^{(3)})^4}^m}\right),
\end{align}
where $\delta Z_2^{(3)} :=  Z_2^{(3)} - \overline{Z_2^{(3)}}$, and $\delta (Z_1^{(3)})^4 :=  (Z_1^{(3)})^4 - \overline{(Z_1^{(3)})^4}$.
The $m=1$ term is trivially zero, and we will evaluate the \(m=2\) corrections:
\begin{align} 
\overline{\log\frac{Z^{(3)}_2}{(Z^{(3)}_1)^4}} = \log\frac{\overline{Z^{(3)}_2}}{\overline{(Z^{(3)}_1)^4}} - \frac{1}{2}\left(\frac{\overline{(\delta Z^{(3)}_2)^2}}{\overline{Z^{(3)}_2}^2} - \frac{\overline{(\delta (Z_1^{(3)})^4)^2}}{\overline{(Z_1^{(3)})^4}^2} \right) + \cdots . 
\end{align}
We need to evaluate the variance 
\(\overline{(\delta Z^{(3)}_2)^2}=\overline{(Z^{(3)}_2)^2}-\overline{Z^{(3)}_2}^2\).
The squared average can be computed by the ``double" replica model, where the replica group is \(S_8\):
\begin{align}
\overline{(Z^{(3)}_2)^2} = \sum_{g\in S_8} d_\text{R1}^{-d(g,g'_1)}d_{\text{R2}}^{-d(g,g'_2)}d_\text{BH}^{-d(g,e)}, 
\end{align}
and the twist operators are simply products of the twist operators in the single replica model:
\begin{align} g'_1 = (12)(34)(56)(78), \quad g'_2 = (13)(24)(57)(68). \end{align}

By using Mathematica for the sum over symmetric group $S_8$, we obtain
\begin{align}
\overline{(Z^{(3)}_2)^2} = \frac{1}{d_{\rm R1R2BH}^7}\Big(& ((d_{\rm{R1}}^7d_{\rm{R2}}^3d_{\rm{BH}}^3+\cdots)+2(d_{\rm{R1}}^5d_{\rm{R2}}^5d_{\rm{BH}}^3+\cdots)) \notag\\
& + (4(d_{\rm{R1}}^6d_{\rm{R2}}^4d_{\rm{BH}}^2+\cdots)+12d_{\rm{R1}}^4d_{\rm{R2}}^4d_{\rm{BH}}^4) \notag\\
& + (26(d_{\rm{R1}}^5d_{\rm{R2}}^3d_{\rm{BH}}^3+\cdots)+4(d_{\rm{R1}}^5d_{\rm{R2}}^5d_{\rm{BH}}+\cdots)) \notag\\
& + (20(d_{\rm{R1}}^6d_{\rm{R2}}^2d_{\rm{BH}}^2+\cdots)+76(d_{\rm{R1}}^4d_{\rm{R2}}^4d_{\rm{BH}}^2)) \notag\\
&+ (80(d_{\rm{R1}}^5d_{\rm{R2}}^3d_{\rm{BH}}+\cdots)+349d_{\rm{R1}}^3d_{\rm{R2}}^3d_{\rm{BH}}^3) \notag\\
&+ (80(d_{\rm{R1}}^4d_{\rm{R2}}^4+\cdots)+644(d_{\rm{R1}}^4d_{\rm{R2}}^2d_{\rm{BH}}^2+\cdots)) \notag\\
&+(124(d_{\rm{R1}}^5d_{\rm{R2}}d_{\rm{BH}}+\cdots)+1452(d_{\rm{R1}}^3d_{\rm{R2}}^3d_{\rm{BH}}+\cdots)) \notag\\
&+(392(  {d_{\rm{R1}}^4d_{\rm{R2}}^2}+\cdots)+5664d_{\rm{R1}}^2d_{\rm{R2}}^2d_{\rm{BH}}^2) \notag\\
&+3300(d_{\rm{R1}}^3d_{\rm{R2}}d_{\rm{BH}}+\cdots)+(200(d_{\rm{R1}}^4+\cdots)+2304(d_{\rm{R1}}^2d_{\rm{R2}}^2+\cdots)) \notag\\
&+4604d_{\rm{R1}}d_{\rm{R2}}d_{\rm{BH}}+664(d_{\rm{R1}}^2+\cdots)+144\Big),
\end{align}
where $d_{\rm R1R2BH}:=d_{\rm{R1}}d_{\rm{R2}}d_{\rm{BH}} = d_{\rm Total}$ and \(\cdots\) means sum over symmetrized products with the same total power (without repeats), for example
\begin{align}
\label{dotexample1}
d_\text{R1}^2d_\text{R2}^4 + \cdots &= d_\text{R1}^2d_\text{R2}^4+d_\text{R2}^2d_\text{R1}^4+d_\text{R2}^2d_\text{BH}^4+d_\text{BH}^2d_\text{R2}^4+d_\text{R1}^2d_\text{BH}^4+d_\text{BH}^2d_\text{R1}^4\,, \\
\label{dotexample2}
d_\text{R1}^4d_\text{R2}^2d_\text{BH}^2 + \cdots &= d_\text{R1}^4d_\text{R2}^2d_\text{BH}^2+d_\text{R1}^2d_\text{R2}^4d_\text{BH}^2+d_\text{R1}^2d_\text{R2}^2d_\text{BH}^4 \,.
\end{align}
In terms of expansion of \(d_\text{BH}\), we have
\begin{align} \overline{(Z^{(3)}_2)^2} = \frac{1}{d_{\rm R1}^4d_{\rm R2}^4}\left(1+\frac{4(d_{\rm R1}^2+d_{\rm R2}^2+5)}{d_{\rm R1}d_{\rm R2}d_{\rm BH}}+\order{d_{\rm BH}^{-2}}\right). \end{align}
Therefore we get
\begin{align}
\overline{(\delta Z^{(3)}_2)^2} = \overline{(Z^{(3)}_2)^2}-\overline{Z^{(3)}_2}^2 = \frac{16}{d^5_\text{R1}d^5_\text{R2}d_\text{BH}} + \order{d_\text{BH}^{-2}},
\end{align}
and
\begin{align} \frac{\overline{(\delta Z^{(3)}_2)^2}}{\overline{Z^{(3)}_2}^2} = \frac{16}{d_{\rm {R1R2BH}}}+\order{d_\text{BH}^{-2}}. \end{align}

For the other term, using the generic expression for \(\overline{(Z^{(3)}_1)^n}\) \footnote{A proof for this expression is given in Appendix~\ref{app:proof}.}, we obtain
\begin{align}
\frac{\overline{(\delta (Z_1^{(3)})^4)^2}}{\overline{(Z_1^{(3)})^4}^2}
&= \frac{d_{\rm {R1R2BH}}(1+d_{\rm {R1R2BH}})(2+d_{\rm {R1R2BH}})\cdots(7+d_{\rm {R1R2BH}})}{\left(d_{\rm {R1R2BH}}(1+d_{\rm {R1R2BH}})(2+d_{\rm {R1R2BH}})(3+d_{\rm {R1R2BH}})\right)^2} - 1\notag\\
&= \frac{16}{d_{\rm {R1R2BH}}} + \order{d_{\text{BH}}^{-2}}
\end{align}
Very nicely, despite that the two terms both contribute in the \(1/d_\text{BH}\) order, their net effect in the final answer cancels out.
Therefore it turns out there is no corrections in the \(1/d_\text{BH}\) order from the variance. 

This result can be understood as follows. The variance $\overline{(\delta Z^{(3)}_2)^2}$ can be thought of as the contribution from the “connected” permutation elements across two replica copies ({\it i.e.,} replica $\{1,2,3,4\}$ and $\{5,6,7,8\}$) of $Z^{(3)}_2$. In the leading order, these elements are simply swaps between two $Z_2^{(3)}$. The number of such swaps is $n^{\mathtt{q}-1}\times n^{\mathtt{q}-1}=16$, which explains why the coefficient is $16$. This counting is true for both $Z_2^{(3)}$ and $(Z_1^{(3)})^{4}$.
Also, since a simple swap across two different replicas must increase the cycle counting by 1, after dividing out the denominator we get the same scaling factor of $1/d_{\rm {R1R2BH}}$. Therefore, in the \(1/d_\text{BH}\) order, the corrections cancel out.

The $m>2$ terms are also zero in the $1/d_{\rm BH}$ order. For example, at $m=3$, we have $\overline{(\delta Z)^3}=\overline{Z^3}-3\overline{Z}\overline{Z^2}+2\overline{Z}^3$, which receives contribution only from connected \footnote{By connected we mean that the permutation does not involve self contractions within any single copy of 4 density matrices.} group permutations in $S_{12}$ (12 since $m=3$ and we need to use the ``triple'' replica model). Note that $1/d_{\rm BH}$ corrections are given by a single swap permutation only and two or more swap permutations always yield higher order corrections. In addition, since any single swap will leave at least one copy of $\delta Z$ for $m>2$, they do not contribute to $\overline{(\delta Z)^3}$. Therefore, we conclude that, in the $1/d_\text{BH}$ order, there are no corrections from $\sum_{m=1}^\infty\frac{(-1)^{m-1}}{m}\left(\frac{\overline{(\delta Z_2^{(3)})^m}}{\overline{Z_2^{(3)}}^m}-\frac{\overline{(\delta (Z_1^{(3)})^4)^m}}{\overline{(Z_1^{(3)})^4}^m}\right)$ in (\ref{avblog}).
For a more detailed argument by mathematical induction, please see Sec.~\ref{sec:org704fc05}.

\section{Black hole multi-entropy curves for other $n \ge 2$ and $\mathtt{q} \ge 3$ cases}
\label{sec:org1e539bc}

In this section, we show more explicit examples of the black hole multi-entropy curves. We focus on two cases: $n=3,\mathtt{q}=3$ and $n=2,\mathtt{q}=4$, where we obtain exact results of R\'enyi multi-entropies by explicitly summing over the replica permutation group. As we will see, these examples exhibit the same generic properties for the multi-entropy curve given in the introduction as well as in Sec.~\ref{q=3n=2multicurve}. 

\subsection{A black hole multi-entropy curve for \(n=3, \mathtt{q}=3\)}
\label{sec:org531c218}
Let us first consider \(n=3, \mathtt{q}=3\) R\'enyi multi-entropy. In this case, there are $3^2=9$ 
reduced density matrices, as shown in Fig.~\ref{fig:contraction-multi} for $Z_n^{(3)}$. 

The permutation group \(S_9\) has \(9!=362880\) elements. It is too complicated to calculate all diagrams by hand. We perform the summation using Mathematica and we get the following expression:
\begin{align}
\overline{Z^{(3)}_3} = \frac{1}{(d_{\text{R1R2BH}})^8}&\bigg(d_{\text{R1R2BH}}^2((d_\text{R1}^6+\cdots)+9(d_\text{R1}^2d_\text{R2}^4+\cdots)+27d_\text{R1}^2d_\text{R2}^2d_\text{BH}^2) \notag\\
&+d_{\text{R1R2BH}}( 9(d_\text{R1}^6d_\text{R2}^2+\cdots)+45(d_\text{R1}^4d_\text{R2}^4+\cdots)+180(d_\text{R1}^4d_\text{R2}^2d_\text{BH}^2+\cdots) ) \notag\\
&+((21d_\text{R1}^6d_\text{R2}^4+\cdots)+162(d_\text{R1}^6d_\text{R2}^2d_\text{BH}^2+\cdots)+1032(d_\text{R1}^4d_\text{R2}^4d_\text{BH}^2+\cdots)) \notag\\
&+d_{\text{R1R2BH}}(18(d_\text{R1}^6+\cdots)+927(d_\text{R1}^4d_\text{R2}^2+\cdots)+7485d_\text{R1}^2d_\text{R2}^2d_\text{BH}^2) \notag\\
&+(129(d_\text{R1}^6d_\text{R2}^2+\cdots)+648(d_\text{R1}^4d_\text{R2}^4+\cdots)+9678(d_\text{R1}^4d_\text{R2}^2d_\text{BH}^2+\cdots)) \notag\\
&+d_{\text{R1R2BH}}(2412(d_\text{R1}^4+\cdots)+15942(d_\text{R1}^2d_\text{R2}^2+\cdots)) \notag\\
&+(66(d_\text{R1}^6+\cdots)+3945(d_\text{R1}^4d_\text{R2}^2+\cdots)+51324d_\text{R1}^2d_\text{R2}^2d_\text{BH}^2) \notag\\
&+(1434(d_\text{R1}^4+\cdots)+16290(d_\text{R1}^2d_\text{R2}^2+\cdots)+25683(d_\text{R1}^2d_\text{R2}d_\text{BH}+\cdots)) \notag\\
&+35499d_\text{R1}d_\text{R2}d_\text{BH}+5340(d_\text{R1}^2+\cdots)+1512\bigg),
\label{overlineZ33}
\end{align}
where $d_{\rm R1R2BH}\equiv d_{\rm R1}d_{\rm R2}d_{\rm BH}$.
Here again, \(\cdots\) means sum over symmetrized products with the same total power (without repeats) as in
\eqref{dotexample1} and \eqref{dotexample2}. A simple consistency check that this $ \overline{{Z}_3^{(3)}}$ comes from the summation over \(9!\) patterns of Wick contraction figures is that, if all dimensions are set to 1 in \eqref{overlineZ33}, $ \overline{{Z}_3^{(3)}}$  becomes \(362880\).

It is useful to organize the above expressions in power series of \(d_\text{BH}\) for later comparison:
\begin{align} 
\overline{Z^{(3)}_3} = (d_\text{R1}d_\text{R2})^{-6} + d_\text{BH}^{-1}(9d_\text{R1}^{-5}d_\text{R2}^{-7}+9d_\text{R1}^{-7}d_\text{R2}^{-5}+18d_\text{R1}^{-7}d_\text{R2}^{-7}) + \order{d_\text{BH}^{-2}} .
\label{overlineZ33expansion}
\end{align}
The replica partition function in the denominator is
\begin{align}
\overline{(Z^{(3)}_1)^9} = \frac{1}{(d_\text{R1}d_\text{R2}d_\text{BH})^8}(1+d_\text{R1}d_\text{R2}d_\text{BH})(2+d_\text{R1}d_\text{R2}d_\text{BH})\cdots(8+d_\text{R1}d_\text{R2}d_\text{BH}). \label{ReplicaPF1}
\end{align}
At leading order it is simply unity, as expected. The form of $\overline{Z_1^n}$ does follow a simple pattern, which we prove in Appendix~\ref{app:proof}.

Then the $n=3$ R\'enyi multi-entropy is given in the leading order as
\begin{align}
S^{(3)}_3 \approx - \frac{1}{2} \log  \frac{\overline{Z^{(3)}_3}}{\overline{(Z^{(3)}_1)^9}},
\label{S33}
\end{align}
where $\overline{Z^{(3)}_3}$ and $\overline{(Z^{(3)}_1)^9}$ are given by \eqref{overlineZ33} and \eqref{ReplicaPF1} respectively. 
In the leading order in $d_{\rm BH} \to \infty$, using \eqref{overlineZ33expansion}, we obtain Hawking's prediction  
\begin{align}
    S_{3 \; \text{Hawking}}^{(3)}:= 
\lim_{d_{\rm BH} \to \infty} S^{(3)}_3 = 3 \log  d_\text{R1}  d_\text{R2}.
\label{S33Hawking}
\end{align}

\begin{figure}[t]
    \centering
    \includegraphics[width=9cm]{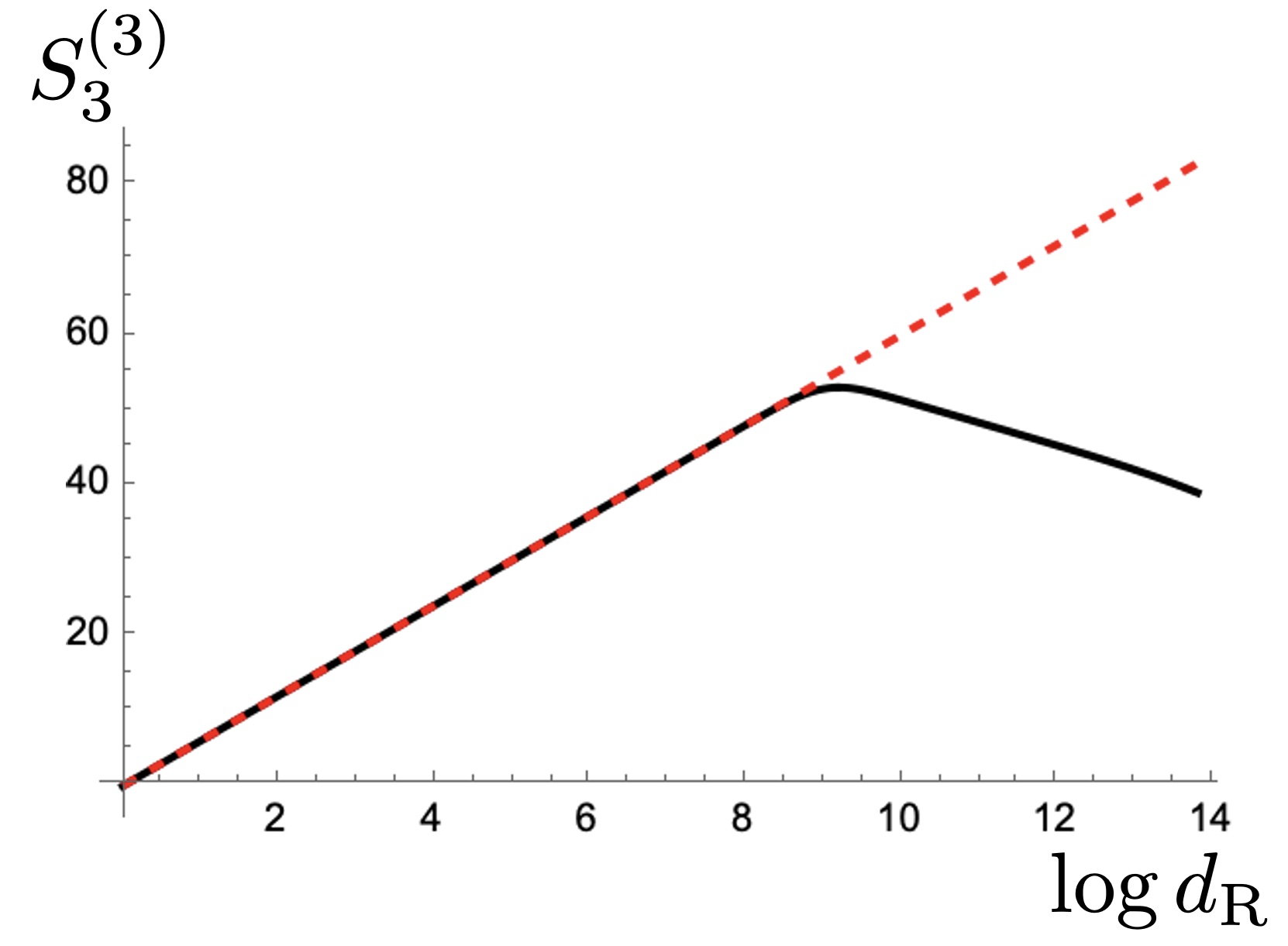}
    \caption{A black solid curve is a black hole multi-entropy curve for the $n=3$ R\'enyi multi-entropy with $\mathtt{q}=3$ case, $S^{(3)}_{3}$, as a function of $\log d_{\rm R}$ for the parameter choice $d_{\rm Total} =d_{\rm R1} d_{\rm R2} d_{\rm BH} =  10^{12}$. This curve is obtained from \eqref{S33}. On the other hand, a red dashed curve is Hawking's prediction given by \eqref{S33Hawking}. They begin to deviate at multi-entropy time, $\log d_{\rm R} = \log \left( d_{\rm Total}\right)^{1/3} = \log 10^4 \approx 9.21$. }
    \label{fig:multiq=3n=3}
\end{figure}

Given these, we can obtain a black hole multi-entropy curve. The result is shown in Fig.~\ref{fig:multiq=3n=3}.  

As we claimed in Sec. \ref{q=3n=2multicurve}, the multi-entropy time, where $S^{(3)}_{3}$ reaches the maximal value, does not depend on R\'enyi $n$ and it deviates significantly from $S^{(3)}_{3\;\text{Hawking}}$ after the multi-entropy time. Again the multi-entropy does not become zero at the end of black hole evaporation, which represents the entanglement between final Hawking radiation.

\subsection{A black hole multi-entropy curve for \(n=2, \mathtt{q}=4\)}
\label{sec:org88a13f8}
Let us also examine the 4-party case. Note that this $\mathtt{q}=4$ case would be an example that cannot be handled by entanglement negativity or reflected entropy. We now need to look at a 4-party random state \(\psi_{ABCD}\). In this case, the replica reduced density matrices $\rho$ form 3-dim cube, $n \times n \times n$. For \(\mathtt{q}=4\) and \(n=2\), $Z^{(4)}_2$ includes \(n^{\mathtt{q}-1}=2^3=8\) density matrices contracted according to the permutations\footnote{Here we label the lower $2\times 2$ replicas as $1 \cdots 4$ and upper replicas as $5 \cdots 8$.}
\begin{align} g_A = (12)(34)(56)(78), \quad g_B = (13)(24)(57)(68), \quad g_C = (15)(26)(37)(48), \quad g_D = e. \end{align}
As in the \(\mathtt{q}=3\) case, the permutation \(g_{A/B/C}\) corresponds to contractions along the rows/columns/hights, respectively.
We must sum over the group \(S_8\) to compute
\begin{align}
\overline{Z^{(4)}_2}=\sum_{g\in S_{8}}d^{-d(g,g_A)}_{A}d^{-d(g,g_B)}_{B}d^{-d(g,g_C)}_{C}d^{-d(g,e)}_{D}.
\end{align}
Again this is too complicated to do by hand. But by using Mathematica, one can obtain the result as
\begin{align}
\hspace{-1mm}\overline{Z^{(4)}_2} &= \frac{1}{(d_{ABCD})^7}\Big( d_{ABCD}^3(d_A^4+\cdots) +d_{ABCD}(2(d_A^4d_B^4d_C^2+\cdots)+4(d_A^5d_B^3d_Cd_D+\cdots)) \notag\\
&\,\,+((12d_A^4d_B^4d_C^4+\cdots)+2(d_A^5d_B^5d_Cd_D+\cdots)+26(d_A^5d_B^3d_C^3d_D+\cdots)+16(d_A^6d_B^2d_C^2d_D^2+\cdots)\notag\\ &\,\, \quad+32(d_A^4d_B^4d_C^2d_D^2+\cdots)+36d_A^3d_B^3d_C^3d_D^3)) \notag\\
&\,\, +(44(d_A^4d_B^4d_C^2+\cdots)+52(d_A^5d_B^3d_Cd_D+\cdots)+312(d_A^3d_B^3d_C^3d_D+\cdots)+384(d_A^4d_B^2d_C^2d_D^2+\cdots))\notag\\
&\,\, +(24(d_A^4d_B^4+\cdots)+224(d_A^4d_B^2d_C^2+\cdots)+70(d_A^5d_Bd_Cd_D+\cdots)+1114(d_A^3d_B^3d_Cd_D+\cdots)\notag\\ 
&\,\, \quad +3632d^2_Ad^2_Bd^2_Cd^2_D)\notag\\
&\,\, +(124(d_A^4d_B^2+\cdots)+1632(d_A^2d_B^2d_C^2+\cdots)+2134(d_A^3d_Bd_Cd_D+\cdots))\notag\\
&\,\, +(52(d_A^4+\cdots)+448(d_A^2d_B^2+\cdots)+2400d_Ad_Bd_Cd_D)\notag\\
&\,\, +92(d_A^2+d_B^2+d_C^2+d_D^2)\Big),
\label{Z42}
\end{align}
where $d_{ABCD}:=d_Ad_Bd_Cd_D$.\footnote{Again, a simple consistency check that this $ \overline{{Z}_2^{(4)}}$ comes from the summation over \(8!\) patterns of Wick contraction figures is that, if all dimensions are set to 1 in \eqref{Z42}, $ \overline{{Z}_2^{(4)}}$  becomes \(40320\). }
At large $d_D$, we get
\begin{align} 
\overline{Z^{(4)}_2} = \frac{1}{d_A^4d_B^4d_C^4}\left(1+\frac{4(d_A^2+d_B^2+d_C^2+4)}{d_{ABCD}}+\order{d_D^{-2}}\right),  \end{align}
and the averaged replica partition function in the denominator is
\begin{align} 
\overline{(Z^{(4)}_1)^8} = \frac{(1+d_{ABCD})(2+d_{ABCD})\cdots(7+d_{ABCD})}{d_{ABCD}^7}.  
\label{Z148}
\end{align}

Then a multi-entropy for $n=2$, $\mathtt{q}=4$ case is given by 
\begin{align}
\label{S42}
S^{(4)}_2 \approx -  \log  \frac{\overline{Z^{(4)}_2}}{\overline{(Z^{(4)}_1)^8}}\,,
\end{align}
where $\overline{Z^{(4)}_2}$ and $\overline{(Z^{(4)}_1)^8}$ are given by \eqref{Z42} and \eqref{Z148} respectively. 

Let us choose one of the indices, say $D$ as BH, and $A, B, C$ as R1, R2, R3.

Hawking's limit $d_{\rm BH} \to \infty$ gives 
\begin{align}
\label{S42Hawking}
S^{(4)}_{2 \;\text{Hawking}}:= \lim_{d_{\rm BH} \to \infty} S^{(4)}_{2}  
= 4 \log  d_\text{R1}   d_\text{R2}   d_\text{R3} 
\end{align}

\begin{figure}[t]
    \centering
    \includegraphics[width=9cm]{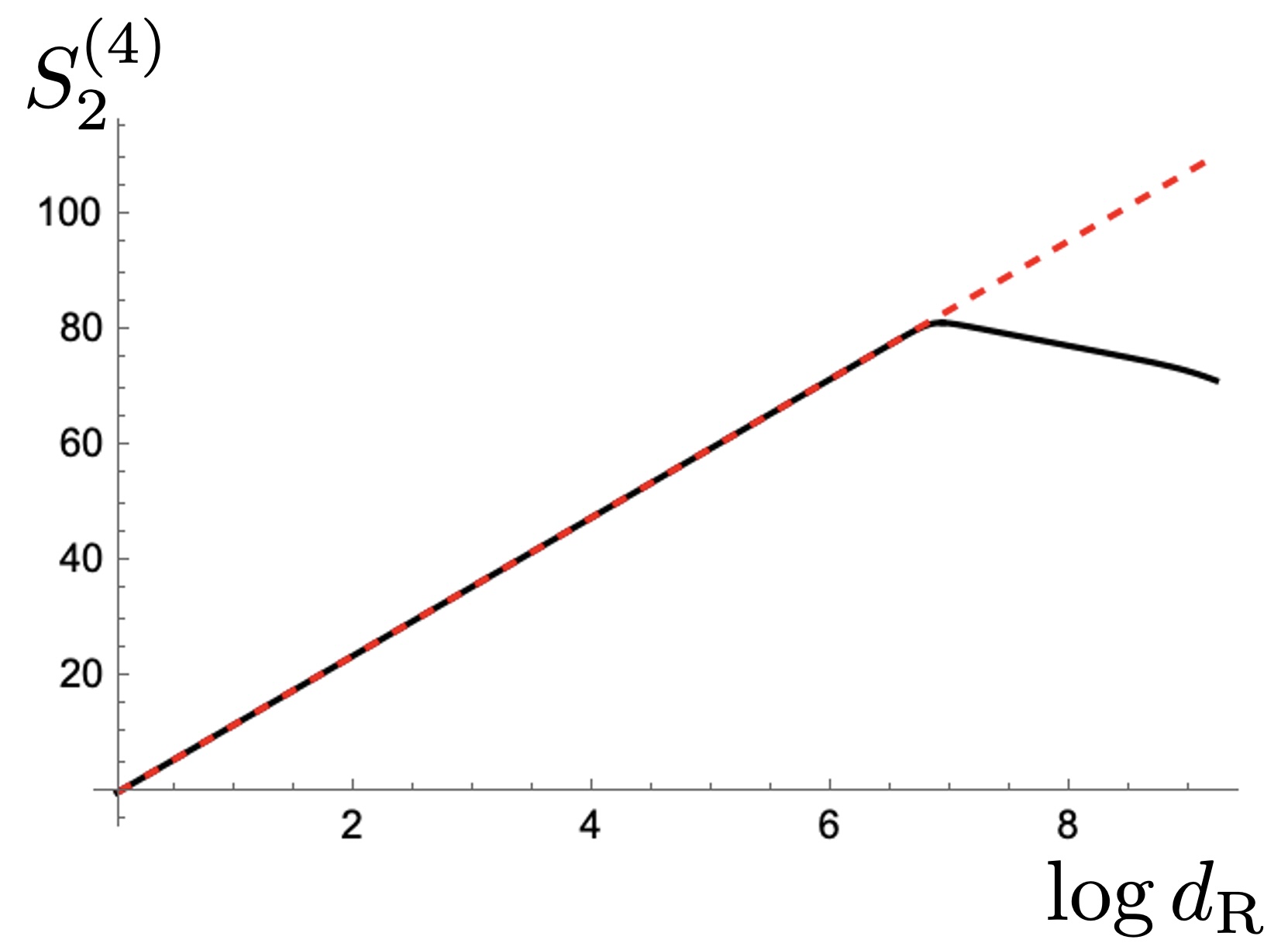}
    \caption{A black solid curve is a black hole multi-entropy curve for the $n=2$ R\'enyi multi-entropy with $\mathtt{q}=4$ case, $S^{(4)}_{2}$, as a function of $\log d_{\rm R}$ for the parameter choice $d_{\rm Total} =d_{\rm R1} d_{\rm R2} d_{\rm BH} =  10^{12}$. 
    Origin is $d_{\rm R}=1$ and it ends at $\log d_{\rm BH} = 0$ and $d_{\rm R} = \log \left( d_{\rm Total} \right)^{1/4} \approx 9.21$. This curve is obtained from \eqref{S42}. 
    On the other hand, a red dashed curve is Hawking's prediction given by \eqref{S42Hawking}. They begin to deviate at multi-entropy time, $\log d_{\rm R} = \log \left( d_{\rm Total}\right)^{1/4} = \log 10^3 \approx 6.91$. }
    \label{fig:multiq=4n=2}
\end{figure}

With these in hand, we can write down a black hole multi-entropy curve for $n=2$, $\mathtt{q}=4$ case. See Fig.~\ref{fig:multiq=4n=2}. Generic properties of the curve are as we commented in Sec. \ref{q=3n=2multicurve}. %Note that the multi-entropy time for $\mathtt{q}=4$ case is later than the multi-entropy time for $\mathtt{q} =3$ case as expected. }

\section{Analytic expressions for general $n$ and $\mathtt{q}$}
\label{sec:analytic}
In this section, we present partial results regarding the multi-entropy curve for arbitrary $n$ and $\mathtt{q}$.
Our main goal in this section is to obtain analytic expressions at either ends of the multi-entropy curve. In Sec.~\ref{sec:subleadinggeneral}, we focus on the early stage of the evaporation, where $d_{\rm BH}\sim d_{\rm Total}$. We obtain an analytic expression of the multi-entropy as an expansion in $d_{\rm BH}$ and its first order correction, which has a straightforward analytic continuation to $n\to 1$.
In Sec.~\ref{sec:latetime} we focus on the end of the evaporation where $d_R\gg d_{\rm BH}$. In contrary to the other case we do not have a complete understanding of this limit yet. Nonetheless we will conjecture an expression for the multi-entropy in leading order of $1/d_R$.

\subsection{Early time limit}
\label{sec:subleadinggeneral}

Here we generalize the computation of (\ref{MElargeBHlimit}) to arbitrary $S^{(\mathtt{q})}_n$. 
Let us assume that the $\mathtt{q}$-partite random tensor has bond dimensions \(\{d_1,d_2,\cdots,d_\mathtt{q}\}\).
We interpret $d_\mathtt{q}=d_{\text{BH}}$ as the dimension of black hole microstates and $d_i$ with $i<q$ as dimensions of Hawking radiation. We are interested in the leading and subleading behaviors at large $d_\mathtt{q}$.

\begin{figure}[t]
  \centering
  \begin{subfigure}[t]{.32\textwidth}
    \centering
    \includegraphics[width=\linewidth]{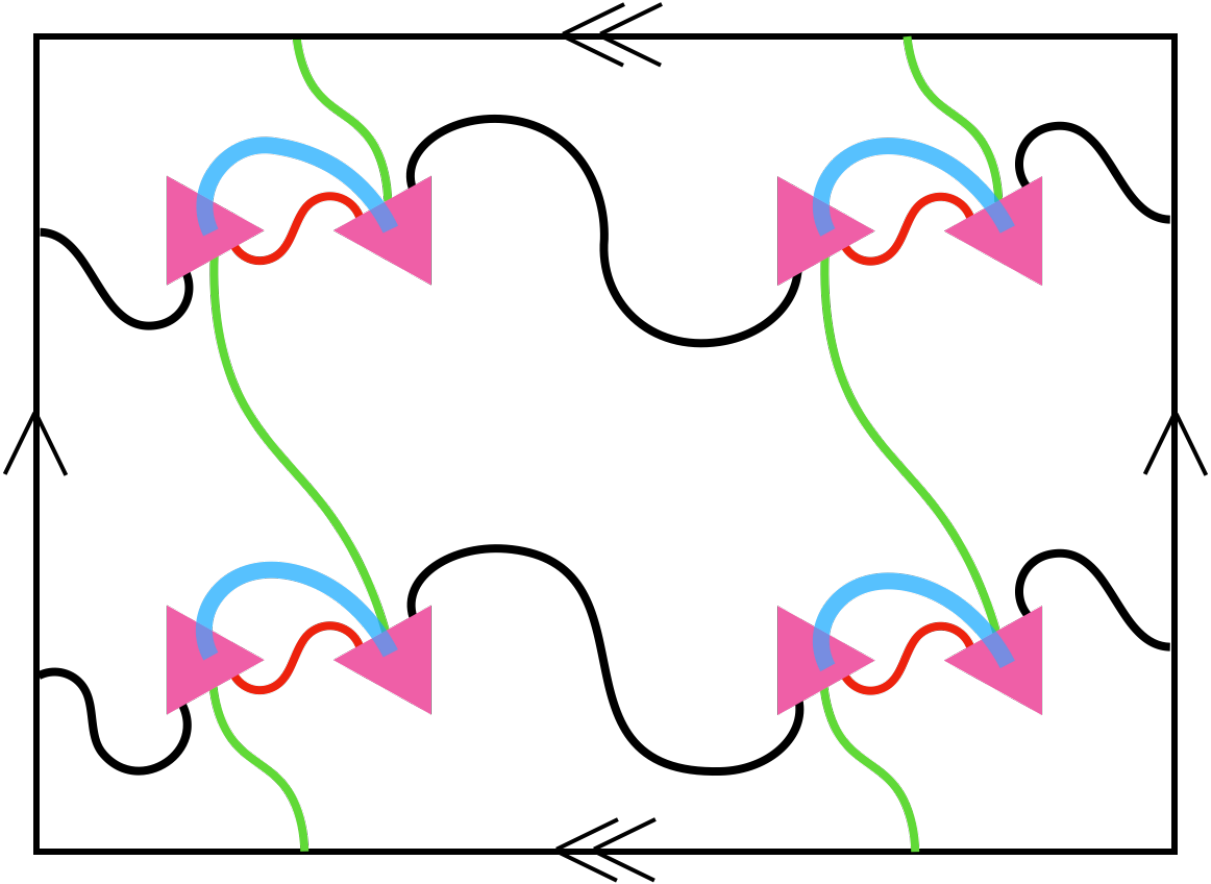}
    \caption{Identity permutation}
    \label{fig:type_0}
  \end{subfigure}
  \begin{subfigure}[t]{.32\textwidth}
    \centering
    \includegraphics[width=\linewidth]{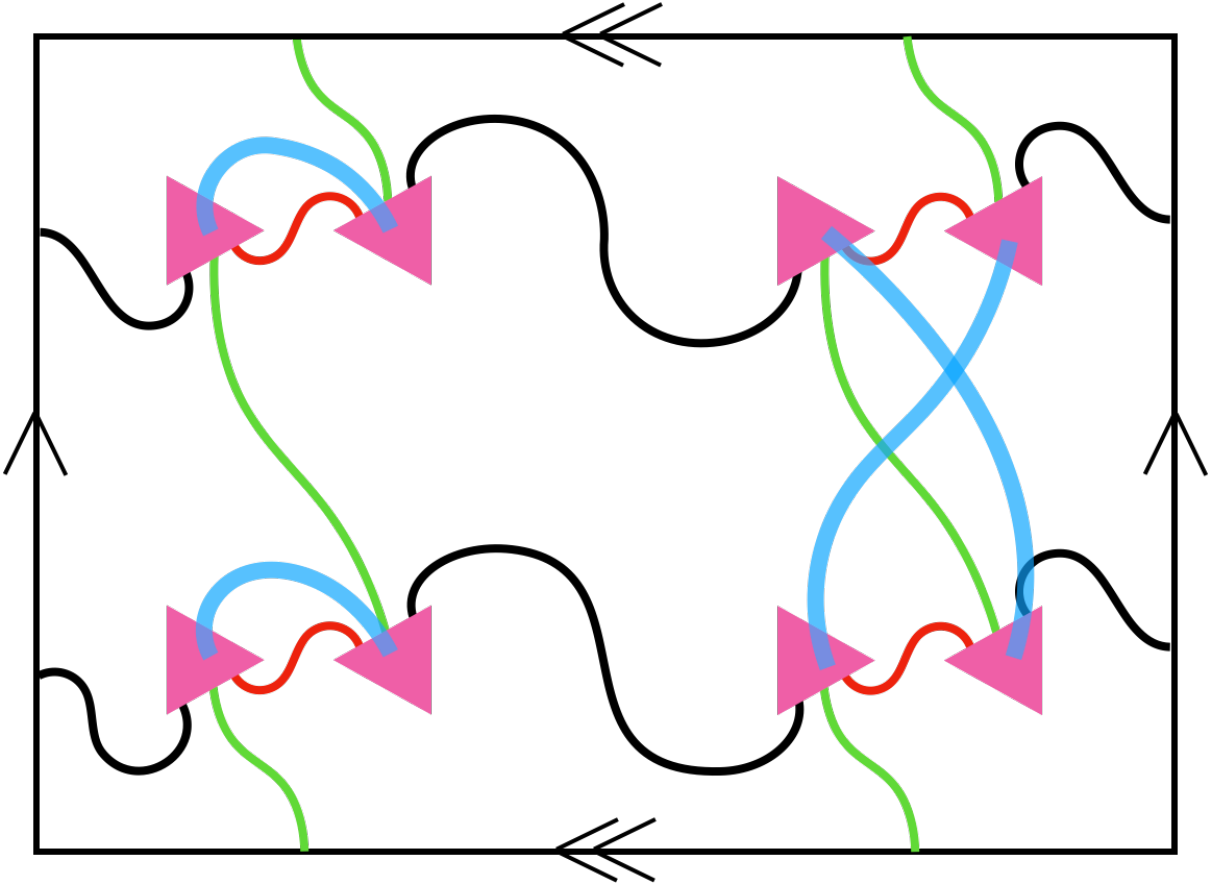}
    \caption{Type I}
    \label{fig:type_I}
  \end{subfigure}
  \begin{subfigure}[t]{.32\textwidth}
    \centering
    \includegraphics[width=\linewidth]{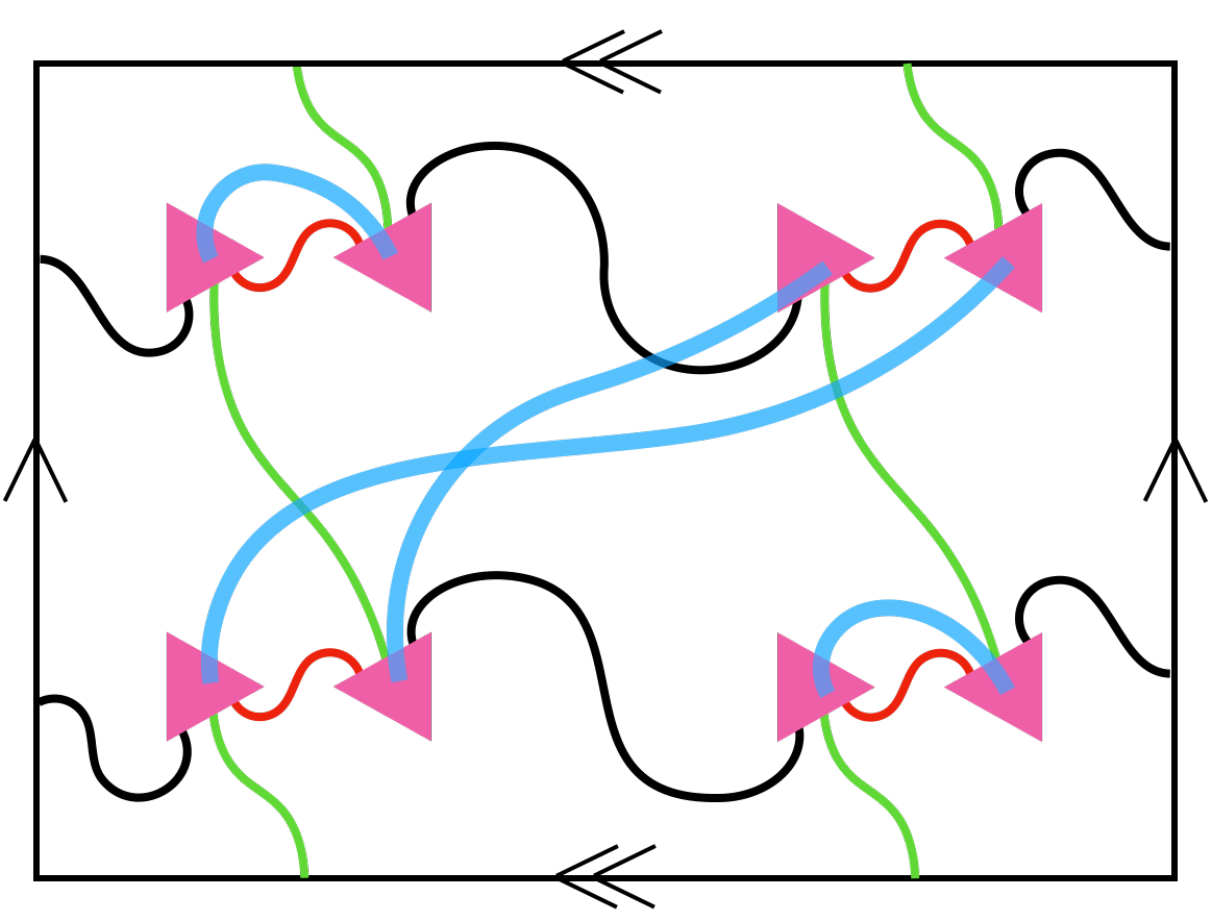}
    \caption{Type II}
    \label{fig:type_II}
  \end{subfigure}
  \caption{Examples of the permutations we consider in this section. We follow the same notation as in Fig.~\ref{fig:twoexamples}. In (a) the total number of loops are (2, 2, 4) for (R1, R2, BH). In (b), $\tau_{ab}$ is chosen from the cycle of column $g_2$ and the total number of loops are  (1, 3, 3), the R1 loops decrease by 1 and R2 increase by 1. In (c), $\tau_{ab}$ is not chosen from the cycle of $g_1$ nor $g_2$. In this case, the total number of loops are (1, 1, 3), where both R1 and R2 loops decrease by 1.}
\end{figure}

The replica group associated to this case is \(S_{n^{\mathtt{q}-1}}\), which can be thought of as the permutation group of a \((\mathtt{q}-1)\)-dimensional lattice hypercube of length $n$. We set $g_\mathtt{q}=e\in S_{n^{\mathtt{q}-1}}$, and the other permutation operators \(g_i\) can be described as cyclic permutations for the elements along one of the axes.
They satisfy
\begin{align} d(g_i,g_j) = n^{\mathtt{q}-2}(n-1),\quad { 1 \le  i\ne j \le \mathtt{q} }\,, \end{align}
where the Cayley distance of $S_{n^{\mathtt{q}-1}}$ is $d(g,h) = n^{\mathtt{q}-1} - \#(gh^{-1})$. Our target is
\begin{align}
    S_n^{(\mathtt{q})}\approx\frac{1}{1-n}\log \frac{\overline{Z_n^{(\mathtt{q})}}}{\overline{(Z_1^{(\mathtt{q})})^{n^{\mathtt{q}-1}}}}=\frac{1}{1-n}\log\left(\frac{\sum_{g\in S_{n^{\mathtt{q}-1}}}d^{-d(g,g_1)}_{1}\dots d^{-d(g,g_{\mathtt{q}-1})}_{\mathtt{q}-1} d^{-d(g,e)}_{\mathtt{q}}}{\sum_{g\in S_{n^{\mathtt{q}-1}}}(d_{1}\dots d_{\mathtt{q}})^{-d(g,e)}}\right).\label{menq}
\end{align}
In the leading order at large $d_\mathtt{q}$, the dominant contribution in the sum comes from the elements $g$ satisfying  $d(g,e)=0$, which is, the identity permutation. Thus the averaged replica partition function is
\begin{align} \lim_{d_\mathtt{q}\to\infty}\overline{Z^{(q)}_n} =(d_1\cdots d_{\mathtt{q}-1})^{n^{\mathtt{q}-2}(1-n)}, \end{align}
See Fig.~\ref{fig:type_0}.

We now examine the $O(1/d_\mathtt{q})$ corrections.
They are characterized by elements $g$ such that $d(g,e)=1$, which are simply two-element transpositions \(g=\tau_{ab}=(ab)\) with $a,b\in\{1,\cdots,n^{\mathtt{q}-1}\}$.
There are in total $n^{\mathtt{q}-1}(n^{\mathtt{q}-1}-1)/2$ of them.
We classify them into two types based on how they interact with the rest of $g_i$'s. Since $\tau_{ab}$  is a transposition, it can only modify the Cayley distance by $1$ compared to $d(e,g_i)$:
\begin{itemize}
\item Type I: \\
\(d(\tau_{ab},g_i)=n^{\mathtt{q}-2}(n-1)-1\) for some \(i<\mathtt{q}\) and \(d(\tau_{ab},g_j)=n^{\mathtt{q}-2}(n-1)+1\) for all \(j<\mathtt{q}, j \ne i\).
They are characterized by \(a,b\) belonging to the same cycle of some \(g_i\). See Fig.~\ref{fig:type_I} for an example.
The total number of type I transpositions are
\begin{align}
 &(\text{\# of } g_i \text{ excluding } g_\mathtt{q})\times(\text{\# of cycles in }g_i)\times\text{(\# of swaps within each cycle)} \notag\\
 =& (\mathtt{q}-1)n^{\mathtt{q}-2}\frac{n(n-1)}{2}.
\end{align}

\item Type II: \\ 
\(d(\tau_{ab},g_i)=n^{\mathtt{q}-2}(n-1)+1\) for all \(i<\mathtt{q}\).
These are the remaining ones. See Fig.~\ref{fig:type_II} for an example.
The total number of type II transpositions are
\begin{align} \frac{n^{\mathtt{q}-1}(n^{(\mathtt{q}-1)}-1)}{2}- (\mathtt{q}-1)n^{\mathtt{q}-2}\frac{n(n-1)}{2} = \frac{n^{2\mathtt{q}-2}-(\mathtt{q}-1)n^\mathtt{q}+(\mathtt{q}-2)n^{\mathtt{q}-1}}{2}.\end{align}
\end{itemize}
Note that not exist swaps that simultaneously decrease the Cayley distance for two or more \(g_i\) since the intersection of any two twist operators \(g_i\cap g_j=e\).

We can now formulate the $\order{d_\mathtt{q}^{-1}}$ correction to the averaged replica partition function:
\begin{align}
\overline{Z^{(\mathtt{q})}_{n}} = (d_1\cdots d_{\mathtt{q}-1})^{n^{\mathtt{q}-2}(1-n)} &\left[1+n^{\mathtt{q}-2}\frac{n(n-1)}{2}\frac{d_1^2+d_2^2+\cdots+d_{\mathtt{q}-1}^2}{d_1d_2\cdots d_\mathtt{q}} \right.\notag\\
&\quad +\left.\frac{n^{2\mathtt{q}-2}-(\mathtt{q}-1)n^\mathtt{q}+(\mathtt{q}-2)n^{\mathtt{q}-1}}{2}\frac{1}{d_1d_2\cdots d_\mathtt{q}} + \order{d^{-2}_\mathtt{q}}\right].
\end{align}
For the denominator, we have (see Appendix~\ref{app:proof} for a proof):
\begin{align} \overline{(Z^{(\mathtt{q})}_1)^{n^{\mathtt{q}-1}}} = \frac{1}{(d_1d_2\cdots d_\mathtt{q})^{n^\mathtt{q}}}\prod_{i=0}^{n^{\mathtt{q}-1}-1} (i+d_1d_2\cdots d_\mathtt{q}) = 1 + \frac{n^{\mathtt{q}-1}(n^{\mathtt{q}-1}-1)}{2}(d_1d_2\cdots d_\mathtt{q})^{-1} + \order{d_\mathtt{q}^{-2}}.\end{align}
Putting them together we obtain
\begin{align} \frac{\overline{Z^{(\mathtt{q})}_n}}{\overline{(Z^{(\mathtt{q})}_1)^{n^{\mathtt{q}-1}}}} = (d_1\cdots d_{\mathtt{q}-1})^{n^{\mathtt{q}-2}(1-n)}\left[1+\frac{n^{\mathtt{q}-1}(n-1)}{2}\frac{d_1^2+d_2^2+\cdots+d_{\mathtt{q}-1}^2-(\mathtt{q}-1)}{d_1d_2\cdots d_\mathtt{q}} + \order{d^{-2}_\mathtt{q}}\right]. \end{align}
By setting $d_\mathtt{q}=d_\text{BH}$ and $d_{i<q}=d_{\text{R}_i}$, the R\'enyi multi-entropy and $n\to 1$ limit is given by
\begin{align}
\label{finalresult}
S^{(\mathtt{q})}_n &= n^{\mathtt{q}-2}\log(d_\text{R1}\cdots d_{\text{R}\mathtt{\mathtt{q}}-1}) - \frac{n^{\mathtt{q}-1}}{2}\frac{d_\text{R1}^2+d_\text{R2}^2+\cdots d_{\text{R}\mathtt{\mathtt{q}}-1}^2-(\mathtt{q}-1)}{d_\text{R1}\cdots d_{\text{R}\mathtt{\mathtt{q}}-1}d_\text{BH}} + \order{d^{-2}_\text{BH}}, \\
S^{(\mathtt{q})}_1 &= \log(d_\text{R1}\cdots d_{\text{R}\mathtt{\mathtt{q}}-1}) - \frac{1}{2}\frac{d_\text{R1}^2+d_\text{R2}^2+\cdots d_{\text{R}\mathtt{\mathtt{q}}-1}^2-(\mathtt{q}-1)}{d_\text{R1}\cdots d_{\text{R}\mathtt{\mathtt{q}}-1}d_\text{BH}} + \order{d^{-2}_\text{BH}}.
\label{finalresult1}
\end{align}
This is one of our main results. 
We have checked the above expressions against various results for specific $(\mathtt{q},n)$ in Sec.~\ref{secRenyimultientropyn2q3} and Sec.~\ref{sec:org1e539bc}.

Note that in \eqref{finalresult} and \eqref{finalresult1} in the leading order, 
\begin{align}
\label{finalresult2}
\lim_{d_{\rm BH} \to \infty} S^{(\mathtt{q})}_n &= n^{\mathtt{q}-2}\log(d_\text{R1}\cdots d_{\text{R}\mathtt{\mathtt{q}}-1}) \,, \\
\lim_{d_{\rm BH} \to \infty}  S^{(\mathtt{q})}_1 &= \log(d_\text{R1}\cdots d_{\text{R}\mathtt{\mathtt{q}}-1}) \,,
\label{finalresult3}
\end{align}
therefore suppose we define our ``time'' as 
\begin{align}
\mbox{time} &\equiv \log(d_\text{R1}\cdots d_{\text{R}\mathtt{\mathtt{q}}-1})  = \log \mbox{(total radiation dimension)} \\
&\propto \mbox{(total number of Hawking particles)}
\end{align}
then the slope of $\lim_{d_{\rm BH} \to \infty} S^{(\mathtt{q})}_n$ with respect to time depends on $n$ and $\mathtt{q}$ but the one of $\lim_{d_{\rm BH} \to \infty}  S^{(\mathtt{q})}_1$ is independent of $\mathtt{q}$. 
Thus, we can draw black hole multi-entropy curves where $x$-axis is our time, and $y$-axis is $S^{(\mathtt{q})}_1$. This is exactly the black hole multi-entropy curve we draw for Fig.~\ref{fig:BHmultiScurve} in the introduction.   
The range of $x$-axis is $0 \le \mbox{time} \le \log {d_{\rm Total}}$, where the Page time and the multi-entropy time is respectively,  
\begin{align}
\mbox{(Page time)} &= \frac{1}{2}  \log {d_{\rm Total}}\,, \\
\mbox{(Multi-entropy time)} &= \frac{\mathtt{q}-1}{\mathtt{q}}  \log {d_{\rm Total}} > \mbox{(Page time)} \,, \quad \mbox{for $\mathtt{q} \ge 3$}
\end{align}

Note that the $x$-axis in Fig.~\ref{fig:BHmultiScurve} is $\log d_{\rm R}^{q-1}$ under \eqref{alldrthesame}, on the other hand, the one in  Fig.~\ref{fig:multiq=3n=2}, \ref{fig:MultiandPage}, \ref{fig:multiq=3n=3}, \ref{fig:multiq=4n=2} are $\log d_{\rm R}$. But they are related just by rescaling. 

\subsubsection{Corrections from the variance}
\label{sec:org704fc05}
As a generalization of Sec.~\ref{sec:orge89da24}, one can compute the $\order{d_\mathtt{q}^{-1}}$ corrections to the R\'enyi multi-entropy from the variance and the $m>2$ terms in the following expression
\begin{align} \label{avglog2}
\overline{\log \frac{Z_n^{(\mathtt{q})}}{(Z_1^{(\mathtt{q})})^{n^{\mathtt{q}-1}}}}=\log \frac{\overline{Z_n^{(\mathtt{q})}}}{\overline{(Z_1^{(\mathtt{q})})^{n^{\mathtt{q}-1}}}}+\sum_{m=1}^\infty\frac{(-1)^{m-1}}{m}\left(\frac{\overline{(\delta Z_n^{(\mathtt{q})})^m}}{\overline{Z_n^{(\mathtt{q})}}^m}-\frac{\overline{(\delta (Z_1^{(\mathtt{q})})^{n^{\mathtt{q}-1}})^m}}{\overline{(Z_1^{(\mathtt{q})})^{n^{\mathtt{q}-1}}}^m}\right).
\end{align}
The same conclusion is drawn as in Sec.~\ref{sec:orge89da24}, only the number of replicas is different.
First, $m=1$ is trivially zero since $\overline{\delta Z} = \overline{Z - \overline{Z}} = 0$. Then let us focus on the $m=2$ case. 
We need to compute the $\order{d_\mathtt{q}^{-1}}$ corrections in 
\begin{align} \frac{\overline{(\delta Z^{(\mathtt{q})}_n)^2}}{\overline{Z^{(\mathtt{q})}_n}^2} -\frac{\overline{(\delta (Z_1^{(\mathtt{q})})^{n^{\mathtt{q}-1}})^2}}{\overline{(Z_1^{(\mathtt{q})})^{n^{\mathtt{q}-1}}}^2}. \end{align}
In the $\order{d_\mathtt{q}^{-1}}$ order, nonzero contributions to the variance come from single swaps between two $Z$ in $(\delta Z)^2$. When $Z$ includes $n^{\mathtt{q}-1}$ replica density matrices, the number of such swaps is $n^{\mathtt{q}-1}\times n^{\mathtt{q}-1}=n^{2(\mathtt{q}-1)}$, where this counting is true for both $Z^{(\mathtt{q})}_n$ and $(Z_1^{(\mathtt{q})})^{n^{\mathtt{q}-1}}$.
And a single swap between two $Z$ increases the Cayley distance by 1 compared to the $g=e$ case, such as Type II in the previous subsection. 
Therefore, we obtain
\begin{align} \frac{\overline{(\delta Z^{(\mathtt{q})}_n)^2}}{\overline{Z^{(\mathtt{q})}_n}^2} -\frac{\overline{(\delta (Z_1^{(\mathtt{q})})^{n^{\mathtt{q}-1}})^2}}{\overline{(Z_1^{(\mathtt{q})})^{n^{\mathtt{q}-1}}}^2}=\frac{n^{2(\mathtt{q}-1)}}{d_1d_2\cdots d_\mathtt{q}}-\frac{n^{2(\mathtt{q}-1)}}{d_1d_2\cdots d_\mathtt{q}} +\order{d_\mathtt{q}^{-2}}=\order{d_\mathtt{q}^{-2}}, \end{align}
{\it i.e.}, the variance does not contribute in the $\order{d_\mathtt{q}^{-1}}$ correction for the $m=2$ term in \eqref{avglog2}.

In addition, one can utilize a similar argument as in Sec.~\ref{sec:orge89da24} to show that all the $m>2$ terms do not contribute in the $\order{d_\mathtt{q}^{-1}}$. To evaluate $\overline{(\delta Z)^m}$ we need to sum over a permutation group with $m\times n^{\mathtt{q}-1}$ replicas. We claim that $\overline{(\delta Z)^m}$ does not receive a contribution that involves self-contractions within any $Z$ in $(\delta Z)^m$. To see this, suppose that the statement is true for $m-1$. We write
\begin{equation}
  \overline{(\delta Z)^m} = \overline{(Z-\overline{Z})(\delta Z)^{m-1}} = \overline{Z(\delta Z)^{m-1}} - \overline{Z}\times\overline{(\delta Z)^{m-1}}.
\end{equation}
The disconnected contribution in $\overline{Z(\delta Z)^{m-1}}$, which means that $Z$ is self-contracted, is canceled by the second term. Since $\overline{(\delta Z)^{m-1}}$ does not involve the self-contractions from the assumption, we conclude that $\overline{(\delta Z)^m}$ also does not receive contributions from the self-contractions.
Since this statement is trivially true for $m=1$, by induction we prove that it is true for all $m$.
The rest of the argument is straightforward. The $\order{d_\mathtt{q}^{-1}}$ correction in $\overline{(\delta Z)^{m}}$ comes from a single swap. Since a single swap must produce at least one self-contraction of $Z$ in $(\delta Z)^{m}$ for $m\ge3$, it cannot contribute to $\overline{(\delta Z)^m}$.
Therefore, in the $\order{d_\mathtt{q}^{-1}}$ order, we conclude that there are no corrections arising from the second term in \eqref{avglog2} for all $m$.

\subsection{Late time limit}
\label{sec:latetime}
One of the defining features of the multi-entropy curve is that after the black hole has completely evaporated $(d_{\rm BH}=1)$, the multi-entropy of the remaining degrees of freedom does not come to zero, as opposed to the usual Page curve $(\mathtt{q}=2)$.
Here we attempt to find an analytic expression for this value for generic $(n,\mathtt{\mathtt{q}})$.
Even though we base our claim on various evidences, we would like remind the reader that the expressions we claim in this subsection are not proven, so it should be taken with a grain of salt.

For simplicity we assume that all the radiation subsystem dimensions are equal, $d_1=d_2=\cdots=d_{\mathtt{q}-1}\equiv d_{\rm R}$ and we work in the late time limit $d_R \gg d_{\rm BH} \equiv d_{\mathtt{q}}\gg 1$. Based on expressions for finite $(\mathtt{q},n)$ in Sec.~\ref{secRenyimultientropyn2q3} and Sec.~\ref{sec:org1e539bc}, as well as theoretical understanding of the permutation group $S_{n^{\mathtt{q}-1}}$,  we {\it conjecture} that in this limit the (R\'enyi) replica partition function scales as
\begin{equation}
  \label{eq:latetime_conjecture}
  Z^{(\mathtt{q})}_n \sim b^{(\mathtt{q})}_n d_{\rm R}^{(\mathtt{q}-2)n^{\mathtt{q}-2}({1-n})} d_{\rm BH}^{n^{\mathtt{q}-2}({1 - n})} \quad \Longrightarrow \quad S^{(\mathtt{q})}_{n} \approx n^{\mathtt{q}-2} ((\mathtt{q}-2) \log d_{\rm R} + \log d_{\rm BH})
\end{equation}
to leading order in $1/d_{\rm R}$. $b^{(\mathtt{q})}_n$ is some multiplicative constant with explicit dependence on $\mathtt{q}$ and $n$. Its effect on $S^{(\mathtt{q})}_n$ is subleading, but may become important when we consider the analytic continuation $n\to1$.
If we look at the end of the evaporation where $d_{\rm BH}=1$, we have a similar behavior but with different multiplicative coefficient:
\begin{equation}
  \label{eq:latetime_conjecture2}
  Z^{(\mathtt{q})}_n \sim c^{(\mathtt{q})}_n d_{\rm R}^{(\mathtt{q}-2)n^{\mathtt{q}-2}({1 - n})}  \quad \Longrightarrow \quad S^{(q)}_{n} \approx n^{\mathtt{q}-2} (\mathtt{q}-2) \log d_{\rm R} 
\end{equation}
General expressions for $b^{(\mathtt{q})}_n$ and $c^{(\mathtt{q})}_n$ are not known except for a few special cases. 
We discuss more details behind our conjectures \eqref{eq:latetime_conjecture}, \eqref{eq:latetime_conjecture2} and the detailed form for $b^{(\mathtt{q})}_n$ and $c^{(\mathtt{q})}_n$ in Appendix~\ref{app:conjecturecoefs}.

\subsection{The shape for the multi-entropy curve}
\begin{figure}[t]
  \centering
  \includegraphics[width=.6\textwidth]{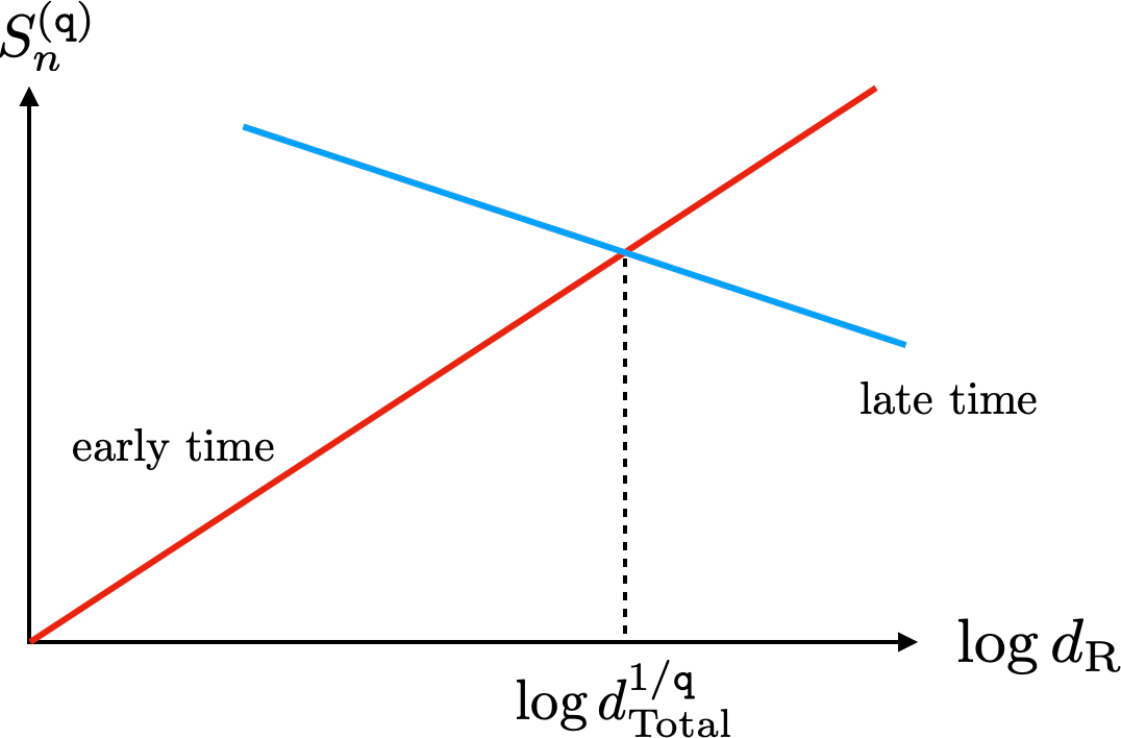}
  \caption{A example illustration of the expressions given in \eqref{eq:asyms}. The red curve is the early time expression $S^{(\mathtt{q})}_{n, \text{early time}}$ and the blue curve is the late time expression $S^{(\mathtt{q})}_{n, \text{late time}}$. The cross-over time of the two curves is given by the multi-entropy time $\log d_{\rm Total}^{1/\mathtt{q}}$.}
  \label{fig:extrapolate}
\end{figure}

So far, we have obtained expressions for the multi-entropy at the early time \eqref{finalresult2} and the late time limit \eqref{eq:latetime_conjecture} in the evaporation process. For completeness, we reproduce them here:
\begin{align}
  \label{eq:asyms}
  \begin{split}
  S^{(\mathtt{q})}_{n, \text{early time}} &\approx n^{q-2}\log(d_{\rm R1}\cdots d_{\rm R\mathtt{q}-1}) = n^{q-2} \left(\log d_{\rm Total} - \log d_{\rm BH}\right), \\
  S^{(\mathtt{q})}_{n, \text{late time}} &\approx n^{q-2}\left((\mathtt{q}-2)\log d_{\rm R} + \log d_{\rm BH}\right) = n^{q-2} \left(\log d_{\rm Total} - \log d_{\rm R}\right),     
  \end{split}
\end{align}
where we have set $d_{\rm R1}=\cdots =d_{\rm R\mathtt{q}-1}=d_{\rm R}$.
A natural question to ask is then: What is the global behavior of the multi-entropy curve? If one naively extrapolate the early and late time expression into the intermediate regime, then one gets a ``inverted-V'' shaped cuve that is in shocking agreement of the actual examples of the multi-entropy curves we obtained in earlier sections. We plot this naive extrapolation curve in Fig.~\ref{fig:extrapolate}.

\begin{figure}[t]
    \centering
    \includegraphics[width=15cm]{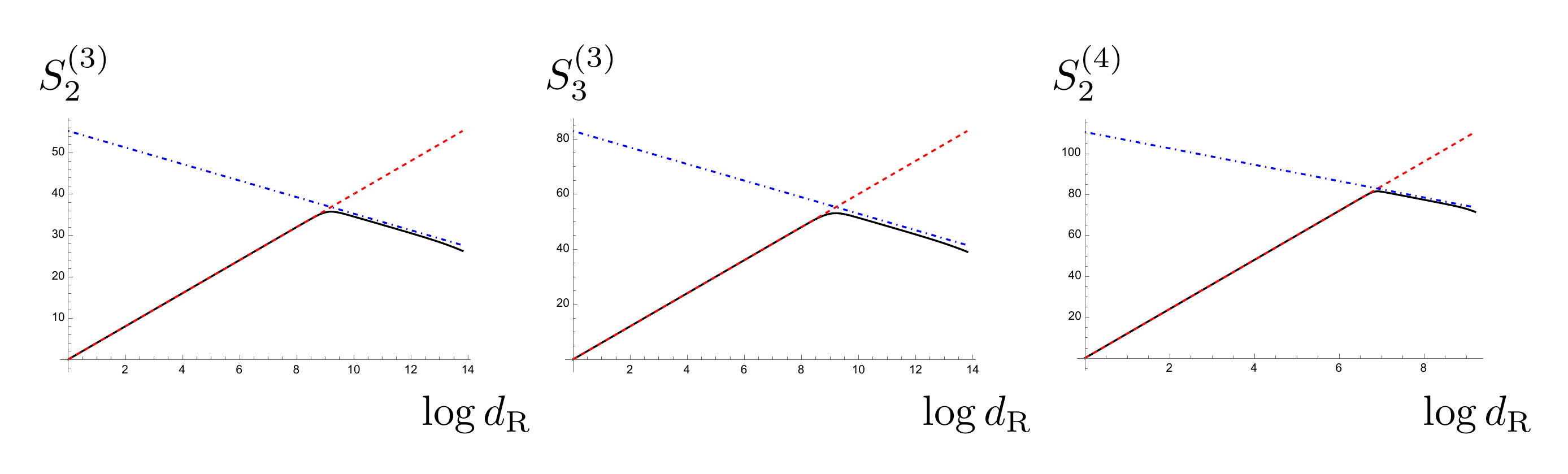}
    \caption{Black solid curves are black hole multi-entropy curves $S^{(\mathtt{q})}_{n}$, as functions of $\log d_{\rm R}$ for the parameter choice $d_{\rm Total} =  10^{12}$. 
    Red and blue dashed curves are $S^{(\mathtt{q})}_{n, \text{early time}}$ and $S^{(\mathtt{q})}_{n, \text{late time}}$ in (\ref{eq:asyms}), respectively.}
\label{fig:RenyiMultiEntropyConjecture}
\end{figure}

The cross-over time between the early and late time behavior is given by
\begin{equation}
  d_{\rm BH} = d_{\rm R} \quad \Leftrightarrow \quad d_{\rm R} = (d_{\rm Total})^{1/\mathtt{q}}.
\end{equation}
This is exactly the multi-entropy time defined in \eqref{multiStime}. Note that when $\mathtt{q}=2$ we reobtain the Page time, and the curve in Fig.~\ref{fig:extrapolate} goes back to the usual Page curve.
The value of the multi entropy at the end of the evaporation is
\begin{equation}
  S^{(\mathtt{q})}_{n}(d_{\rm BH}=1) \approx n^{\mathtt{q}-2}\frac{\mathtt{q}-2}{\mathtt{q}-1}\log d_{\rm Total},
\end{equation}
which is also in agreement with the numerical results in Sec.~\ref{secRenyimultientropyn2q3} and Sec.~\ref{sec:org1e539bc} in a good approximation.
See Fig.~\ref{fig:RenyiMultiEntropyConjecture} for explicit examples on the conjectured curves (\ref{eq:asyms}), where red and blue dashed curves are $S^{(\mathtt{q})}_{n, \text{early time}}$ and $S^{(\mathtt{q})}_{n, \text{late time}}$, respectively. The conjectured curves approximate the slope of black solid curves $S^{(\mathtt{q})}_{n}$ well at early and late times.  

The implication here is that like in the Page curve, the shape of the multi-entropy curve seems to be controlled entirely by its early-time and late-time behavior. While this observation is largely speculative since we do not have analytic control over the intermediate time regime of the evaporation, we believe this is indeed a true statement in the large dimension limit $d_{\rm Total}\to \infty$ due to the fact that all the numerical examples we've seen in this paper follow this behavior perfectly.

\section{Conclusion and discussion}\label{sec:Conclusion}

The main results in this paper are the black hole multi-entropy curves. The black hole multi-entropy curves are the natural generalization of the Page curve.  Page curve measures bi-partite entanglement, {\it i.e.,} between an evaporating black hole and whole Hawking radiation. The black hole multi-entropy curve measures multi-partite entanglement, by dividing the whole radiation into $\mathtt{q} -1$ subsystems. Thus, $\mathtt{q} =2 $ multi-entropy curve reduces to the Page curve.
Even though we have already mentioned the key properties of the multi-entropy curve, let us emphasize the most important aspects again. Just as the Page curve, the black hole multi-entropy curves keep increasing until the multi-entropy time, where the dimensions of all the subsystems are the same. After the multi-entropy time, the curves start to decrease, but different from the Page curve, it does not become zero at the end of the black hole evaporation. This non-zero entropy reflects the secret entanglement between Hawking particles, which one misses in Hawking's semi-classical approximation.

We end this discussion section with several open issues and questions.  

One of the most curious points in the multi-entropy curve is that at the late stage of the black hole evaporation process, the multi-entropy of the radiation is non-zero. 
Not only does this imply that there is non-trivial multi-partite entanglement among the final state Hawking radiation, it also implies that the amount of entanglement is considerably large. Based on various proposals of ER=EPR \cite{VanRaamsdonk:2010pw,Maldacena:2013xja}, one is naturally led to the question of whether this large amount of entanglement between radiation implies classical connectivity for these asymptotic spacetime regions. 
The final state from the $\mathtt{q} \to \infty$ limit is suggested as follows. Since multi-entropy becomes maximal at $\mathtt{q} \to \infty$, the final entanglement seems to appear between single particles. Therefore the large amount of entanglement might be made up by minimal entanglement between single particles. This implies that there are large amounts of microscopic scale wormholes as Planck scale and many of them are not necessarily traversable. 
However, it is desirable to study the ER=EPR aspect in the final evaporation state in more detail. We leave this as one of the open questions.

Recently, there has been progress in deriving the Page curve from a bulk gravitational viewpoint using ``island'' or replica wormholes \cite{Penington:2019npb, Almheiri:2019psf, Almheiri:2019hni, Penington:2019kki, Almheiri:2019qdq}. Given the fact that the multi-entropy is a natural generalization of the entanglement entropy into multi-partite subsystems, it would sound very reasonable to expect that one can derive the multi-entropy curves in similar methods. In fact, our approach parallels the replica wormhole calculations in the west coast model \cite{Penington:2019kki}, in the sense that we both use the replica group technique in our calculations. %the group summation method and we both obtain the leading order result by considering dominant saddles in the sum. 
The big difference is that their model comes from an explicit physical setup containing JT gravity, 
while we start from a random tensor perspective. 
On the other hand, the east coast model \cite{Almheiri:2019qdq} is more technically different from ours.
This is a direction worth investigating in the future.

We also comment on the differences between our paper and \cite{Penington:2022dhr}. In that paper, R\'enyi multi-entropy in random tensor networks with large bond dimensions was studied, where the boundary vertices of random tensor networks are partitioned into three subsystems. One of their motivations is to show the equality between R\'enyi multi-entropy and the size of the minimal tripartition of random tensor networks geometrically. In our paper, the geometrical interpretation of R\'enyi multi-entropy is not the main focus. Instead, we are interested in how different subsystems of Hawking radiation get entangled as the black hole evaporates, which is not necessarily represented geometrically.

Another interesting direction involves the multi-entropy itself. As of now, there is no clear understanding of the operational meaning of, nor is there a good quantum-informatic picture for the multi-entropy. Part of the difficulty stems from the fact that, the definition of the multi-entropy involves an analytic continuation for the index $n$. Such continuation can only be done in a case-by-case basis, severely impeding a more general understanding of the quantity. This is different from other multi-partite measures such as the (R\'enyi) reflected entropy and entanglement negativity, which can be thought of as the moments of a particular density matrix, and thus is completely determined by its spectrum. We hope that by studying the multi-entropy for random tensors (and possibly for tensor networks in the near future), we can gain a better understanding of this quantity. 

At the technical and mathematical level, there are several unsolved problems. 
For example, we have explicitly drawn several multi-entropy curves for R\'enyi case $n\ge 2$. It would be desirable to have an explicit examples of the multi-entropy curves for $n \to 1$ limit even though we do expect qualitative behavior of the multi-entropy curves does not change much in $n \to 1$ limit\footnote{For example, this is indeed true for the entanglement entropy ($\mathtt{q}=2$), where all the R\'enyi entropies are the same. this is because random tensor networks have a flat entanglement spectra \cite{Hayden:2016cfa}.}.
On the other hand, although we have provided analytic expressions of the multi-entropy in early and late time, the behavior in the intermediate regime, especially around the phase transition at the multi-entropy time, is poorly understood. While we believe our expressions already describe most features of the curve, a more rigorous treatment is pending. Because of the simplicity of the model we considered in this paper, we expect this should be doable with a better understanding of the replica permutation group. 
Also in this paper, a single random tensor model was investigated. However, by extending it into multi-tensor models or even random tensor networks, one is expected to capture more holographic aspects.  It would be very interesting to extend our analysis to random tensor networks and see the significance of it. 

Finally, we do not have a clear understanding of where Hawking's argument breaks down and the semiclassical approximation fails.  These are the most urgent questions we have yet to understand, and we hope to come back to these questions in the near future.

\acknowledgments
The work of N.I. was supported in part by JSPS KAKENHI Grant Number 18K03619, MEXT KAKENHI Grant-in-Aid for Transformative Research Areas A “Extreme Universe” No. 21H05184. 
S.L. would like to thank National Tsing Hua University (NTHU), Okinawa Institute of Science and Technology (OIST) and the organizers of the workshop ``Quantum Extreme Universe: Matter, Information, and Gravity'' for their hospitality where this work was initiated. 
\appendix

\section{Comparison with R\'{e}nyi entanglement negativity and R\'{e}nyi reflected entropy}
%Averaged replica partition functions for R\'{e}nyi entanglement negativity, R\'{e}nyi reflected entropy, and their comparison to the R\'{e}nyi multi-entropy curve}
\label{negativityappendix}

We compare $\mathtt{q}=3$ R\'enyi multi-entropy with R\'enyi entanglement negativity and R\'enyi reflected entropy. Note that multi-entropy can be easily generalized to $\mathtt{q}$-partite cases, but entanglement negativity and reflected entropy are measures for the $\mathtt{q}=3$ case only\footnote{There are several proposals for multi-partite generalizations of reflected entropy \cite{Bao:2019zqc, Chu:2019etd, Yuan:2024yfg}, but these measures are not symmetric with respect to subsystems in pure states.}.

\subsection{Averaged replica partition functions for R\'enyi entanglement negativity}
\begin{figure}[ht]
  \centering
  \includegraphics[width=.9\textwidth]{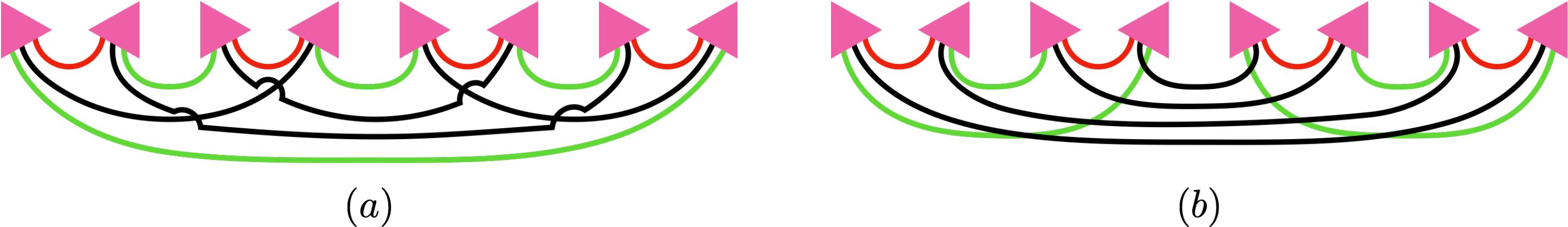}
  \caption{(a) The contraction pattern for R\'enyi entanglement negativity of the state $\rho_{AB} = \tr_C\ket{\psi}_{ABC}\bra{\psi}_{A^*B^*C^*}$ for $n=4$. The green lines represent subsystem $A$ and the black lines subsystem $B$. (b) We also include the contraction pattern for the $n=2$ tripartite multi-entropy for comparison.}
  \label{negativitymulti4replica}
\end{figure}

In this appendix, we estimate averaged replica partition functions for entanglement negativity \cite{Peres:1996dw, Horodecki:1996nc,   Zyczkowski:1998yd, Eisert:1998pz, Simon:1999lfr, Vidal:2002zz, Plenio:2005cwa} of a single tripartite random tensor. Let us consider a pure state $|\psi\rangle$ on $\mathcal{H}_{A}\otimes\mathcal{H}_B\otimes\mathcal{H_C}$
\begin{align}
|\psi\rangle=\sum_{i=1}^{d_A}\sum_{j=1}^{d_B}\sum_{k=1}^{d_C}c_{ijk}| A_i\rangle\otimes | B_j\rangle\otimes |C_k\rangle,\label{3randomstate}
\end{align}
where $c_{ijk}$ are complex Gaussian random variables, and $|A_i\rangle, |B_j\rangle, |C_k\rangle$ are orthonormal bases of $\mathcal{H}_A, \mathcal{H}_B, \mathcal{H}_C$, respectively. The dimensions of each Hilbert space are
\begin{align}
\dim\mathcal{H}_{A}=d_{A},\;\;\;\dim\mathcal{H}_{B}=d_{B},\;\;\;\dim\mathcal{H}_C=d_{C}.
\end{align}
Define a reduced density matrix $\rho_{AB}$ as
\begin{align}
\rho_{AB}:=\Tr_C |\psi\rangle\langle\psi|=\sum_{i_1, i_2, j_1, j_2}\rho_{i_1j_1i_2j_2}| A_{i_1}\rangle| B_{j_1}\rangle\langle A_{i_2}|\langle B_{j_2}|,
\end{align}
where $\rho_{i_1j_1i_2j_2}$ is an index representation of the reduced density matrix. Its partial transpose $\rho_{AB}^{T_B}$ is defined by exchanging the indices of $B$ as in
\begin{align}
\rho_{AB}^{T_B}:=\sum_{i_1, i_2, j_1, j_2}\rho_{i_1j_1i_2j_2}| A_{i_1}\rangle| B_{j_2}\rangle\langle A_{i_2}|\langle B_{j_1}|.
\end{align}
To evaluate R\'enyi entanglement negativity, we need to compute the following ratio of averaged replica partition functions
\begin{align}\label{RatioAPFNegativity}
\frac{\overline{\Tr \left(\rho_{AB}^{T_B}\right)^n}}{\overline{\left(\Tr \rho_{AB}\right)^n}}=\frac{\sum_{g\in S_{n}}d^{-d(g,g_A)}_{A}d^{-d(g,g_B)}_{B}d^{-d(g,e)}_{C}}{\sum_{g\in S_{n}}(d_{A}d_{B}d_{C})^{-d(g,e)}}.
\end{align}
The group elements $g_A,g_B\in S_n$ are
\begin{align}
g_A &= (12\cdots n), \\
g_B &= (n(n-1)\cdots1),
\end{align}
where $g_B=g_A^{-1}$ means the partial transpose. In our convention, R\'enyi entanglement negativity $\mathcal{E}_n$ of the random state $|\psi\rangle$ is defined by 
\begin{align}
\mathcal{E}_n:=-\overline{\log \frac{\Tr \left(\rho_{AB}^{T_B}\right)^n}{\left(\Tr \rho_{AB}\right)^n}}\approx -\log \frac{\overline{\Tr \left(\rho_{AB}^{T_B}\right)^n}}{\overline{\left(\Tr \rho_{AB}\right)^n}}.
\end{align}
Fig.~\ref{negativitymulti4replica} shows the contraction patterns for the R\'enyi entanglement negativity $\mathcal{E}_4$ and the R\'enyi multi-entropy $S_2^{(3)}$.
Entanglement negativity $\mathcal{E}$,\footnote{Sometimes, $\mathcal{E}$ is called logarithmic negativity.} can be computed by an analytic continuation of $\mathcal{E}_{2 n}$ 
\begin{align}
\label{deflogneg}
\mathcal{E}:=-\lim_{n\to\frac{1}{2}}\mathcal{E}_{2 n}.
\end{align}

By counting the number of cycles or using Mathematica, one can compute the sum over $S_n$ in (\ref{RatioAPFNegativity}) explicitly. Concrete examples with $n=3,4$ are given by
\begin{align}
\frac{\overline{\Tr \left(\rho_{AB}^{T_B}\right)^3}}{\overline{\left(\Tr \rho_{AB}\right)^3}}=\frac{d_A^2+d_B^2+d_C^2+3d_Ad_Bd_C}{(d_Ad_Bd_C+1)(d_Ad_Bd_C+2)} \,,
\end{align}
which was derived in \cite{Bhosale:2012ldd}, and 
\begin{align}
\frac{\overline{\Tr \left(\rho_{AB}^{T_B}\right)^4}}{\overline{\left(\Tr \rho_{AB}\right)^4}}=\frac{2d_A^2d_B^2d_C+d_Ad_B(d_A^2+d_B^2+6d_C^2+4)+4d_A^2d_C+4d_B^2d_C+d_C^3+d_C}{(d_Ad_Bd_C+1)(d_Ad_Bd_C+2)(d_Ad_Bd_C+3)} \,.
\label{negativityfor4replica}
\end{align}
which is, as far as we are aware, not explicitly written in the literature. 

\subsubsection{Leading and subleading terms at large $d_C$}
Let us compute the leading and subleading terms in eq.~(\ref{RatioAPFNegativity}) at large $d_C$. The leading term comes from the identity element $g=e$. Since the Cayley distance becomes
\begin{align}
d(e,g_A)=d(e,g_B)=n-1, \;\;\; d(e,e)=0,
\end{align}
we obtain
\begin{align}
\frac{\overline{\Tr \left(\rho_{AB}^{T_B}\right)^n}}{\overline{\left(\Tr \rho_{AB}\right)^n}}=(d_Ad_B)^{1-n}+\order{d_C^{-1}},
\end{align}
which was derived in \cite{Shapourian:2020mkc}. The subleading term comes from two element swaps $g=\tau_{ab}$ with $a,b\in \{1,\dots,n\}$. The number of such swaps is $n(n-1)/2$, and the Cayley distance for such swaps is
\begin{align}
d(\tau_{ab},g_A)=d(\tau_{ab},g_B)=n-2, \;\;\; d(\tau_{ab},e)=1.
\end{align}
Therefore, we obtain
\begin{align}
\frac{\overline{\Tr \left(\rho_{AB}^{T_B}\right)^n}}{\overline{\left(\Tr \rho_{AB}\right)^n}}&=\frac{(d_Ad_B)^{1-n}+\frac{n(n-1)}{2}\frac{(d_Ad_B)^{2-n}}{d_C}}{1+\frac{n(n-1)}{2}\frac{1}{d_Ad_Bd_C}}+\order{d_C^{-2}}\notag\\
&=(d_Ad_B)^{1-n}\left[1+\frac{n(n-1)}{2}\frac{d_A^2d_B^2-1}{d_Ad_Bd_C}+\order{d_C^{-2}}\right].
\end{align}

\subsection{Averaged replica partition functions for R\'{e}nyi reflected entropy}

Let us explain how to define R\'enyi reflected entropy \cite{Dutta:2019gen}. For a given reduced density matrix $\rho_{AB}$, we consider the canonical purified state $|\sqrt{\rho_{AB}}\rangle$ \footnote{This pure state $|\sqrt{\rho_{AB}}\rangle$ implicitly has indices of $\mathcal{H}_A\otimes \mathcal{H}_B\otimes \mathcal{H}^*_A\otimes \mathcal{H}^*_B$.}, where
\begin{align}
\Tr_{A^*B^*} |\sqrt{\rho_{AB}}\rangle\langle\sqrt{\rho_{AB}}|=\rho_{AB}.
\end{align}
As a R\'enyi generalization of $|\sqrt{\rho_{AB}}\rangle$, we also consider the canonical purified state $|\rho_{AB}^{m/2}\rangle\in \mathcal{H}_A\otimes \mathcal{H}_B\otimes \mathcal{H}^*_A\otimes \mathcal{H}^*_B$, where 
\begin{align}
\Tr_{A^*B^*} |\rho_{AB}^{m/2}\rangle\langle\rho_{AB}^{m/2}|=\rho_{AB}^m.
\end{align}
Then, R\'enyi reflected entropy $S_{R}^{(m,n)}$ is defined by
\begin{align}
S_{R}^{(m,n)}&:=\frac{1}{1-n}\log \frac{Z_{m,n}}{(Z_{m,1})^n},\\
Z_{m,n}&:=\Tr_{AA^*}\left(\Tr_{BB^*}|\rho_{AB}^{m/2}\rangle\langle\rho_{AB}^{m/2}|\right)^n,
\end{align}
where the normalization by $(Z_{m,1})^n$ is required for $\lim_{n\to1}\frac{Z_{m,n}}{(Z_{m,1})^n}=1$.

The R\'enyi reflected entropy $S_{R}^{(m,n)}$ of the random state $|\psi\rangle$ (\ref{3randomstate}) is given by \cite{Akers:2021pvd}
\begin{align}
  S_{R}^{(m,n)}&=  \frac{1}{1-n}\overline{\log \frac{Z_{m,n}}{(Z_{m,1})^n}}\approx \frac{1}{1-n}\log \frac{\overline{Z_{m,n}}}{\overline{(Z_{m,1})^n}}\,,\\
  \frac{\overline{Z_{m,n}}}{\overline{(Z_{m,1})^n}}&=\frac{\sum_{g\in S_{nm}}d^{-d(g,g_A)}_{A}d^{-d(g,g_B)}_{B}d^{-d(g,e)}_{C}}{\sum_{g\in S_{nm}}(d_{A}d_{B})^{-d(g,g_B)}d^{-d(g,e)}_{C}} \,.
\end{align}
For $(m, n) =(2, 2)$ case, the group elements $g_A,g_B\in S_{nm}$ are given in Fig.~\ref{negativitymulti4replica} (b). 
Then the explicit expression of $S_{R}^{(2,2)}$ is given by
\begin{align}\label{REfor4replica}
    S_{R}^{(2,2)}&=-\log \frac{\overline{Z_{2,2}}}{\overline{(Z_{2,1})^2}}=-\log \frac{\sum_{g\in S_{4}}d^{-d(g,g_A)}_{A}d^{-d(g,g_B)}_{B}d^{-d(g,e)}_{C}}{\sum_{g\in S_{4}}(d_{A}d_{B})^{-d(g,g_B)}d^{-d(g,e)}_{C}}\notag\\
    &=-\log \frac{d_Ad_Bd_C(9+d_A^2+d_B^2+d_C^2)+2(d_A^2+d_B^2+d_C^2+d_A^2d_B^2+d_B^2d_C^2+d_C^2d_A^2)}{2d_A^2 d_B^2 d_C^2+d_A d_B d_C(d_A^2d_B^2+d_C^2+10)+4d_A^2 d_B^2+4d_C^2+2}.
\end{align}
The contraction pattern in the numerator is the same as in (\ref{men2q3}), but the ones in the denominator are different. Thus, the R\'enyi multi-entropy $S_{2}^{(3)}$ and the R\'enyi reflected entropy $S_R^{(2,2)}$ have different behaviors due to the difference of the denominator.

\subsection{Comparison of curves}

We compare three numerical plots of $S_{2}^{(3)}$, $\mathcal{E}_4$, $S_{R}^{(2,2)}$, which are tri-partite measures with four replicas. The expression of $S_{2}^{(3)}$ is given by eq.~(\ref{BHmultiScurveq=3n=2}). By using eqs.~(\ref{negativityfor4replica}) and (\ref{REfor4replica}), we obtain
\begin{align}
\mathcal{E}_4&=-\log \frac{d_\text{BH}^3+6d_\text{BH}^2d_\text{R}^2+d_\text{BH}(2d_\text{R}^4+8d_\text{R}^2+1)+2d_\text{R}^4+4d_\text{R}^2}{(1+d_\text{BH}d_\text{R}^2)(2+d_\text{BH}d_\text{R}^2)(3+d_\text{BH}d_\text{R}^2)},\\
S_{R}^{(2,2)}&=-\log  \frac{ d_{\rm R}^2 d_{\rm BH}  (9 + 2 d_{\rm R}^2 + d_{\rm BH}^2 ) +  2 ( 2 d_{\rm R}^2 + d_{\rm BH}^2 + d_{\rm R}^4 + 2 d_{\rm R}^2 d_{\rm BH}^2 )}{d_\text{BH}^3d_\text{R}^2+d_\text{BH}^2(2d_\text{R}^4+4)+d_\text{BH}(d_\text{R}^6+10d_\text{R}^2)+4d_\text{R}^4+2},
\end{align}
where we set $d_A=d_B=d_\text{R}$ and $d_C=d_\text{BH}$.

\begin{figure}[t]
    \centering
        \includegraphics[width=11.3cm]{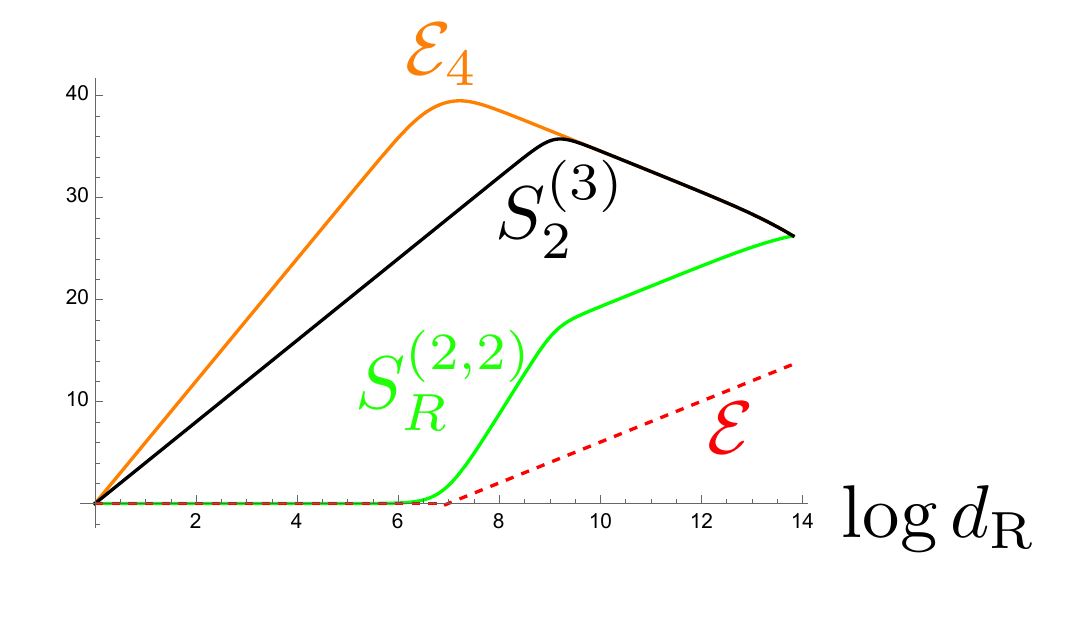}
    \caption{Numerical comparison of four measures for tri-partite systems with $d_\text{Total}=d_\text{R}^2 d_\text{BH}=10^{12}$. Black curve: $S_{2}^{(3)}$. Orange curve: $\mathcal{E}_4$. Green curve: $S_{R}^{(2,2)}$. Red dashed curve: $\mathcal{E}$.}
    \label{fig:Comparison}
\end{figure}

Fig.~\ref{fig:Comparison} shows numerical plots of the four measures with $d_\text{Total}=d_\text{R}^2 d_\text{BH}=10^{12}$. The black curve is $S_{2}^{(3)}$, the orange curve is $\mathcal{E}_4$, and the green curve is $S_{R}^{(2,2)}$. Just for a reference, we also plot the negativity $\mathcal{E}$ defined in \eqref{deflogneg}, which is studied in \cite{Shapourian:2020mkc,Bhosale:2012ldd, Lu:2020jza}. The R\'enyi entanglement negativity $\mathcal{E}_4$ starts to decrease from the Page time $\log d_{\rm R} =\log \left( d_{\rm Total}\right)^{1/4} \approx 6.91$ and agrees with $S_{2}^{(3)}$ from the multi-entropy time $\log d_{\rm R} = \log \left( d_{\rm Total}\right)^{1/3} \approx 9.21$. On the other hand, the negativity $\mathcal{E}$ is almost zero till the Page time, and then it grows linearly. The R\'enyi reflected entropy $S_{R}^{(2,2)}$ has three different behaviors, which are consistent with three approximate behaviors derived in (3.96) of \cite{Akers:2021pvd}. 
Until the Page time, $S_{R}^{(2,2)}$ is almost zero. From the Page time, $S_{R}^{(2,2)}$ starts to increase, and its increasing behavior changes at the multi-entropy time $\log d_{\rm R} = \log \left( d_{\rm Total} \right)^{1/3}$. Interestingly, the three measures coincide when the black hole evaporates at $\log d_{\rm R} =\log \left( d_{\rm Total} \right)^{1/2}\approx 13.8$, but the negativity $\mathcal{E}$ is smaller than the others.

\section{Late time limit formulae} 
\label{app:conjecturecoefs}
\subsection{Regarding our conjectures \eqref{eq:latetime_conjecture}, \eqref{eq:latetime_conjecture2}}

We formulate our conjectures \eqref{eq:latetime_conjecture}, \eqref{eq:latetime_conjecture2} base on several observations:
\begin{enumerate}
\item The replica partition function in this setting is
  \begin{equation}
    \label{eq:Z_latetime}
    \overline{Z^{(\mathtt{q})}_n} = \sum_{g\in S_{n^{\mathtt{q}-1}}} d_{\rm R}^{-(d(g,g_1)+d(g,g_2)+\cdots+d(g,g_{\mathtt{q}-1}))} d_{\rm BH}^{-d(g,e)}.
  \end{equation}
  Since we $d_R\gg d_{BH}$, the sum is dominated by the first factor. Thus we are looking for $g$ such that it minimizes the sum
  \begin{equation}
    \label{eq:min_g}
    \min_{g\in S_{n^{\mathtt{q}-1}}} \left(d(g,g_1)+d(g,g_2)+\cdots+d(g,g_{\mathtt{q}-1})\right)
  \end{equation}
  
\item Consider the extreme case where the dimension of one particular radiation subsystem, say R1, to be much larger, i.e. $d_1\gg d_2,\cdots ,d_{\mathtt{q}-1}$. In such case we expect that the replica partition function is dominated by $g=g_1$. 
  Consider the process which we slowly change $d_1$ from $d_1\gg d_i$ to $d_1\sim d_i$.
  If one can assume $g_1$ stays dominant along this process, then $g_1$ should be a minimizer for \eqref{eq:min_g}. For symmetric reasoning all the other $g_i$'s (except for $g_\mathtt{q}=e$) are also minimizers for \eqref{eq:min_g}.

\item We have assumed that there is no phase transition when we smoothly vary R1 across a large parameter space. For a arbitrary set of twist operators this will not be true. For example, if there is a common geodesic $\Gamma_{g_i,g_j}$\footnote{A geodesic $\Gamma_{g_1,g_2}$ between two permutation group elements $g_1,g_2$ is the set of elements $\{h\}$ such that the triangle inequality $d(g_1,h)+d(h,g_2)\ge d(g_1,g_2)$ is saturated.} across all pairs $(g_i,g_j)$, then any such element will be a new minimizer at $d_{\rm R1}=d_{\rm R2}=\cdots=d_{\rm R_{\mathtt{q}-1}}$ and there will be a phase transition.\footnote{Such effects, often referred to in the literature as ``replica symmetry breaking'', is particularly important in the case of reflected entropy \cite{Dutta:2019gen,Akers:2021pvd,Akers:2022zxr,Akers:2024pgq} and entanglement negativity \cite{Shapourian:2020mkc,Kudler-Flam:2021efr,Dong:2021clv}.}
  
  However, the twist operators for the multi-entropy has the special property that the geodesics of any two distinct pairs of $g_i$ do not intersect, i.e. $\Gamma_{g_i,g_j}\cap\Gamma_{g_k,g_l}=\emptyset$ for $i\ne j\ne k\ne l$. Together with the property that $g_i\cap g_j=e$, we conjecture that they are sufficient to guarantee that such a phase transition do not happen. Therefore $g_i$s give the dominant contribution in the replica partition function \eqref{eq:Z_latetime}, giving rise to the scaling behavior in \eqref{eq:latetime_conjecture} and \eqref{eq:latetime_conjecture2}. We check that they are consistent with finite $(\mathtt{q},n)$ results in Sec.~\ref{secRenyimultientropyn2q3} and \ref{sec:org1e539bc}.

  \item Our previous argument only shows that $g_i$ minimizes \eqref{eq:min_g}. But since $d_1=d_2=\cdots=d_{\mathtt{q}-1}$ is a phase transition point, in general there will also be other elements that are equally dominant, and indeed there are plenty. These degeneracies are counted by the multiplicative constant $b^{(\mathtt{q})}_n$ and $c^{(\mathtt{q})}_n$. In general $c^{(\mathtt{q})}_n> b^{(\mathtt{q})}_n$.  This is because $d_{\rm BH}\ne1$ for $b^{(\mathtt{q})}_n$, we must also take into account the effect of $d(g,e)$, lifting some of the degeneracies.  %For more details on them please see Appendix~\ref{app:coefs}.

\end{enumerate}

\subsection{Form of the coefficients $b^{(\mathtt{q})}_n$ and $c^{(\mathtt{q})}_n$}

In this appendix, we give more detailed information regarding the multiplicative coefficient $b^{(\mathtt{q})}_n$ and $c^{(\mathtt{q})}_n$, which appears in \eqref{eq:latetime_conjecture} and \eqref{eq:latetime_conjecture2} respectively.
Consider the minimizers for the following ``energy functionals'' in $S_{n^{\mathtt{q}-1}}$:
\begin{align}
  s_1 &\equiv \underset{h\in S_{n^{\mathtt{q}-1}}}{\text{argmin}}~ [d(h,g_1)+d(h,g_2)+\cdots+d(h,g_{\mathtt{q}-1})] \\
  s_2 &\equiv \underset{h\in s_1}{\text{argmin}}~ [d(h,e)]
\end{align}
$b^{(\mathtt{q})}_n$ and $c^{(\mathtt{q})}_n$ are defined as the cardinality of the minimizer sets:
\begin{equation}
  c^{(\mathtt{q})}_n = |s_1|, \quad   b^{(\mathtt{q})}_n = |s_2|
\end{equation}
By definition we have $c^{(\mathtt{q})}_n\ge b^{(\mathtt{q})}_n$.

We can say a little more about $c^{(\mathtt{q})}_n$. Using the fact that the Cayley distance is invariant under right translation, we can write
\begin{align*}
d(h,g_1)+d(h,g_2)+\cdots+d(h,g_{\mathtt{q}-1}) = d(h',e)+d(h',g_2g_1^{-1})+\cdots+d(h',g_{\mathtt{q}-1}g^{-1}_1),
\end{align*}
where \(h'=hg_1^{-1}\). We will omit the prime in the following since it does not affect the minimization.
Now, since the permutation elements satisfies the symmetric property, we can do a relabeling of replica indices such that
\begin{equation}
  g_2g_1^{-1} \to g_1, ~~ g_3g_1^{-1} \to g_2,  ~~ \cdots  , ~~ g_{\mathtt{q}-1}g_1^{-1} \to g_{\mathtt{q}-2}.
\end{equation}
Note that this procedure leaves the identity permutation \(e\) invariant. Hence
\begin{align}
   c^{(\mathtt{q})}_n =  \left| \underset{h\in S_{n^{\mathtt{q}-1}}}{\text{argmin}}~ [d(h,e)+d(h,g_1)+\cdots+d(h,g_{\mathtt{q}-2})]  \right|.
\end{align}
We don't have the \(g_{\mathtt{q}-1}\) permutation element in the sum (because we've set \(d_{\rm BH}=1\)).
What this means is that there are no elements acting in the \((\mathtt{q}-1)\)-th axis of the hypercube, and thus we have the factorization
\begin{align}
   c^{(\mathtt{q})}_n =  \left| \underset{h\in S_{n^{\mathtt{q}-2}}}{\text{argmin}}~ [d(h,e)+d(h,g_1)+\cdots+d(h,g_{\mathtt{q}-2})]  \right|^n = (a^{(\mathtt{q})}_n)^n,
\end{align}
where
\begin{equation}
  a^{(\mathtt{q})}_n = \left| \underset{h\in S_{n^{\mathtt{q}-2}}}{\text{argmin}}~ [d(h,e)+d(h,g_1)+\cdots+d(h,g_{\mathtt{q}-2})]  \right|
\end{equation}
is the number of elements for the energy functional in $\mathtt{q}-1$ parties.
Note that this amounts to calculating the  multi-entropy for \(\mathtt{q}-1\) parties (at large \(d_{\rm R}\)), with the dimension of all subsystems equal.
What we have shown here is that the coefficient $c^{(\mathtt{q})}_n$ is a perfect power.
We do not know the number $a^{(\mathtt{q})}_n$ in general, except for the case of \(\mathtt{q}=3\), where one can write down an general expression for \(a^{(3)}_n\).
We simply have \(S_{n^{\mathtt{q}-2}} = S_n\) and \(g_1 = \tau_n\), the maximal cyclic permutation. Thus
\begin{equation}
  a^{(3)}_n = \left| \underset{h\in S_{n^{\mathtt{q}-2}}}{\text{argmin}}~ [d(h,e)+d(h,\tau_n)]  \right|
\end{equation}
The elements that minimize the sum is known as {\it non-crossing permutations}, whose counting is given by the {\it Catalan numbers}:
\begin{equation}
  C_n = \frac{(2n)!}{(n+1)!n!},\quad \{C_n\}=\{1,2,5,14,\cdots\}
\end{equation}
Thus we've found
\begin{equation}
  a^{(3)}_n = C_n 
\end{equation}

On the other hand, not much is known about the generic form of $b^{(\mathtt{q})}_n$ since it involves an additional minimization step.
We list the known cases from explicit expressions in Table~\ref{table:coefs} for reader's reference. 

\begin{table}[ht]
  \centering
\begin{tabular}{|c|c c|}
\hline
\(b^{(\mathtt{q})}_n\) & \(n=2\) & \(n=3\) \\
\hline
\hline
\(\mathtt{q}=3\) & 2 & 20 \\
\hline
\(\mathtt{q}=4\) & 3 & ? \\
\hline
\(\mathtt{q}>4\) & $(\mathtt{q}-1)$? & ? \\
\hline
\end{tabular}

\vspace{.5cm}

\begin{tabular}{|c|c c c|}
\hline
\(c^{(\mathtt{q})}_n\) & \(n=2\) & \(n=3\) & higher \(n\)\\
\hline
\hline
\(\mathtt{q}=3\) & 4 & \(125=5^3\) & \((C_n)^n\) (proven)\\
\hline
\(\mathtt{q}=4\) & 9 & \(84^3\) (through \(a^{(4)}_3\)) & ?\\
\hline
\(\mathtt{q}=5\) & 16 (through \(a^{(5)}_2\)) & ? & ?\\
\hline
\(\mathtt{q}>5\) & \((\mathtt{q}-1)^2\)? & ? & ?\\
\hline
\end{tabular}
\caption{The known cases for the multiplicative coefficients $b^{(\mathtt{q})}_n$ (top) and $c^{(\mathtt{q})}_n$ (bottom). It does seem that at least for the $n=2$ column, the coefficients follow a simple pattern $(a^{(\mathtt{q})}_n=b^{(\mathtt{q})}_n=\mathtt{q}-1)$. Although we currently do not have a proof for this claim.}
\label{table:coefs}
\end{table}

\section{Proof of (\ref{ReplicaPF1})}
\label{app:proof}
Note that \(S_{n+1}\) can be decomposed into the group \(S_n\) itself and \(S_n\) followed by a transposition with the \((n+1)\)-th element, i.e.
\begin{align} S_{n+1} = \{g|g\in S_n\} \cup \{g\tau_{a,n+1}|g\in S_n, a\in \{1,\cdots,n\}\}. \end{align}
The elements of first subset has cycle count \(\#(g)+1\) while the elements of the second subset has the same cycle count. Thus, we have the following recursion relation:
\begin{align}
\sum_{g\in S_{n+1}} d^{\#(g)} = (d+n)\sum_{g\in S_n} d^{\#(g)}.
\end{align}
Given that \(Z_1=1\), it is trivial to see 
\begin{align} \overline{Z_1^n} =\sum_{g\in S_{n}} d^{\#(g)-n}= \frac{(1+d)(2+d)\cdots(n-1+d)}{d^{n-1}}, \end{align}
which proves our claim for $d=d_\text{R1}d_\text{R2}d_\text{BH}$ and $n=9$.

In fact one can solve for the generating function exactly. Let us consider
\begin{align} \zeta(d;\lambda) := \sum_{n=1}^\infty \lambda^n\left(\sum_{g\in S_n}d^{\#(g)}\right). \end{align}
Using the recursion relation above one can write
\begin{align}
\zeta(d;\lambda) &= \lambda d+\sum_{n=2}^\infty \lambda^n\left(\sum_{g\in S_n}d^{\#(g)}\right) \notag\\
&= \lambda d + \sum_{n=2}^\infty \lambda^n (d+n-1) \left(\sum_{g\in S_{n-1}}d^{\#(g)}\right) \notag\\
&= \lambda d +  \sum_{n=1}^\infty \lambda^{n+1} (d+n) \left(\sum_{g\in S_n}d^{\#(g)}\right) \notag\\
&= \lambda d +  \lambda(d+\lambda\partial_\lambda) \zeta(d;\lambda).
\end{align}
This is a first order differential equation, and its solution with \(\zeta(0;\lambda)=0\) is
\begin{align} \zeta(d;\lambda) = -d e^{-1/\lambda}(-1/\lambda)^d \,\Gamma(-d,-1/\lambda)),\end{align}
where
\begin{align} \Gamma(a,z) = \int_z^\infty t^{a-1} e^{-t} dt \end{align}
is the incomplete gamma function.
The generating function has the Taylor series
\begin{align} \zeta(d;\lambda) = \sum_{n=1}^\infty \lambda^n \prod_{i=0}^{n-1} (i+d), \end{align}
as expected.

\bibliography{Refs}

\providecommand{\href}[2]{#2}\begingroup\raggedright\begin{thebibliography}{10}

\bibitem{Hawking:1976ra}
S.~W. Hawking, \emph{{Breakdown of Predictability in Gravitational Collapse}},
  \href{http://dx.doi.org/10.1103/PhysRevD.14.2460}{\emph{Phys. Rev. D} {\bf
  14} (1976) 2460--2473}.

\bibitem{Page:1993df}
D.~N. Page, \emph{{Average entropy of a subsystem}},
  \href{http://dx.doi.org/10.1103/PhysRevLett.71.1291}{\emph{Phys. Rev. Lett.}
  {\bf 71} (1993) 1291--1294}, [\href{http://arxiv.org/abs/gr-qc/9305007}{{\tt
  gr-qc/9305007}}].

\bibitem{Page:1993wv}
D.~N. Page, \emph{{Information in black hole radiation}},
  \href{http://dx.doi.org/10.1103/PhysRevLett.71.3743}{\emph{Phys. Rev. Lett.}
  {\bf 71} (1993) 3743--3746}, [\href{http://arxiv.org/abs/hep-th/9306083}{{\tt
  hep-th/9306083}}].

\bibitem{Hawking:1975vcx}
S.~W. Hawking, \emph{{Particle Creation by Black Holes}},
  \href{http://dx.doi.org/10.1007/BF02345020}{\emph{Commun. Math. Phys.} {\bf
  43} (1975) 199--220}.

\bibitem{Walter:2016lgl}
M.~Walter, D.~Gross and J.~Eisert, \emph{{Multi-partite entanglement}},
  \href{http://arxiv.org/abs/1612.02437}{{\tt 1612.02437}}.

\bibitem{Iizuka:2013ria}
N.~Iizuka and D.~Kabat, \emph{{Mutual information in Hawking radiation}},
  \href{http://dx.doi.org/10.1103/PhysRevD.88.084010}{\emph{Phys. Rev. D} {\bf
  88} (2013) 084010}, [\href{http://arxiv.org/abs/1308.2386}{{\tt 1308.2386}}].

\bibitem{Hollowood:2021nlo}
T.~J. Hollowood, S.~P. Kumar, A.~Legramandi and N.~Talwar, \emph{{Islands in
  the stream of Hawking radiation}},
  \href{http://dx.doi.org/10.1007/JHEP11(2021)067}{\emph{JHEP} {\bf 11} (2021)
  067}, [\href{http://arxiv.org/abs/2104.00052}{{\tt 2104.00052}}].

\bibitem{Hollowood:2021lsw}
T.~J. Hollowood, S.~P. Kumar, A.~Legramandi and N.~Talwar, \emph{{Grey-body
  factors, irreversibility and multiple island saddles}},
  \href{http://dx.doi.org/10.1007/JHEP03(2022)110}{\emph{JHEP} {\bf 03} (2022)
  110}, [\href{http://arxiv.org/abs/2111.02248}{{\tt 2111.02248}}].

\bibitem{Vidal:2002zz}
G.~Vidal and R.~F. Werner, \emph{{Computable measure of entanglement}},
  \href{http://dx.doi.org/10.1103/PhysRevA.65.032314}{\emph{Phys. Rev. A} {\bf
  65} (2002) 032314}, [\href{http://arxiv.org/abs/quant-ph/0102117}{{\tt
  quant-ph/0102117}}].

\bibitem{Dutta:2019gen}
S.~Dutta and T.~Faulkner, \emph{{A canonical purification for the entanglement
  wedge cross-section}},
  \href{http://dx.doi.org/10.1007/JHEP03(2021)178}{\emph{JHEP} {\bf 03} (2021)
  178}, [\href{http://arxiv.org/abs/1905.00577}{{\tt 1905.00577}}].

\bibitem{Takayanagi:2017knl}
T.~Takayanagi and K.~Umemoto, \emph{{Entanglement of purification through
  holographic duality}},
  \href{http://dx.doi.org/10.1038/s41567-018-0075-2}{\emph{Nature Phys.} {\bf
  14} (2018) 573--577}, [\href{http://arxiv.org/abs/1708.09393}{{\tt
  1708.09393}}].

\bibitem{Tamaoka:2018ned}
K.~Tamaoka, \emph{{Entanglement Wedge Cross Section from the Dual Density
  Matrix}}, \href{http://dx.doi.org/10.1103/PhysRevLett.122.141601}{\emph{Phys.
  Rev. Lett.} {\bf 122} (2019) 141601},
  [\href{http://arxiv.org/abs/1809.09109}{{\tt 1809.09109}}].

\bibitem{Zou:2020bly}
Y.~Zou, K.~Siva, T.~Soejima, R.~S.~K. Mong and M.~P. Zaletel, \emph{{Universal
  tripartite entanglement in one-dimensional many-body systems}},
  \href{http://dx.doi.org/10.1103/PhysRevLett.126.120501}{\emph{Phys. Rev.
  Lett.} {\bf 126} (2021) 120501}, [\href{http://arxiv.org/abs/2011.11864}{{\tt
  2011.11864}}].

\bibitem{Gadde:2022cqi}
A.~Gadde, V.~Krishna and T.~Sharma, \emph{{New multipartite entanglement
  measure and its holographic dual}},
  \href{http://dx.doi.org/10.1103/PhysRevD.106.126001}{\emph{Phys. Rev. D} {\bf
  106} (2022) 126001}, [\href{http://arxiv.org/abs/2206.09723}{{\tt
  2206.09723}}].

\bibitem{Gadde:2023zzj}
A.~Gadde, V.~Krishna and T.~Sharma, \emph{{Towards a classification of
  holographic multi-partite entanglement measures}},
  \href{http://dx.doi.org/10.1007/JHEP08(2023)202}{\emph{JHEP} {\bf 08} (2023)
  202}, [\href{http://arxiv.org/abs/2304.06082}{{\tt 2304.06082}}].

\bibitem{Balasubramanian:2014hda}
V.~Balasubramanian, P.~Hayden, A.~Maloney, D.~Marolf and S.~F. Ross,
  \emph{{Multiboundary Wormholes and Holographic Entanglement}},
  \href{http://dx.doi.org/10.1088/0264-9381/31/18/185015}{\emph{Class. Quant.
  Grav.} {\bf 31} (2014) 185015}, [\href{http://arxiv.org/abs/1406.2663}{{\tt
  1406.2663}}].

\bibitem{Cui:2018dyq}
S.~X. Cui, P.~Hayden, T.~He, M.~Headrick, B.~Stoica and M.~Walter, \emph{{Bit
  Threads and Holographic Monogamy}},
  \href{http://dx.doi.org/10.1007/s00220-019-03510-8}{\emph{Commun. Math.
  Phys.} {\bf 376} (2019) 609--648},
  [\href{http://arxiv.org/abs/1808.05234}{{\tt 1808.05234}}].

\bibitem{Bao:2018gck}
N.~Bao and I.~F. Halpern, \emph{{Conditional and Multipartite Entanglements of
  Purification and Holography}},
  \href{http://dx.doi.org/10.1103/PhysRevD.99.046010}{\emph{Phys. Rev. D} {\bf
  99} (2019) 046010}, [\href{http://arxiv.org/abs/1805.00476}{{\tt
  1805.00476}}].

\bibitem{Bao:2019zqc}
N.~Bao and N.~Cheng, \emph{{Multipartite Reflected Entropy}},
  \href{http://dx.doi.org/10.1007/JHEP10(2019)102}{\emph{JHEP} {\bf 10} (2019)
  102}, [\href{http://arxiv.org/abs/1909.03154}{{\tt 1909.03154}}].

\bibitem{Balasubramanian:2024ysu}
V.~Balasubramanian, M.~J. Kang, C.~Murdia and S.~F. Ross, \emph{{Signals of
  multiparty entanglement and holography}},
  \href{http://arxiv.org/abs/2411.03422}{{\tt 2411.03422}}.

\bibitem{Maldacena:1997re}
J.~M. Maldacena, \emph{{The Large N limit of superconformal field theories and
  supergravity}}, \href{http://dx.doi.org/10.1023/A:1026654312961,
  10.1023/A:1026654312961}{\emph{Adv.Theor.Math.Phys.} {\bf 2} (1998)
  231--252}, [\href{http://arxiv.org/abs/hep-th/9711200}{{\tt
  hep-th/9711200}}].

\bibitem{Sekino:2008he}
Y.~Sekino and L.~Susskind, \emph{{Fast Scramblers}},
  \href{http://dx.doi.org/10.1088/1126-6708/2008/10/065}{\emph{JHEP} {\bf 10}
  (2008) 065}, [\href{http://arxiv.org/abs/0808.2096}{{\tt 0808.2096}}].

\bibitem{Shenker:2013pqa}
S.~H. Shenker and D.~Stanford, \emph{{Black holes and the butterfly effect}},
  \href{http://dx.doi.org/10.1007/JHEP03(2014)067}{\emph{JHEP} {\bf 03} (2014)
  067}, [\href{http://arxiv.org/abs/1306.0622}{{\tt 1306.0622}}].

\bibitem{Maldacena:2015waa}
J.~Maldacena, S.~H. Shenker and D.~Stanford, \emph{{A bound on chaos}},
  \href{http://dx.doi.org/10.1007/JHEP08(2016)106}{\emph{JHEP} {\bf 08} (2016)
  106}, [\href{http://arxiv.org/abs/1503.01409}{{\tt 1503.01409}}].

\bibitem{Witten:1998qj}
E.~Witten, \emph{{Anti-de Sitter space and holography}}, {\emph{Adv. Theor.
  Math. Phys.} {\bf 2} (1998) 253--291},
  [\href{http://arxiv.org/abs/hep-th/9802150}{{\tt hep-th/9802150}}].

\bibitem{Witten:1998zw}
E.~Witten, \emph{{Anti-de Sitter space, thermal phase transition, and
  confinement in gauge theories}}, {\emph{Adv.Theor.Math.Phys.} {\bf 2} (1998)
  505--532}, [\href{http://arxiv.org/abs/hep-th/9803131}{{\tt
  hep-th/9803131}}].

\bibitem{Srednicki:1994mfb}
M.~Srednicki, \emph{{Chaos and Quantum Thermalization}},
  \href{http://dx.doi.org/10.1103/PhysRevE.50.888}{\emph{Phys. Rev. E} {\bf 50}
  (3, 1994) }, [\href{http://arxiv.org/abs/cond-mat/9403051}{{\tt
  cond-mat/9403051}}].

\bibitem{Cotler:2016fpe}
J.~S. Cotler, G.~Gur-Ari, M.~Hanada, J.~Polchinski, P.~Saad, S.~H. Shenker
  et~al., \emph{{Black Holes and Random Matrices}},
  \href{http://dx.doi.org/10.1007/JHEP05(2017)118}{\emph{JHEP} {\bf 05} (2017)
  118}, [\href{http://arxiv.org/abs/1611.04650}{{\tt 1611.04650}}].

\bibitem{Shapourian:2020mkc}
H.~Shapourian, S.~Liu, J.~Kudler-Flam and A.~Vishwanath, \emph{{Entanglement
  Negativity Spectrum of Random Mixed States: A Diagrammatic Approach}},
  \href{http://dx.doi.org/10.1103/PRXQuantum.2.030347}{\emph{PRXQuantum} {\bf
  2} (2021) 030347}, [\href{http://arxiv.org/abs/2011.01277}{{\tt
  2011.01277}}].

\bibitem{Kudler-Flam:2021efr}
J.~Kudler-Flam, V.~Narovlansky and S.~Ryu, \emph{{Negativity spectra in random
  tensor networks and holography}},
  \href{http://dx.doi.org/10.1007/JHEP02(2022)076}{\emph{JHEP} {\bf 02} (2022)
  076}, [\href{http://arxiv.org/abs/2109.02649}{{\tt 2109.02649}}].

\bibitem{Akers:2021pvd}
C.~Akers, T.~Faulkner, S.~Lin and P.~Rath, \emph{{Reflected entropy in random
  tensor networks}},
  \href{http://dx.doi.org/10.1007/JHEP05(2022)162}{\emph{JHEP} {\bf 05} (2022)
  162}, [\href{http://arxiv.org/abs/2112.09122}{{\tt 2112.09122}}].

\bibitem{Akers:2022max}
C.~Akers, T.~Faulkner, S.~Lin and P.~Rath, \emph{{The Page curve for reflected
  entropy}}, \href{http://dx.doi.org/10.1007/JHEP06(2022)089}{\emph{JHEP} {\bf
  06} (2022) 089}, [\href{http://arxiv.org/abs/2201.11730}{{\tt 2201.11730}}].

\bibitem{Akers:2022zxr}
C.~Akers, T.~Faulkner, S.~Lin and P.~Rath, \emph{{Reflected entropy in random
  tensor networks. Part II. A topological index from canonical purification}},
  \href{http://dx.doi.org/10.1007/JHEP01(2023)067}{\emph{JHEP} {\bf 01} (2023)
  067}, [\href{http://arxiv.org/abs/2210.15006}{{\tt 2210.15006}}].

\bibitem{Akers:2023obn}
C.~Akers, T.~Faulkner, S.~Lin and P.~Rath, \emph{{Entanglement of purification
  in random tensor networks}},
  \href{http://dx.doi.org/10.1103/PhysRevD.109.L101902}{\emph{Phys. Rev. D}
  {\bf 109} (2024) L101902}, [\href{http://arxiv.org/abs/2306.06163}{{\tt
  2306.06163}}].

\bibitem{Akers:2024pgq}
C.~Akers, T.~Faulkner, S.~Lin and P.~Rath, \emph{{Reflected entropy in random
  tensor networks III: triway cuts}},
  \href{http://arxiv.org/abs/2409.17218}{{\tt 2409.17218}}.

\bibitem{Hayden:2016cfa}
P.~Hayden, S.~Nezami, X.-L. Qi, N.~Thomas, M.~Walter and Z.~Yang,
  \emph{{Holographic duality from random tensor networks}},
  \href{http://dx.doi.org/10.1007/JHEP11(2016)009}{\emph{JHEP} {\bf 11} (2016)
  009}, [\href{http://arxiv.org/abs/1601.01694}{{\tt 1601.01694}}].

\bibitem{Harrow:2013nib}
A.~W. Harrow, \emph{{The Church of the Symmetric Subspace}},
  \href{http://arxiv.org/abs/1308.6595}{{\tt 1308.6595}}.

\bibitem{Lubkin:1978nch}
E.~Lubkin, \emph{{Entropy of an n-system from its correlation with a
  k-reservoir}}, \href{http://dx.doi.org/10.1063/1.523763}{\emph{J. Math.
  Phys.} {\bf 19} (1978) 1028}.

\bibitem{10.1093/acprof:oso/9780199535255.001.0001}
S.~Boucheron, G.~Lugosi and P.~Massart, \emph{{Concentration Inequalities: A
  Nonasymptotic Theory of Independence}}.
\newblock Oxford University Press, 02, 2013,
  \href{http://dx.doi.org/10.1093/acprof:oso/9780199535255.001.0001}{10.1093/acprof:oso/9780199535255.001.0001}.

\bibitem{PhysRevLett.86.5184}
M.~A. Nielsen and J.~Kempe, \emph{Separable states are more disordered globally
  than locally},
  \href{http://dx.doi.org/10.1103/PhysRevLett.86.5184}{\emph{Phys. Rev. Lett.}
  {\bf 86} (May, 2001) 5184--5187}.

\bibitem{2006math.ph...9050M}
F.~{Mezzadri}, \emph{{How to generate random matrices from the classical
  compact groups}},
  \href{http://dx.doi.org/10.48550/arXiv.math-ph/0609050}{\emph{arXiv e-prints}
  (Sept., 2006) math--ph/0609050},
  [\href{http://arxiv.org/abs/math-ph/0609050}{{\tt math-ph/0609050}}].

\bibitem{Penington:2022dhr}
G.~Penington, M.~Walter and F.~Witteveen, \emph{{Fun with replicas:
  tripartitions in tensor networks and gravity}},
  \href{http://dx.doi.org/10.1007/JHEP05(2023)008}{\emph{JHEP} {\bf 05} (2023)
  008}, [\href{http://arxiv.org/abs/2211.16045}{{\tt 2211.16045}}].

\bibitem{VanRaamsdonk:2010pw}
M.~Van~Raamsdonk, \emph{{Building up spacetime with quantum entanglement}},
  \href{http://dx.doi.org/10.1007/s10714-010-1034-0,
  10.1142/S0218271810018529}{\emph{Gen. Rel. Grav.} {\bf 42} (2010)
  2323--2329}, [\href{http://arxiv.org/abs/1005.3035}{{\tt 1005.3035}}].

\bibitem{Maldacena:2013xja}
J.~Maldacena and L.~Susskind, \emph{{Cool horizons for entangled black holes}},
  \href{http://dx.doi.org/10.1002/prop.201300020}{\emph{Fortsch. Phys.} {\bf
  61} (2013) 781--811}, [\href{http://arxiv.org/abs/1306.0533}{{\tt
  1306.0533}}].

\bibitem{Penington:2019npb}
G.~Penington, \emph{{Entanglement Wedge Reconstruction and the Information
  Paradox}}, \href{http://dx.doi.org/10.1007/JHEP09(2020)002}{\emph{JHEP} {\bf
  09} (2020) 002}, [\href{http://arxiv.org/abs/1905.08255}{{\tt 1905.08255}}].

\bibitem{Almheiri:2019psf}
A.~Almheiri, N.~Engelhardt, D.~Marolf and H.~Maxfield, \emph{{The entropy of
  bulk quantum fields and the entanglement wedge of an evaporating black
  hole}}, \href{http://dx.doi.org/10.1007/JHEP12(2019)063}{\emph{JHEP} {\bf 12}
  (2019) 063}, [\href{http://arxiv.org/abs/1905.08762}{{\tt 1905.08762}}].

\bibitem{Almheiri:2019hni}
A.~Almheiri, R.~Mahajan, J.~Maldacena and Y.~Zhao, \emph{{The Page curve of
  Hawking radiation from semiclassical geometry}},
  \href{http://dx.doi.org/10.1007/JHEP03(2020)149}{\emph{JHEP} {\bf 03} (2020)
  149}, [\href{http://arxiv.org/abs/1908.10996}{{\tt 1908.10996}}].

\bibitem{Penington:2019kki}
G.~Penington, S.~H. Shenker, D.~Stanford and Z.~Yang, \emph{{Replica wormholes
  and the black hole interior}},
  \href{http://dx.doi.org/10.1007/JHEP03(2022)205}{\emph{JHEP} {\bf 03} (2022)
  205}, [\href{http://arxiv.org/abs/1911.11977}{{\tt 1911.11977}}].

\bibitem{Almheiri:2019qdq}
A.~Almheiri, T.~Hartman, J.~Maldacena, E.~Shaghoulian and A.~Tajdini,
  \emph{{Replica Wormholes and the Entropy of Hawking Radiation}},
  \href{http://dx.doi.org/10.1007/JHEP05(2020)013}{\emph{JHEP} {\bf 05} (2020)
  013}, [\href{http://arxiv.org/abs/1911.12333}{{\tt 1911.12333}}].

\bibitem{Chu:2019etd}
J.~Chu, R.~Qi and Y.~Zhou, \emph{{Generalizations of Reflected Entropy and the
  Holographic Dual}},
  \href{http://dx.doi.org/10.1007/JHEP03(2020)151}{\emph{JHEP} {\bf 03} (2020)
  151}, [\href{http://arxiv.org/abs/1909.10456}{{\tt 1909.10456}}].

\bibitem{Yuan:2024yfg}
M.-K. Yuan, M.~Li and Y.~Zhou, \emph{{Reflected multi-entropy and its
  holographic dual}},  \href{http://arxiv.org/abs/2410.08546}{{\tt
  2410.08546}}.

\bibitem{Peres:1996dw}
A.~Peres, \emph{{Separability criterion for density matrices}},
  \href{http://dx.doi.org/10.1103/PhysRevLett.77.1413}{\emph{Phys. Rev. Lett.}
  {\bf 77} (1996) 1413--1415},
  [\href{http://arxiv.org/abs/quant-ph/9604005}{{\tt quant-ph/9604005}}].

\bibitem{Horodecki:1996nc}
M.~Horodecki, P.~Horodecki and R.~Horodecki, \emph{{On the necessary and
  sufficient conditions for separability of mixed quantum states}},
  \href{http://dx.doi.org/10.1016/S0375-9601(96)00706-2}{\emph{Phys. Lett. A}
  {\bf 223} (1996) 1}, [\href{http://arxiv.org/abs/quant-ph/9605038}{{\tt
  quant-ph/9605038}}].

\bibitem{Zyczkowski:1998yd}
K.~Zyczkowski, P.~Horodecki, A.~Sanpera and M.~Lewenstein, \emph{{On the volume
  of the set of mixed entangled states}},
  \href{http://dx.doi.org/10.1103/PhysRevA.58.883}{\emph{Phys. Rev. A} {\bf 58}
  (1998) 883}, [\href{http://arxiv.org/abs/quant-ph/9804024}{{\tt
  quant-ph/9804024}}].

\bibitem{Eisert:1998pz}
J.~Eisert and M.~B. Plenio, \emph{{A Comparison of entanglement measures}},
  \href{http://dx.doi.org/10.1080/09500349908231260}{\emph{J. Mod. Opt.} {\bf
  46} (1999) 145--154}, [\href{http://arxiv.org/abs/quant-ph/9807034}{{\tt
  quant-ph/9807034}}].

\bibitem{Simon:1999lfr}
R.~Simon, \emph{{Peres-Horodecki Separability Criterion for Continuous Variable
  Systems}}, \href{http://dx.doi.org/10.1103/PhysRevLett.84.2726}{\emph{Phys.
  Rev. Lett.} {\bf 84} (2000) 2726--2729},
  [\href{http://arxiv.org/abs/quant-ph/9909044}{{\tt quant-ph/9909044}}].

\bibitem{Plenio:2005cwa}
M.~B. Plenio, \emph{{Logarithmic Negativity: A Full Entanglement Monotone That
  is not Convex}},
  \href{http://dx.doi.org/10.1103/PhysRevLett.95.090503}{\emph{Phys. Rev.
  Lett.} {\bf 95} (2005) 090503},
  [\href{http://arxiv.org/abs/quant-ph/0505071}{{\tt quant-ph/0505071}}].

\bibitem{Calabrese:2012ew}
P.~Calabrese, J.~Cardy and E.~Tonni, \emph{{Entanglement negativity in quantum
  field theory}},
  \href{http://dx.doi.org/10.1103/PhysRevLett.109.130502}{\emph{Phys. Rev.
  Lett.} {\bf 109} (2012) 130502}, [\href{http://arxiv.org/abs/1206.3092}{{\tt
  1206.3092}}].

\bibitem{Bhosale:2012ldd}
U.~T. Bhosale, S.~Tomsovic and A.~Lakshminarayan, \emph{{Entanglement between
  two subsystems, the Wigner semicircle and extreme-value statistics}},
  \href{http://dx.doi.org/10.1103/physreva.85.062331}{\emph{Phys. Rev. A} {\bf
  85} (2012) 062331}.

\bibitem{Lu:2020jza}
T.-C. Lu and T.~Grover, \emph{{Entanglement transitions as a probe of
  quasiparticles and quantum thermalization}},
  \href{http://dx.doi.org/10.1103/PhysRevB.102.235110}{\emph{Phys. Rev. B} {\bf
  102} (2020) 235110}, [\href{http://arxiv.org/abs/2008.11727}{{\tt
  2008.11727}}].

\bibitem{Dong:2021clv}
X.~Dong, X.-L. Qi and M.~Walter, \emph{{Holographic entanglement negativity and
  replica symmetry breaking}},
  \href{http://dx.doi.org/10.1007/JHEP06(2021)024}{\emph{JHEP} {\bf 06} (2021)
  024}, [\href{http://arxiv.org/abs/2101.11029}{{\tt 2101.11029}}].

\end{thebibliography}\endgroup
\bibliographystyle{JHEP}

\end{document}